\pdfoutput=1
\documentclass[aps,pra,twocolumn,superscriptaddress,nofootinbib,floatfix,longbibliography]{revtex4-2}
\usepackage{amsmath}
\usepackage{latexsym}
\usepackage{amssymb}
\usepackage{bm}
\usepackage{color}
\usepackage[usenames,dvipsnames,svgnames]{pstricks} 
\usepackage{mathtools}
\usepackage{amsfonts}
\usepackage{amsthm}
\usepackage{graphicx}
\usepackage{hyperref}
\usepackage{tikz}
\usepackage{soul,xcolor}
\setstcolor{red}

\usepackage{braket}
\usepackage{ dsfont }
\usepackage[utf8]{inputenc}

\usepackage{lmodern}
\DeclarePairedDelimiter\floor{\lfloor}{\rfloor}

\begin{document}
\title{Revisiting thermodynamics in computation and information theory}
\author{Pritam Chattopadhyay}
\email{pritam.cphys@gmail.com}
\affiliation{Cryptology and Security Research Unit, R.C. Bose Center for Cryptology and Security, Indian Statistical Institute,
Kolkata 700108, India}
\author{Goutam Paul}
\email{goutam.paul@isical.ac.in}
\affiliation{Cryptology and Security Research Unit, R.C. Bose Center for Cryptology and Security, Indian Statistical Institute,
Kolkata 700108, India}



\begin{abstract}
One of the primary motivations of the research in the field of computation is to optimize the cost of computation. The major ingredient that a computer needs is the energy to run a process, i.e.,  the thermodynamic cost. The analysis of the thermodynamic cost of computation is one of the prime focuses of research. It started back in the seminal work of Landauer where it was commented that the computer spends $k_BT \, \ln 2$  amount of energy to erase a bit of information (here $T$ is the temperature of the system and $k_B$ represents Boltzmann's constant). The advancement of statistical mechanics has provided us with the necessary tools to understand and analyze the thermodynamic cost for the complicated processes that exist in nature, even the computation of modern computers. The advancement of physics has helped us to understand the connection of statistical mechanics (the thermodynamics cost) with computation. Another important factor that remains a matter of concern in the field of computer science is the error correction of the error that occurs while transmitting information through a communication channel.

Here in this article, we have reviewed the progress of the thermodynamics of computation starting from Landauer's principle to the latest model, which simulates the modern complex computation mechanism. After exploring the salient parts of computation in computer science theory and information theory, we have reviewed the thermodynamic cost of computation and error correction. We have also discussed the alternative computation models that have been proposed with thermodynamically cost-efficient. 
\end{abstract}

\maketitle 

\tableofcontents
\newpage
\section{Introduction}\label{sec1}
 In the 19th century, it was Carnot, whose desire to develop a better steam engine that would help France to win the  Napoleonic Wars gave birth to thermodynamics~\cite{clausius1960motive}, and this led in a way to explain further inventions in this direction.  According to Landauer, every real-world computational process has some thermodynamic cost~\cite{landauer1991information} which means that, when we perform computation in the real world, some physical changes (entropy production or some heat generation) occur in the world. Landauer's limit uncovered the physical consequences of computation. All naturally occurring processes like biological computers and even man-made computers have thermodynamic costs. It is quite fascinating to analyze the difference in the thermodynamic cost of the naturally occurring process and the artificial ones that are created. Translation of RNAs into amino acids is one such natural biological process where one encounters energy costs for the execution of the process. Work in this direction~\cite{kempes2017thermodynamic}, has shown that the thermodynamics cost of this biological process is more efficient than the artificial process.

Out of all such artificial processes, one of the important ones is the existing digital computer system (where a computer can be conveyed as an array of bits- $0$'s and $1$'s and a process to map one such configuration to another). It can be considered an engine that dissipates energy for performing mathematical and logical tasks. According to the earlier scientists, they had a notion that there must be some fundamental thermodynamic limit to the efficiency of such engines irrespective of their hardware structure.  According to  Von-Neumann~\cite{von1966theory}, a computer operated at a temperature $T$ must dissipate at least $k_BT\, \ln 2$ amount of heat (where $k_B$ represents the Boltzmann's constant). The thought experiment of Brillouin~\cite{brillouin2013science} boils down to the same conclusion as that of Von-Neumann with some error probability. But now, it is a well-known fact that today's computers can perform a large amount of reliable computation per $k_BT$ of energy dissipation.  Though reliable computation can be executed per $k_BT$, but today's computer dissipates a vast quantity of energy compared to $k_BT$. Volatile memory devices (such as TTL flip-flops) are the reasons behind this huge waste. The volatile devices dissipate energy even when they are not being used. The macroscopic nature of the existing computers is one of the basic reasons for the inefficiency in the context of energy. Due to its macroscopic nature, the amount of energy required to trigger the system is quantitatively high, and this energy is dissipated instead of reused for the next pulse. It is similar to the case of applying brakes to stop a moving car by saving its kinetic energy. One of the spectacular thermodynamically reversible computation models is the ballistic model, proposed by Fredkin and Toffoli~\cite{fredkin1982conservative}. Other models~\cite{bennett1973logical,keyes1970minimal,likharev1982classical} were developed which were more physically realistic than that of Fredkin's version.

From Landauer's principle, we know that when a bit of information is written on a memory device the entropy of the system decreases by $k_B T \, \ln 2$, so at least the same amount of work has to be done. In the works~\cite{bennett1982thermodynamics,bennett1989time}, they have stated that we can have systems that are able to compute with no lower bound with respect to the thermodynamic cost. With the theoretical advancement of stochastic thermodynamics~\cite{seifert2012stochastic} and information thermodynamics~\cite{parrondo2015thermodynamics}, a better understanding of the physical foundation of information processing is possible. These frameworks are though concentrated on proto-computation~\cite{diana2013finite,lan2012energy,ouldridge2017thermodynamics,barato2015thermodynamic}, few attempts are taken to apply information thermodynamics to computer science. However, the works~\cite{wolpert2015extending,strasberg2015thermodynamics}, suggest that one can consider the fundamental limit to the thermodynamics cost of computation while analyzing the system in the quasi-static limit. Though it sounds interesting, it overlooked the practical aspect of the performance measure in real-world computation. In the article~\cite{chu2018thermodynamically}, they have explored beyond this quasi-static model of computation. In their work, they have taken into account a particular model, i.e., finite-state machine to probe the thermodynamic cost of computation, which is executed within a finite time and produces the correct output with probability near to one.

One can also describe the thermodynamic cost for a computation process as a sum of the energy that is required to provide an extra bit in the due course of the computation process, plus the energy required to destroy~\cite{bennett1982thermodynamics} the generated garbage bits. This part of the computation is non-reversible, and so according to Landauer’s principle, this part of the computation process dissipates heat. The ``Fundamental Theorem"~\cite{maroney2010does} provides the upper and the lower bounds over the thermodynamic cost of the computation process. It is also stated that a slow computational process releases less amount of energy~\cite{bennett1982thermodynamics}. In the article~\cite{li1992mathematical}, they have provided proof of the existence of this statement mathematically, that with the increase of the time of computation, there exists a time-energy trade-of hierarchy for diminishing the energy cost of the process. From real-life examples, we know that garbage needs to be compressed and this needs time.

A different aspect of the computational process was formulated by Schulman. Schulman~\cite{schulman2005computer}, along with the help of Landauer~\cite{landauer1961irreversibility}, was able to convey that the intrinsic arrow of the computational processes is aligned with the thermodynamic arrow. So, one can consider the computer as a system without an independent arrow of time having the capability of retaining the past/future distinction with a thermodynamic arrow in a particular direction. Similar to Schulman's proposed idea, suggestions in the same direction can be found in~\cite{hawking1985arrow,hawking1993no}. According to these previous works, the total entropy of the system increases when the computer records something in memory. So it can be concluded that computers remember things in the direction of time where the system encounters an entropy increase. Whereas in the work~\cite{hawking1994no}, they have expressed that in a universe where entropy decreases with time the computer memories will work in a backward direction.

Another aspect of computation is the analysis of the Turing machine (TM). Turing machines have a deep impact in quantum information~\cite{cubitt2015undecidability,cubitt2012extracting} and quantum computation~\cite{deutsch1985quantum,benioff1982quantum,nielsen2010quantum}. Efforts to model Turing machines~\cite{strasberg2015thermodynamics}, and to develop limit on computation~\cite{wolpert2015extending,hani1998trna,boyd2017transient}, has been analyzed. Computing machines in the real world will belong to the regime, which is bounded by the infinite dissipation and the zero-energy limit. The intuitive idea that the foundation of physics can be restricted due to some properties of TM~\cite{dawson2012kurt,aaronson2005guest} has been explored.  In addition to that, it has been thoroughly investigated whether all functions that are implemented by physical systems can also be computed by TM~\cite{arrighi2012physical,piccinini2011physical,pitowsky1990physical,rescher1968many}, which is so-called ``physical Church-Turing thesis". Even the thermodynamic properties considered for the physical system that implements TM are also studied. Two different physical perceptions of TM~\cite{kolchinsky2020thermodynamic} are analyzed. The first perception is based on thermodynamic reversibility. Whereas the second perception is to produce, that much amount of heat which is allowed by the laws of physics, assuming that it holds the ``Church-Turing thesis".  For every perception, three thermodynamic costs are taken into consideration. The first thermodynamic cost is the heat that is produced when a TM is working. This term is coined as ``heat function". The second one is the ``thermodynamic complexity" which refers to the minimum heat that is produced when TM runs with an input, and the third one is the heat generation where it is considered that a TM is working to compute a distribution of inputs. For both, the perception one encounters that, the thermodynamic complexity for an output of the TM computation is bounded by a constant. One can also assume that deterministic computing machines exist which dissipate heat without violating the thermodynamic laws. Works in this direction~\cite{chu2018thermodynamics} describe that deterministic computation is thermodynamically consistent. Deterministic computation means it will execute computation and will provide a correct output that is indistinguishable in nature from other processes. The process will be executed within a finite time and is universal in nature.

In the complexity theory, it is a well-known fact that the deterministic polynomial time (P) complexity classes are not equal to the non-deterministic polynomial-time (NP)~\cite{cook2023complexity,lipton2013people} complexity classes. The existence of the one-way bijection~\cite{razborov1994natural,goldreich2001foundations} proves this statement. The existence of such bijection arises due to the self-referential character of P $=?$ NP~\cite{fortnow2003p}. Different approaches to prove the existence of one-way bijection mathematically have resulted in a debacle. The existence of one-way-ness has a great impact on the complexity class problems, as one can link this to the physical constraints rather than its mathematical limitations~\cite{aaronson2005guest,zurek1989thermodynamic,aaronson2013philosophers}. In the seminal work~\cite{de2014one}, they have proposed an approach to bridge a connection between the one-way permutation to thermodynamic computation. They have shown that a quantum circuit maps a target bit to itself~\cite{raussendorf2001one}. During the execution of this process, it encounters entropy constraints which can be easily inverted when the system is immersed in an adiabatic heat bath.

Along with the increase of interest in investigating the connection of thermodynamics with computation, the connection with information-theoretic notions can gain importance.  Now, information theory is treated as physical science, and the bridge between information theory and thermodynamics provides a trade route between them. The investigation to connect thermodynamics with information theory is highly pursued. Claude Shannon in his seminal paper~\cite{shannon1948mathematical,shannon1957certain}, proposed the modern aspect of information theory, where he developed a mathematical definition for information and then proved his noiseless and noisy coding theorem.  The quantum version of Shannon's theory is provided by Schumacher in this work~\cite{schumacher1995quantum}, where it is conveyed that entropy plays an important part in quantum information theory. Maxwell in his seminal work coined the term ``Maxwell demon"~\cite{knott1911life}, who has the capability of monitoring the position and momentum of individual molecules in a chamber, which is divided into two parts by a partition. The mean velocity of the air particles is uniform whereas for the individual molecules it varies. The demon has the power to open the aperture of the partition of the two-chamber to allow the fast-moving molecules into one chamber and the slower ones into the alternative chamber. So, the question that emerges in one's mind is that, whether it is possible to develop a machine that will exploit statistical fluctuations to convert heat into work. Smoluchowski~\cite{smoluchowski1927experimentell,smoluchowski1967gultigkeitsgrenzen} proposed a way to implement the automatic mechanism with the help of a spring-loaded trapdoor at the aperture between the two chambers. The prime objective of the seminal work of Szilard~\cite{szilard1929entropieverminderung} was to investigate the conditions that would allow the construction of a motion machine. Landauer~\cite{landauer1961irreversibility} and Bennett~\cite{bennett1982thermodynamics} in their work have conveyed that the measurement can be considered as a reversible process and it can be executed without any entropy cost. This had a great impact on the thermodynamics of computation. Works~\cite{bennett1973logical,fredkin1982conservative,toffoli1980reversible} in this direction have shown that efficient computers are reversible which is equivalent to a Carnot engine.

Earman and Norton in their work~\cite{earman1998exorcist,earman1999exorcist}, propose the rejection of the information-theoretic notion of Maxwell's demon. They suggest that if we consider the demon as a thermodynamic system supervised by the second law, then no such assumption about information and entropy is required to save the second law of thermodynamics. The alternative way is that one can take the demon not to be a part of the system. In this type of situation, no information-theoretic assumption is required to protect the second law of thermodynamics.  The argument of Earman and Norton is wrong. The process to execute a measurement does not require any entropy cost, whereas, for the case of erasure of information from the memory register, it requires a minimum entropy cost~\cite{bub2001maxwell}.

Reprocessing of information using the mechanism of copying is a fundamental process in the natural world. This is even true for all living systems. Devices that we use also need to copy information to its better functioning. While replicating the information, there is a high possibility of committing an error. The accuracy of a copy thus relies on its accurate reproduction. It can be quantified by counting the wrongly copied bits while executing. Though the error can be reduced at the macro level, at the molecular level, perfect copying is not achievable due to thermal fluctuations. It is the primary source of error at the molecular level. The replication process is limited by the thermal noise, so it must be interpreted in terms of thermodynamics as proposed by Von Neumann~\cite{von1956probabilistic}. Now, one fundamental question pokes into one's mind, whether one can develop a connection between the errors that occur when copying with the physical quantities that describe the copying process.  Generally, while executing a copying process, it has to undergo various intermediate steps to control the accuracy and speed of the process. This is true for artificial as well as natural scenario~\cite{johnson1993conformational}.  As we are trying to explain the errors from thermodynamics laws (second law), we have to take into account that the copying process is cyclically repeated than a single shot~\cite{bennett1973logical}.

One can develop a framework to understand the thermodynamics where one replicates information. In this framework, we consider the replication process which is highly complex in characteristics~\cite{hopfield1974kinetic,ninio1975kinetic,murugan2012speed,murugan2014discriminatory}, and the operation of replication is executed in a cyclic way, following the method developed in previous works~\cite{bennett1979dissipation,andrieux2008nonequilibrium,cady2009open,andrieux2009molecular,esposito2010extracting,sartori2013kinetic,andrieux2013information,gaspard2014kinetics}. Works in the direction~\cite{sartori2015thermodynamics,korepin2002thermodynamic} to bridge a connection of thermodynamics with errors, will show that the error in copying protocols has a relation with thermodynamic observables, which can characterize the errors. An alternative approach to connect thermodynamics and information theory is built on the basis of using entanglement in quantum systems~\cite {horodecki2001balance,popescu1997thermodynamics,rohrlich2001thermodynamical}.

In this article, we will present an overview of the impact of thermodynamics in the field of computation, mainly the artificial process (computer science theory)~\cite{hopcroft2001introduction,lewis1998elements,hromkovivc2003theoretical} and the information theory~\cite{cover1999elements,kullback1997information,gallager1968information}. For the information theory, our primary focus will be on the error-correcting codes. To get a clear and vivid idea of the topics considered in this work, it is worth mentioning the related topics that are not considered in this review article. In this work, we have considered some specific computational models. The effect of thermodynamics on analog computers and the dynamical systems (e.g. cellular automata) is not studied here. If the readers are interested in this direction they are suggested to go through the works~\cite{diamantini2016landauer} for the analysis in the analog system and works~\cite{baez2012algorithmic} which investigates the application of statistical physics in dynamical systems. Interested readers can go through various surveys in the field of computation and information. A general survey in quantum computation is done in~\cite{lo1998introduction}, whereas~\cite{deffner2017quantum} reviews quantum speed limits. Discussion for the analysis of the fundamental limitation of the physical process is shown in~\cite{pour1982noncomputability,moore1990unpredictability,lloyd2000ultimate,lloyd2017uncomputability}. Thermodynamical analysis of biochemical processes is not analyzed in this work, one can see~\cite{ouldridge2017fundamental,ouldridge2018importance,brittain2019biochemical,sartori2014thermodynamic,hasegawa2018multidimensional,mehta2012energetic,mehta2016landauer} to get an idea of this direction. Similarly, we have not considered the modeling of computational machines in the field of computer science with biochemical models. See~\cite{prohaska2010innovation,bryant2012chromatin,benenson2012biomolecular,angluin2006stably,chen2014deterministic,dong2012bisimulation,soloveichik2008computation,rondelez2016dna} for works in this direction. Thermodynamics of the system that interacts with the environment~\cite{touchette2004information,touchette2000information,barato2017thermodynamic,sagawa2008second,sagawa2012nonequilibrium,wilming2016second,large2021stochastic,gingrich2016near,horowitz2017information} is investigated, which is out of the scope of this review article. So, in this article, our primary focus is to visualize the impact that thermodynamics has on computation and information theory.

This article is structured as follows: in Section~\ref{sec2}, we have described some basic notations that are followed throughout the article. In section~\ref{sec3}, we have reviewed some salient features of computation in computer science theory. We have described the finite automata with a suitable example. Then we have described the basic definition of the Turing machine and the universal  Turing machine. In section~\ref{sec4}, we have explored the salient features of information theory, and then we have moved on to the analysis of classical and quantum error correction. Section~\ref{sec5} is focused on the review of the alternative models of computation that were proposed, which are thermodynamically cost-efficient. Ballistic Computer and Brownian computer are the two alternative models of computation that are discussed in detail in this section. We have dedicated the section~\ref{sec6} to explore the mathematical modeling of the thermodynamics of computation. There was a nice debate on whether the thermodynamic arrow of time is equivalent to the computational arrow of time. Schulman in his work conveyed that they are equivalent but, in recent work, the authors have shown this is not necessary that the thermodynamic arrow of time should be equivalent to the computational arrow of time without violating any physical principles. In section~\ref{sec12}, we have thoroughly discussed this. In section~\ref{sec14}, we have thoroughly analyzed the thermodynamic interpretation for finite-state automata. In this model,  the author has considered a modified Bennett algorithm to analyze the thermodynamics of finite automata. Then we moved on to explore the thermodynamic cost for the computation in a Turing machine. In section~\ref{sec16}, we have conveyed a seminal work that provided a thermodynamic proof for the existence of one-way computation. In the next section~\ref {sec18}, we have reviewed the thermodynamic interpretation of error correction. We have first explored the seminal work which has explained the classical as well as the quantum error correction using Maxwell's demon model. Then we have discussed an alternative method for the analysis of error correction. The authors have considered the spin chain model as an alternative model to analyze and relate quantum error correction with thermodynamics. Section~\ref{sec20} is dedicated to the investigation of the thermodynamic cost of the alternative model of computation. In their work, they have conveyed that Brownian computers are not thermodynamically reversible ones. Some miscellaneous topics in this direction have been reviewed in section~\ref{sec22}. We have concluded our article in section~\ref{sec24} with some discussion about the progress in this direction and its future prospects.

\section{Notations}\label{sec2}
For any defined set $\mathbb{S}$, the collections of the set of a finite string of the elements from $\mathbb{S}$ is denoted by $\mathbb{S}^\star$. In computer, we use $\mathbb{S}=\{0,1\}$. Here $\mathbb{S}^\star$ denotes the set of all binary strings.

The Kronecker delta function is defined as 
\begin{eqnarray}\label{eq1}
\delta_{ij}=\left\{
\begin{array}{lc}
  1, &\  i = {j} ,\\
  0, & \  otherwise.
\end{array}\right.
\end{eqnarray} 
This shows that the function takes value  $1$ or $0$ based on the fact that $\beta$ has the same properties with $\alpha$ or not. The symbols $ \alpha$ and $\beta$ represent two variables.

We will consider the output $1$ as Boolean `TRUE' and consider the $0$ output as `FALSE'. For any Boolean function $g(a)$ the indicator function is expressed as 
 \begin{eqnarray}\label{eq2}
\mathbb{I}(g)=\left\{
\begin{array}{lc}
  1, &\  g(a) = {1} ,\\
  0, & \  otherwise.
\end{array}\right.
\end{eqnarray} 

We will denote the length of a finite string $\mathbb{M}$ as $l(\mathbb{M})$. The concatenation of two strings $\mathbb{M}$ and $\mathbb{M'}$ is defined as $\mathbb{M}\mathbb{M'}$. For ceiling operators, we will use $\lceil \bullet \rceil$,  and for flooring operators $\floor*{\bullet}$ wherever required.

In probability theory, the random variables are described by upper case letters, and the instances of these random variables are expressed by the lower case letters. Given a set $A$, the distribution over this set can be expressed as $p(a)$. Now, there exists any  $A' \subseteq A$ for which we can write $p(A')= \sum_{a\in A'} p(a)$. For a distribution $p$ over a set $A$ the conditional distribution $\pi(y|a)$ is defined as
\begin{equation}\label{eq3}
(\pi p)\, y := \sum_{a\in A} \pi(y|a) \, p(a),
\end{equation}
where $\pi p$ is the distribution over $Y$.

\section{Some Basic Aspects of Computation Theory}\label{sec3} 

Theoretical computer science covers different areas starting from the algorithm, data structure, and computation to computational number theory.  In this section, we are going to study about two of the basic computation aspects of computer science. In the further section, we will see the impact of thermodynamics on these aspects. From a computational perspective, the system is defined in this form to describe that it is one of the members of the Chomsky hierarchy~\cite{hopcroft2001introduction}.

\subsection{Finite Automata}

The formal definition of the finite automaton is:

\textbf{Definition 1.} A finite automaton is a 5-tuple $M = (Q,\, \Sigma,\, \delta,\, q, \, F)$. Here,
\begin{itemize}
\item $Q$ represents the finite set. The elements of this set are called the states.

\item $\Sigma$ also represents a finite set. The elements of this set is called the alphabet.  

\item $\delta: \,\, Q \times \Sigma \rightarrow Q$ represents a function. It is called the transition function. 

\item $q \in Q$. It represents the start state.

\item $F \subseteq Q$ describes the set of final accepting states. 
\end{itemize}

One can infer from the definition that, the transition function denoted by $\delta$ represents the program of the finite automaton $M = (Q,\, \Sigma,\, \delta,\, q, \, F)$. The transition function describes how the system evolves for each step. Let us consider a state of $Q$ as $a$, and $b$ be the alphabet from the set $\Sigma$. When the state of the system (i.e. finite automaton) $M$ is at $a$ and it reads the alphabet $b$, then $\delta (a,b)$ describes the switch over of the state $a$ to the state $b$. The input words accepted by the automaton specify a language that is described as a regular language, i.e., any finite language (words of finite length) can be called a regular language. So, we can conclude that a finite automaton has the power to compute input words of arbitrary length.  

\begin{figure}[h]
  \includegraphics[width=1.0\columnwidth]{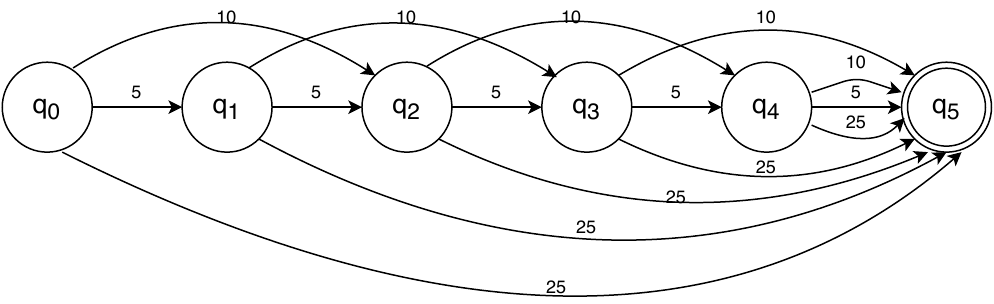}
  \caption{Transition diagram of the system.}
  \label{fig1}
  \end{figure}

We will first consider an example of an automaton that appears naturally. Let us consider a \textit{toll gate}. Now, we will design a computer that will control this \textit{toll gate}. The gate from the toll pass remains closed until the vehicle pays the required amount. For example, we consider that when the driver pays $25$ bucks the toll gate opens, and the driver is allowed to go. We assume that there exist only three sets of coins: $5$, $10$, and $25$ bucks. Whenever a driver appears at the toll gate, he will insert coins into the machine. So, the machine decides whether it will open the gate or not, in other words, whether the driver has fulfilled the payment of the required amount (i.e. $25$ or more). The machine can be in any of the six states based on which it will decide whether it will open the gate or not. The six states of the machine during the execution of the process are:

\begin{itemize}
\item If the machine does not collect any money, the machine state is described by $q_0$.

\item When the machine receives 5 bucks, the machine moves to the state $q_1$.

\item Now, when it gets 10 bucks, the state of the machine is $q_2.$

\item When it has 15 bucks,  the state is $q_3$.

\item When it has 20 bucks,  the state is $q_4$.

\item When it has 25 bucks,  the state is $q_5$.
\end{itemize}

Now, let us consider a situation where a vehicle comes to the toll gate, then the machine will be at its initial state. Now the driver inserts 25 bucks in the sequence (5, 5, 10, 5). The state of the machines evolves through the process as follows:

\begin{itemize}
\item After the driver inserts 5 bucks into the machine, the machine evolves from the state $q_0$ to $q_1$.

\item Similarly, the driver inserts the next 5 bucks into the machine, and the machine now moves from the state $q_1$ to the state $q_2$.

\item After that the driver inserts 10 bucks, the machine jumps by two steps and moves to the state $q_4$ from state $q_2$.

\item Finally, the driver inserts its last 5 bucks, and the machine moves to its final state $q_5$. The toll gate opens up. We assume that the driver provides that exact amount else the machine responds to abort and restart the process. 
\end{itemize}

The state diagram with all the combinations is shown in Fig.~\ref{fig1}. Here, the state $q_5$ is depicted by two circles. The two-circle state represents the final state of the system (or the halt state). Now, if the machine reaches this state, the gate will open otherwise the gate remains closed. In this case, one can observe that the machine only has to remember the state where it belongs at any instant of time.

Finite automata have a great impact on different fields of study which include computer science, biology, mathematics, logic, linguistics, engineering, and even philosophy. Finite automaton has a wide scale of applications in computer science like designing hardware, designing compilers, network protocols, and in computation.

Finite automata have different forms like deterministic and stochastic (alternatively called ``probabilistic automaton") finite automaton. In the stochastic automaton, the single-valued transition function $\delta$ will be replaced by a conditional distribution. One can observe multiple accept states which are described in the literature as `terminal states' for the system. Even one can encounter multiple start states in the process.

\subsection{Turing Machine}
Alan Turing in his seminal work~\cite{turing1936computable}, first coined the term Turing Machine. He conveyed that it is an abstract computation device, which will help to investigate the extent and as well the limitation of what we can compute. It was devised mainly for the computation of real numbers. So, the renowned form of the computational machine studied in computer science is Turing machines~\cite{hopcroft2001introduction,savage1998models}. It is generally conveyed in the literature that one can model every computational machine using a Turing machine. Church-Turing thesis conveys the statement a bit formally, it states that ``A function on the natural numbers is computable by a human being following an algorithm, ignoring resource limitations, if and only if it is computable by a Turing machine." It was further modified in the ` Physical Church-Turing thesis', where it has been conveyed that the set of functions that one can compute by utilizing the mechanical algorithmic method, which abides by the laws of physics~\cite{pour1982noncomputability,moore1990unpredictability,wolpert2019stochastic,arrighi2019overview,wuthrich2015quantum}, is also computable with the help of Turing machine.

Various forms of definitions of the Turing machine exist in the literature which are computationally equivalent to each other. As the various definitions are equivalent to each other, computation done in one type of Turing machine can be executed equivalently in other forms of the Turing machine. The formal definition of the Turing machine is

\textbf{Definition 2.}  A Turing machine is defined by 7-tuple ($Q$, $\Lambda$, $\Gamma$, $\delta$, $q_0$, $q_a$, $q_r$). Here, 

\begin{itemize}
\item $Q$ is a finite set which describes the set of states.

\item $\Lambda$ is also a finite set which depicts the input alphabets.

\item $\Gamma$ represents a finite set of tape alphabet and $\Lambda \,\subset \,\Gamma$.

\item $\delta: Q\times  \Gamma \, \rightarrow \, Q \times \Gamma \times \{L,R, S\}$ is called the transition function. 

\item $q_0$ represents the start state of the Turing machine.

\item $q_a$ is called the accepted state or in other words halting state.

\item $q_r$ is called the rejected state.
\end{itemize}

Here, $\{L, R, S\}$ describes the direction of the movement of the head of the tape. Based on the command, the head moves left, right, or stays in the same position of the tape. In other equivalent computational definitions of the Turing machine, one can encounter multiple sets of accepting states, in other words, halting states.

\begin{figure}[h]
  \includegraphics[width=1.0\columnwidth]{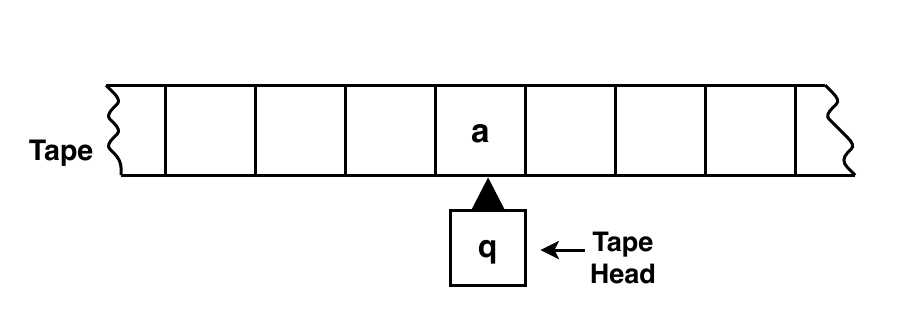}
  \caption{A schematic of TM. The tape of a TM is an infinite tape whose state is specified by $q$. The infinite tape is divided into equidistance square boxes filled with tape alphabets. The TM will scan the tape with its tape head. The tape head has access to move to the left or right of the tape.}
  \label{fig2}
  \end{figure} 

At each step of the computation, the state of the TM reads the alphabet in the square where the tape head is placed and subsequently moves on to a new state $q'$. It writes a new alphabet ($a'$) on the tape, and then it moves its tape head either to the left or to the right. This process is repeated until the system attains the accepted state. Mathematically this map can be expressed as $(q, a) \rightarrow \, (q', a',  d)$, where $d$ denotes the right or left movement of the tape head.  For a given TM, the arguments of the transition states are called ``instantaneous descriptions" (IDs) of the TM. One can also sometimes encounter TM's with no halts.

A function $f$ is called recursive, if there exists a TM with input alphabets $x$ where $x \in \Gamma^{\star}$, for which the TM computes $f(x)$. Similarly, the function $f$ is a partial recursive function, if we are able to compute $f(x)$ using a TM, where $x$ is the input alphabets of the TM. Turing has conveyed in his work that all functions are not recursive, which remains a fundamental limitation. 

Turing machine has a great impact on the analysis of computational complexity~\cite{hopcroft2001introduction,moore2011nature}. One of the profound open problems in mathematics that remains a concern for the exploration of the Turing machine is whether $P = \, NP$. The limitations of mathematics, like G\"odel's incompleteness theorem, have a deep intimacy with the computational device theory of Turing machine, and even in some parts of philosophy as well~\cite{copeland2013computability}. As all devices are physical, it has been argued in some works~\cite{dawson2012kurt,aaronson2005guest} that one might bring some restriction to the foundation of physics by utilizing some properties of Turing machines.

It is a well-known fact that there exist many variants of TM. Here, we have described the formal definition of a single tape. One of the popular variants of TM is the multiple tapes, where one of these tapes contains the input of the computational model, and one of them contains the output of the TM when it reaches its halting state. Other intermediate tapes in the multiple tape TM are called the `work tapes' that are used as scratch pads. It is a more complicated variant than that of the single tape. It is used in the literature as it is easier to prove theorems than single-tape TMs. The computational power for both of these variants is equivalent. The interesting fact about these two variants is that one can convert a multiple tape TM into a single tape TM and similarly a single tape TM into a multiple one~\cite{sipser1996introduction,arora2009computational,li2008introduction}. One of the variants of TM is the universal TM, which has the computational power to compute any other TM. In general, it is considered that TM is a formal structure of algorithm whereas UTM (which takes other TM's as input) is a formal structure of a computer.

\section{Basics Aspects of Information Theory}\label{sec4}
In this section, we will discuss some basic concepts of information theory that are required for our analysis, and then we move on to discuss the error correction theory for the classical system as well as the quantum system. Thermodynamics has an impact on various aspects of quantum information theory. Here we are concerned about the error correction as it has a great impact on computational as well as communication aspects.

\subsection{Notion on Information Theory}
Here, we will guide ourselves through some of the basic aspects of information theory. The Shannon entropy over a set $T$ is defined as
\begin{equation}\nonumber \label{aw1}
S(T) = -\sum_{t\in T} \, p(t) \,\, log_b \, p(t),
\end{equation}
where for the cases when $b=e$ (nats) it will be generally conveyed as $\ln$ throughout the article. For other conditions like $b=2$ (bits) and so on it will be specifically mentioned in the analysis.

Using the above definition, we can describe the conditional entropy of a random variable $X_r$, which is conditioned on another random variable $Z_r$. The mathematical representation of conditional entropy is 
\begin{eqnarray}\nonumber \label{aw2}
S(X_r|Z_r) & = & \sum_{z\in Z_r} \, p(z) \,\, S(X_r|z)\\ \nonumber
& = &\sum_{z\in Z_r,\,  x \in X} \, p(z) \,\,p(x|z)\,\, \ln\, p(x|z).
\end{eqnarray}

Similarly, one can define the mutual information for two random variables $X_r$ and $Z_r$ as 
\begin{eqnarray}\nonumber \label{aw3}
I_p(X_r;Z_r) & =  & S (X_r) + \, S (Z_r) - S (X_r,Z_r) \\ \nonumber
& = & S (X_r) - \, S (X_r|Z_r).
\end{eqnarray}

We have described some of the basic definitions of information theory. Readers who are interested in having a deep understanding of information theory (both in classical and quantum systems) can go through the referred books~\cite{cover1999elements,gray2011entropy,mackay2003information,wilde2013quantum,hayashi2016quantum}.

\subsection{Classical Error Correction}

In a communication process, the data is transmitted from the sender to the receiver end through a channel that is prone to noise, i.e., it is transmitted through a noisy channel. The data string is a sequence of $0$'s and $1$'s. The string to be communicated is encoded with an additional number of bits (redundant bits). In the receiver's end, the receivers reconstruct the actual message by decoding and examining the corrupted message. This reconstruction process is conveyed as decoding.

In the late 40's of the 20th century, the seminal work of Shannon~\cite{shannon1949mathematical} led to the foundation of this field and was extended by Hamming in his work~\cite{hamming1950error}. Since then, this field has gained its importance for developing better communication.  The extent to which error correction of the missing bits is possible depends on the design of the error-correcting code. Generally, there exist two types of error-correcting codes, they are \textbf{block code} and \textbf{convolutional code}. We will be mainly focusing on the linear code of the block code. There are other models of error correction codes that are not covered here, interested readers can go through~\cite{macwilliams1977theory,hoffman1991coding} for further information.

The formal definition of the error correcting code is defined as: 
\\
\textbf{Definition 3.} The error-correcting code can be defined as an injecting map from $n$ symbols (messages bits) to $m$ symbols:
\begin{equation}\nonumber
Enc: \Lambda_n \rightarrow \Lambda_m,
\end{equation}
where $\Lambda$ represents the set of symbols. 
\begin{itemize}
\item We define a variable $a$ which describes the cardinality of our alphabet set ($a = |\Lambda|$). For binary code system $a=2$. 

\item The domain of the set, i.e., $\Lambda_n$ represents the message space, and $\Lambda_m$ represents the message that will be transmitted from the channel. Here $n$ denotes the message length.

\item Block length: Block length is denoted by $m$, it describes the message which is mapped to $m$-bit strings. 

\item Code: The message to be transmitted will be encoded with a codeword. This codeword is called the code. In general, $m\geq n$.

\item Rate: It is defined as the ratio of $n$ over $m$. it describes the efficiency of the protocol. 
\end{itemize}

Now we move on to analyze the linear code protocol. A linear code is a type of error correction code, where the linear combination of codewords also represents a codeword. The channel through which we will communicate is a binary channel, generally known as a binary symmetric channel. In this type of noisy channel, the bit gets affected independently. The symmetric nature of the channels conveys that the channels cause errors $0 \rightarrow 1$ and $1 \rightarrow 0$ with equal probability. Let us consider a message of $n$ symbols as $u= u_1 u_2 \dots u_n$ which are encoded into a codeword $x = x_1 x_2 \dots x_m$. Here, the first part of the codeword is the original message (i.e. $x_{1\rightarrow m} =u_1 u_2 \dots u_n$) to be transmitted and the remaining are the check symbols (i.e. $x_{n+1}\dots x_m$) which satisfies the condition 
\begin{equation}\nonumber \label{a1}
H_p\, x = 0,
\end{equation} 
where the matrix $H_p$ ($(m-n) \times m$) is called the parity check matrix for the code.

\begin{figure}[h]
  \includegraphics[width=1.0\columnwidth]{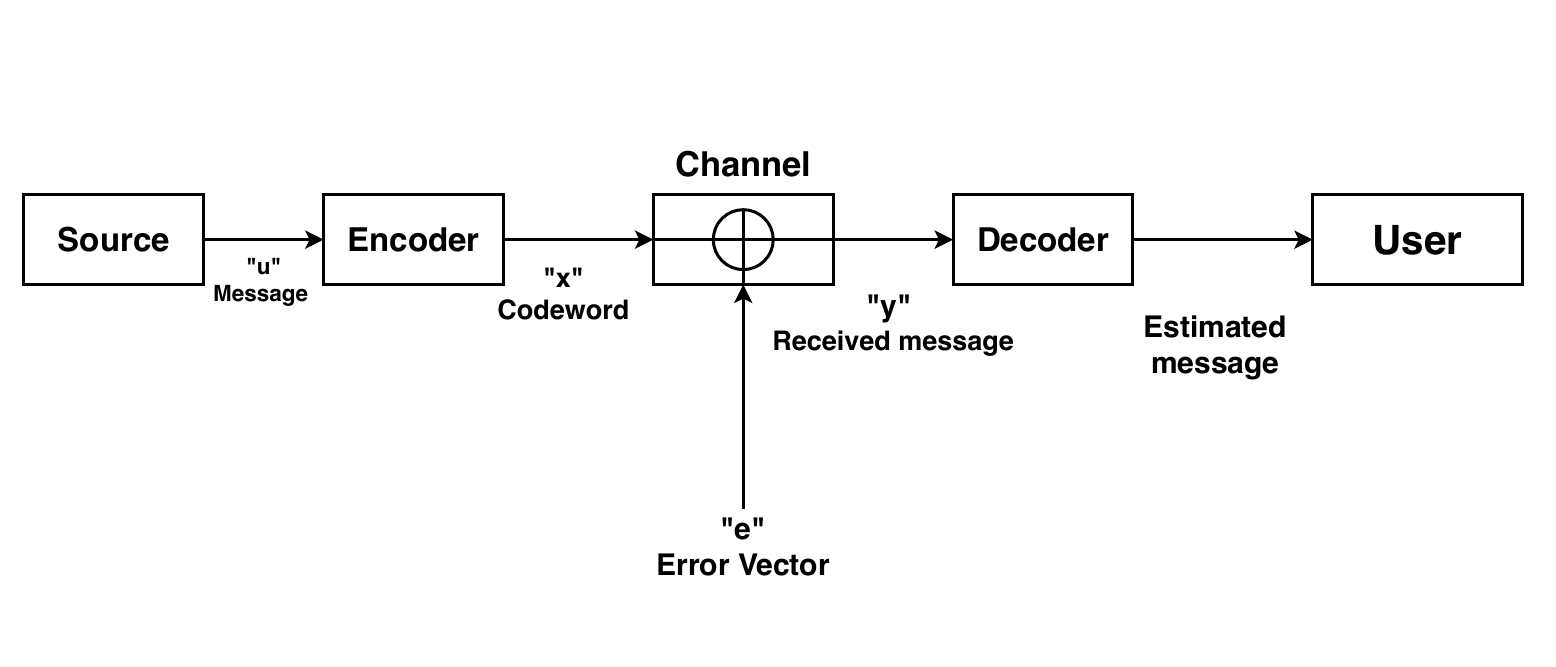}
  \caption{A schematic representation of the whole process in a communication system is depicted here. The channel is the communication medium through which the message is transmitted from the source to the receiver end.}
  \label{fig3}
  \end{figure}

This codeword is now transmitted through the noisy channel. The receiver at the receiver's end, receives a different message say $y = y_1 y_2 \dots y_m$ which is quite different from $x$. Let us denote the error in the message as $e = y-x=e_1 e_2 \dots e_m$. Here, $e_j =0$ with probability $1-p$ will depict that the $j$ symbols in the message are correct whereas, $e_j=1$ with probability $p$ conveys that the $j$ symbol is incorrect. So in the decoding process, the receiver has to identify the message $u$ from the received message $y$, which actually boils down to identifying the error vector $e$. The decoder will try to choose the most likely error vector so that it can reduce the probability of making a mistake in decoding the correct message.  This is generally defined as maximum likelihood decoding. The whole communication process is shown in Fig~\ref{fig3}.

So let us analyze how the decoder chooses the most likely error vector when it receives the message. Before we proceed we will discuss two important definitions. 

\textbf{Definition 4.} \textit{Hamming distance}: The hamming distance between two given vectors $x= x_1 x_2 \dots x_m$ and $y= y_1y_2 \dots y_m$ is described by number of positions the corresponding vectors differs. It is mathematically denoted by 

\begin{equation}\nonumber \label{a2}
Dis(x,y) = |{k: y_k \neq x_k}|.
\end{equation}

\textbf{Definition 5.} \textit{Hamming weight}: The hamming weight is defined as the number of non-zero $x_j$ in a vector $x$. It is denoted by $wt(x)$.

From the above two definitions, we can infer that $Dis(x,y) = wt(y-x)$. Based on the above given definitions we will formalize a parameter. The parameter is called the minimum hamming distance, which is defined as $Dis_{min} = min \, Dis(x,y) = min \, wt(x-y)$. So the strategy of the decoder is to take that error $e$, which will have the least weight. This process is called nearest neighbor decoding. The decoder will compare the message $y$ with all the possible combinations of the codewords, and then select the closest one. This brute force process is fine for small $n$. If $n$ is very large this process becomes tedious, and coding theory aims to develop new schemes to decode these messages faster.

A linear code is generally called a $[m,n,d]$ code, where $m$ describes the length of the codeword, $n$ denotes the length of the message string, and $d$ describes the minimum hamming distance.

\subsection{Quantum Error Correction}

Here in this section, we will discuss error correction in the quantum regime. Classical error correction is a well-developed theory based on the demand for better communication systems. One-to-one mapping from classical to quantum error correction is not possible as the quantum world has some constraints of its own. Qubit (quantum bits) are bounded by the no-cloning principle, which states that in quantum information theory one cannot copy a state which is possible in the case of classical information theory. Another aspect that we encounter in the quantum regime is that the wave function collapses when a measurement is performed over the state. These constraints were the reasons that make the quantum world unique from the classical, and these pose a challenge to the feasibility of quantum computing. It was the seminal work of Peter Shor~\cite{shor1995scheme}, where they proposed the first quantum error correction protocol. Shor in his work has demonstrated that quantum information can be encoded by exploiting the idea of entanglement of qubits. Works in this direction~\cite{preskill1998reliable,kitaev1997quantum,knill1998resilient,gottesman1998theory}, have demonstrated that one can suppress the error rate in the quantum regime provided the qubits meet some physical conditions.  Interested readers can go through the reviews~\cite{gottesman2010introduction,devitt2013quantum,lidar2013quantum,terhal2015quantum} in this direction which covers quantum error correction and its subfields. In this article, we will describe the basic intuition of quantum error correction that we will explore in the latter half from a thermodynamics point of view.

\vspace{0.2in}
\begin{center}
\textit{Quantum errors in digitalize form}
\end{center}

\vspace{.15in}
In the classical world, the units of information are bits, which belong to set $\mathbb{S}$ whereas, in the case of quantum the units of information are the qubits. The general definition of a qubit state  is
\begin{equation}\label{b1}
|\psi\rangle = \alpha |0\rangle + \beta |1\rangle,
\end{equation}
where $\alpha$ and $\beta$ represents complex number satisfying the condition $|\alpha|^2 + |\beta|^2 = 1$. So, a qubit has the power to encode in the superposition of the computational basis states which are denoted by  $|0\rangle$ and  $|1\rangle$. The only error that is possible in the classical regime is the bit flip, i.e., $0 \rightarrow 1$ or viz versa. In the quantum realm, the qubit state is continuous in nature. So, this property of the qubits is the main challenge in developing the error correction code in the quantum world.  The qubits are subjected to an infinite number of errors.

To get a clear idea about the errors let us  describe the qubit state defined in Eq.~\eqref{b1} as:
\begin{equation}\label{b2}
|\psi\rangle = \cos \frac{\theta}{2} \,|0\rangle + e^{i\phi} \, \sin \frac{\theta}{2} \, |1\rangle.
\end{equation}
This is the geometric representation of the qubits. The condition of the probability amplitude is maintained as defined.  The qubit state in Eq.~\eqref{b2} is described by a point in the Bloch sphere.

There can be errors due to a variety of physical processes. The rotation of the qubit from one point to another in a Bloch sphere will cause an error. Mathematically one can denote this type of error by a unitary operator $U(\Delta \theta, \Delta \phi)$ which will evolve the qubit state expressed in Eq.~\eqref{b2} to 

\begin{equation}\label{b3}
U(\Delta \theta, \Delta \phi)\, |\psi\rangle= \cos \frac{\theta + \Delta \theta}{2} \,|0\rangle + e^{i(\phi+\Delta \phi)} \, \sin \frac{\theta+ \Delta \theta}{2} \, |1\rangle.
\end{equation}

The error that is caused by the rotation of the qubits shows that it is continuous in its nature. But we are quite lucky that we can digitalize this error with the help of the Pauli operators (i.e. Pauli matrices). The rotation process described in Eq.~\eqref{b3} can be described in terms of the Pauli matrices as

\begin{equation}\label{b4}
U(\Delta \theta, \Delta \phi)\, |\psi\rangle= a_0 \sigma_0 \, |\psi\rangle + a_1 \sigma_1 \, |\psi\rangle + a_2 \sigma_2 \, |\psi\rangle + a_3 \sigma_3 \, |\psi\rangle, 
\end{equation} 
where $a_i$ ($i=0,1,2,3$) represents the coefficients, and $\sigma_i$ ($i=0,1,2,3$) the respective Pauli operators. So, any error due to this rotation of the qubit can be described by the Pauli operators $[\mathbb{I}, \sigma_x, \sigma_y (\sigma_x\sigma_z), \sigma_z]$. The error correction code, which will have the computational power to correct the error caused by the Pauli matrices will make the message flawless. This procedure causes the digitalization of the error, which will have a great impact on the quantum error correction code.

\vspace{0.2in}
\begin{center}
\textit{Types of Quantum error}
\end{center}

\vspace{.15in}
The error counts in the quantum realm boil down to two fundamental errors due to the digitization of the errors. The quantum code has to encounter these two types of errors. One is the $X$-type error ($X$ describes the Pauli matrix $\sigma_x$), and the other is the $Z_{pau}$-type error ($Z_{pau}$ describes the Pauli matrix $\sigma_z$). The $X$-type error is a bit flip error, which is similar to the classical errors where the state $|0\rangle \rightarrow |1\rangle$ when operated by the Pauli $X$ operator and vice versa.

The other error, i.e., $Z_{pau}$-type error, causes a phase flip of the qubits. There is no classical counterpart of this type of error. Phase flip of the qubit state is described as $Z_{pau} |0\rangle = |0\rangle$ and $Z_{pau} |1\rangle = - |1\rangle$. 

Though the digitalization of the errors has reduced the error counts, we still have some challenges in the quantum world which are unique and have no classical analog. One of these challenges is that we cannot clone (xerox) a quantum state, i.e., we cannot construct a universal unitary operator (universal xerox machine) $U_{c}$, which can xerox or copy a state as 
\begin{equation}\nonumber \label{b5}
U_{c} |\psi\rangle \, |0\rangle = |\psi\rangle \,|\psi\rangle.
\end{equation}
Whereas, in the classical realm one has the power to copy a state as required.

The second challenge is that the message to be transmitted from quantum channels is exposed to both bit-flip and phase-flip errors simultaneously. So the quantum error correction code should be able to detect both these errors simultaneously. Along with these challenges, one has to keep in mind that in the quantum world measurement of a state causes wavefunction collapse, which we have no counterpart in classical theory. In classical systems, one has access to measure arbitrary properties without compromising information loss. 

\vspace{0.2in}
\begin{center}
\textit{Stabilizer Code}
\end{center}

\vspace{.15in}
In this section, we will study how to create a $[[m,n,d]]$ stabilizer code.  Here $m$ represents the total count of qubits, the count of the logical qubits is given by $n$, and $d$ describes the code distance. A stabilizer code encodes $n$ logical qubits into $m$ physical qubits. The stabilizer represents an abelian subgroup of the $m$-fold Pauli group. We have represented the notation of the quantum codes in double brackets to differentiate it from the classical code, which is shown by a single bracket. 

\begin{figure}[h]
  \includegraphics[width=1.0\columnwidth]{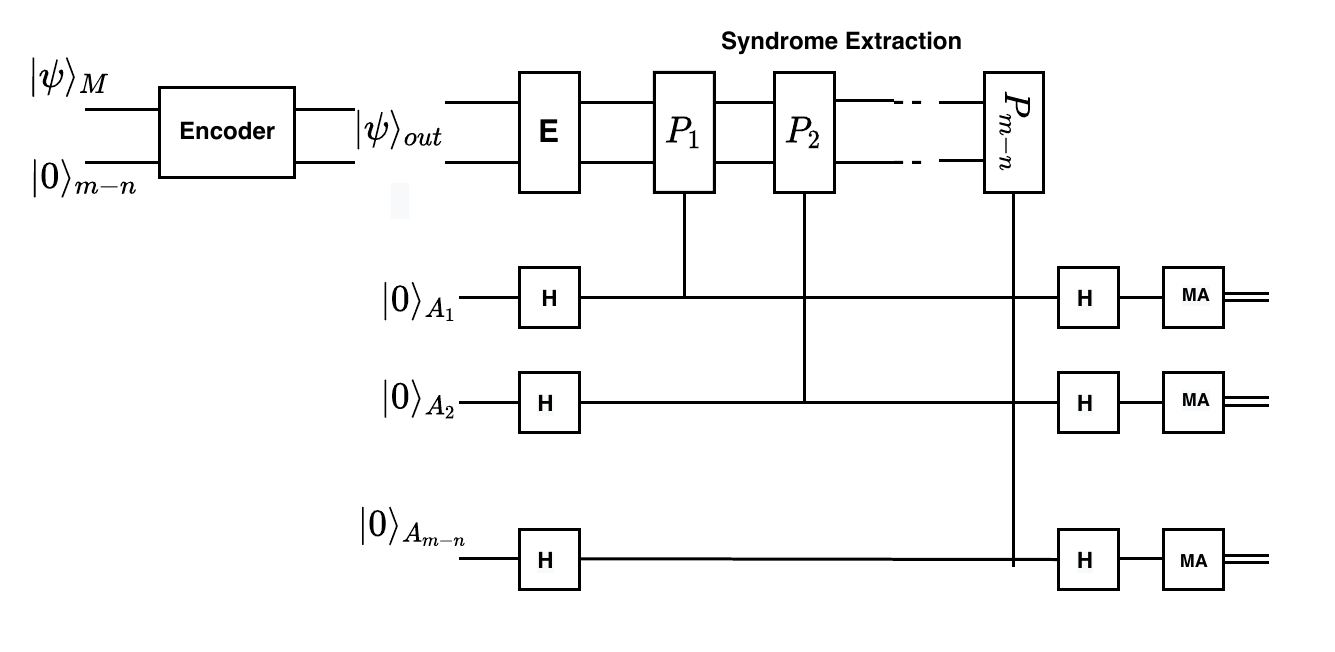}
  \caption{A schematic representation of the circuital analysis of $[[m,n,d]]$  stabilizer code is depicted here.  $|\psi\rangle_M$ represents the message, and $|0\rangle_{m-n}$ represents the ancilla qubits. $H$ is the Hadamard operator, and $MA$ denotes the measurement. }
  \label{fig4}
  \end{figure}

The structure of an $[[m,n,d]]$ stabilizer code is depicted in Fig.~\ref{fig4}. From the Fig.~\ref{fig4}, we can infer that $n$ qubits $|\psi\rangle_M$ are entangled with $m-n$ qubits $|0\rangle_{m-n}$ by an encoding operation.The output state after this process is $|\psi\rangle_{Out}$. So the data that was confined in $|\psi\rangle_M$ is now distributed in the expanded Hilbert space. To detect the error that occurs during the communication, $(m-n)$ stabilizer measurements are performed. For every stabilizer $P_j$ shown in Fig.~\ref{fig4}, the syndrome extraction process is described as

\begin{eqnarray}\label{b6} \nonumber
|\Psi\rangle \, |0\rangle_{m-n} & \xrightarrow[]{\text{syndrome extraction}} & \dfrac{1}{2}(\openone^{\otimes n} + P_j) |\Psi\rangle |{0}\rangle_{A_j} \\ \nonumber
& + & \dfrac{1}{2}(\openone^{\otimes n} - P_j)|\Psi\rangle |{1}\rangle_{A_j},
\end{eqnarray}
where $|\Psi\rangle = E |{\psi}\rangle_{M}$.
The commutation relation of the stabilizer $P_j$ with the error $E$ provides the measurement outcomes. If $P_j$ commutes with $E$ the ancilla returns $``0"$, whereas when it anti-commutes we get $``1"$  as the measurement outcome. The fabrication of a good code boils down to the fact that we have to find stabilizers that anti-commute with the errors.

\vspace{0.2in}
\begin{center}
\textit{Error correction with Stabilizer code}
\end{center}

\vspace{.15in}
So we are left with the process of decoding the message. The process is to find the best-fitted unitary operation $\mathbb{K}$, which would return the state to the codespace. The process of decoding the message will be successful if the action of the error $E$, and $\mathbb{K}$ on the codes state return the exact state, i.e., $\mathbb{K}E |\psi\rangle_{out} = (+1) |\psi\rangle_{out}$, and fails if it returns $\mathbb{K}E |\psi\rangle_{out} = S_a |\psi\rangle_{out}$, where $S_a$ represents a logical operator.

\section{Alternative Computation Model}\label{sec5}
 The computational models that are being explored in section~\ref{sec3} are not thermodynamically efficient ones.  In this section, we are going to analyze some alternative computational systems that are thermodynamically more efficient, i.e., the thermodynamic cost for the computation is quite less due to its reversible nature. We will first discuss the Ballistic computer, which was proposed by Fredkin and Toffoli~\cite{fredkin1982conservative}, and then describe its limitations. Following that, we will analyze the Brownian computer which utilizes thermal fluctuation to perform a computational process.

For the analysis of the ballistic and Brownian computer, the programming style requires a change in its form from irreversible operation to reversible one. The erasure principle of overwriting of data by other data cannot be addressed for these models. Ballistic computer fails totally to operate with irreversible operations but the Brownian computers have the power to tolerate a small amount of irreversibility in the logical operation. But it fails miserably for a large amount of irreversible operations.

\subsection{Ballistic Computer}

\begin{figure}[h]
  \includegraphics[width=1.0\columnwidth]{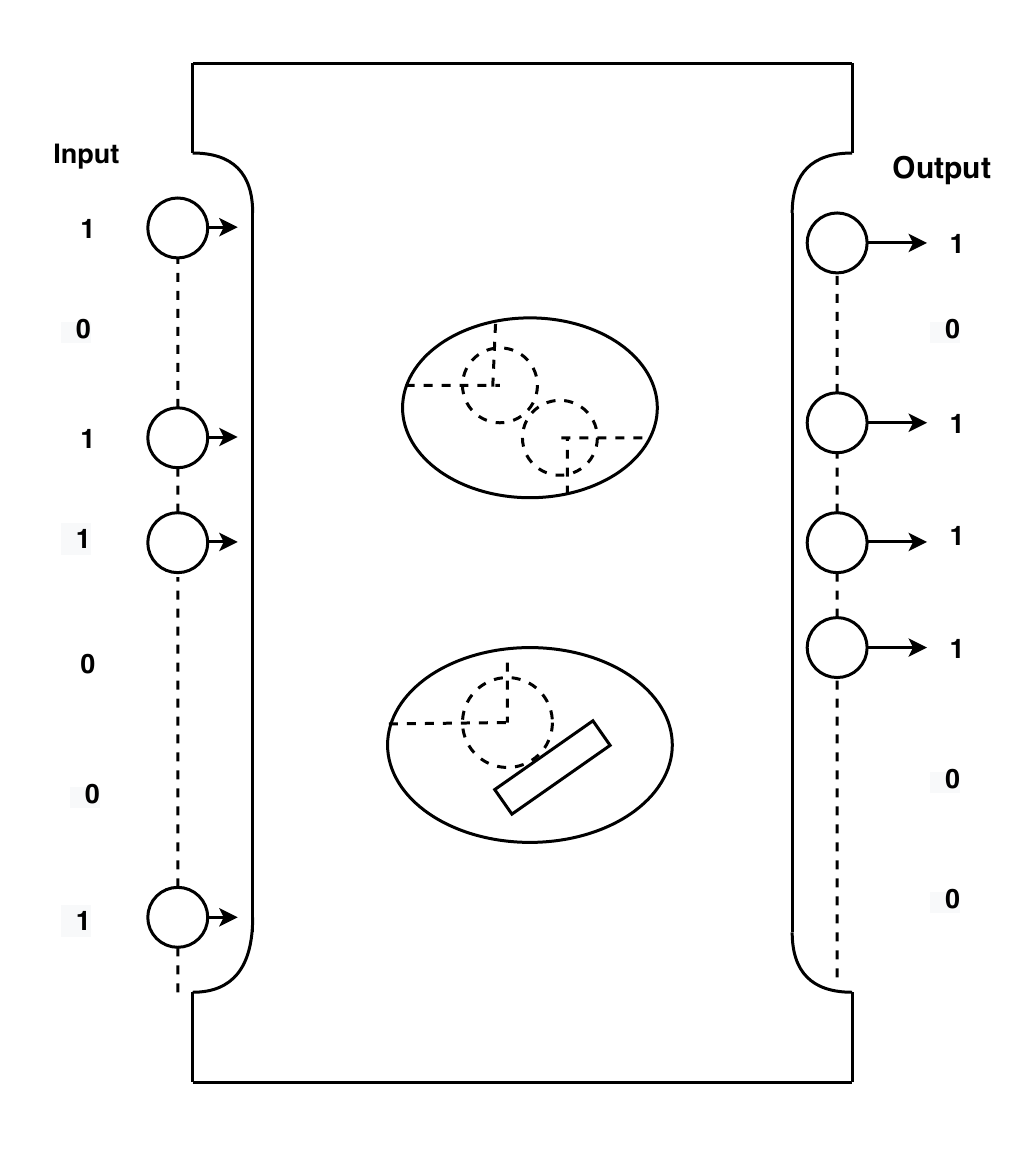}
  \caption{A schematic representation of a Ballistic computer proposed by Toffoli is depicted. The condition of having the same number of 1's in the input and the output should be satisfied, i.e., the boolean function should be conservative and reversible in nature.}
  \label{fig5}
  \end{figure}

The simple basic idea, that an idealized machine has the power to compute without dissipating any amount of kinetic energy forms the principle of the ``ballistic" computation model~\cite{fredkin1982conservative}. This model consists of a hard-sphere that collides with a fixed reflective barrier. In the input side of this computer model, we have a ``starting line" from which a huge number of hard spheres are fired with equal velocity. We will be considering a ball in the starting line if we encounter a `1' in the input, and in the cases where we get `0' there will be no ball in the starting line.  The computer has some mirrors inside it with which it collides. Due to this collision process, the ball changes its direction and collides with the other balls. All the collision processes are considered to be elastic in nature, and between the collisions, the ball moves in a straight line path with a constant velocity. The balls after a finite number of collisions reach their finishing point. This signifies the output of the computer. The presence of a ball in the output line is considered as $1$ in the output and the absence of it as $0$.   The mirror of this computer is equivalent to the logic gates of our digital computers, and the balls are equivalent to the signals. The pictorial representation of this computer is shown in Fig.~\ref{fig5}.

We can infer that this computer is unable to compute non-conservative Boolean functions. It has the power to implement functions whose output has an equal number of ones as in input, and it can implement bijective functions. But Toffoli has shown that one can transform any Boolean function into a conservative, bijective function. So, this model has the power to compute all functions. Though it conveys a decrease in the amount of cost in energy, we encounter some drawbacks of this setup. Two main drawbacks of this system are its sensitivity to small perturbative change, and the second one is related to the collision of the balls. It is quite difficult to make each collision elastic. A small error in the positions and velocities gets amplified after each collision by a factor of 2, so the trajectory of the balls becomes unpredictable after a few numbers of collisions. The stronger the noise source the sooner the trajectory is spoiled. To get better performance we have to eliminate these noises.

One of the ways to overcome this collision problem is to correct the instability in the velocity and position of the ball after the execution of each collision process. Though this provides a solution it makes the system thermodynamically irreversible, but due to its low energy cost it has some practical importance.  To culminate the effect of noise one can think of considering square balls instead of using spherical. It culminates in the exponential growth of the errors as the balls are always parallel to the wall and to each other. Although it seems possible it is quite unnatural as there is no proof of the existence of square atoms in nature. Quantum effects can stabilize the system from noise, but it will bring some new instability. The wave-packet spreading causes instability in the system in the quantum realm. Benioff~\cite{benioff1982quantum} in his work has discussed a quantum version of the ballistic computer, where he has proposed a way to culminate the effect of the noise due to the wave packet spreading by utilizing a time-independent Hamiltonian.

\subsection{Brownian computer}
 We can infer from the previous section that thermal randomization is inevitable, so it is unavoidable. If it is unavoidable one can exploit this. Brownian computers~~\cite{bennett1982thermodynamics} are such a model that exploits these properties for computation. In this model, the trajectory of the dynamical part of the system is influenced by thermal randomization in such a way that it attains Maxwell velocity. The trajectory becomes equivalent to a random walk. Despite its chaotic nature, the Brownian computer is able to execute useful computations. The high potential barriers prevent the trajectory from escaping from the system. So, within these confined walls created by the potential barriers, the system performs a random walk in the forward direction of the computation.

The state transition for the Brownian computer happens due to the random thermal movement of the part that carries the information with it. Due to its random nature, the transition can backtrace (move backward) in the computational process, undoing the transition executed recently. In the macro regime, the execution of computation using a Brownian computer seems counter-intuitive, but this is an obvious situation in the micro regime. In the case of chemical reactions, we encounter such things where the Brownian motion of the particle of the reactants orients the reactants as required for the execution of the reaction. This is equivalent to the transition state for the computational process.

Bennett has shown that one can execute a Turing machine using this thermal randomness. It is made up of clockwork which is frictionless, and rigid in form. The parts of the clockwork Turing machine should be interlocked so that they have the freedom to jiggle around locally, but restricted from moving an appreciable amount for the execution of a logical transition. In computational complexity, Reif~\cite{reif1979complexity} considered a similar model to analyze \textbf{P = PSPACE}.

\begin{figure}[h]
  \includegraphics[width=1.0\columnwidth]{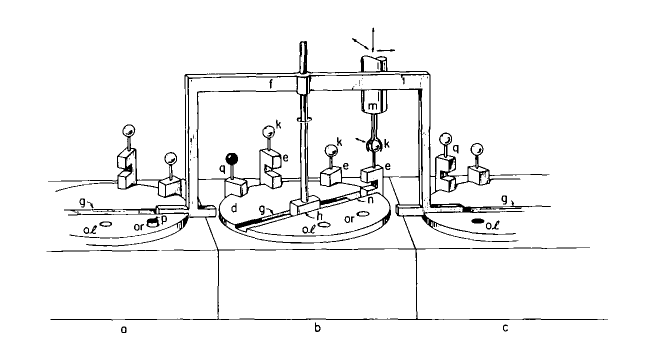}
  \caption{Pictorial representation of Brownian Turing machine. The symbols $(a, \, b,\, c)$ depict the Turing machine tape, and the read, write, and shift equipment is shown in the figure. [Adapted from Ref.~\cite{bennett1982thermodynamics}. Copyright 1982, Springer Nature]}
  \label{fig6}
  \end{figure}

In Fig.~\ref{fig6}, the framework of the Brownian Turing machine has been depicted. The head of the Turing machine is scanning the square $b$ of the tape. In this configuration, each tape is interlocked with some E-shaped bit storage device denoted by (e) in the figure. If it is placed up then it represents 1, if down then 0. The device which is shown in the figure as (f) restricts the movement of the two adjacent squares of the tape. To change the bits of the system the knob (k) guides the way to slide the bit storage device (e) up or down with the help of the manipulator (m). To constrain the disks on the square blocks of the tape that are not scanned, a special knob (q) is attached to each square of the tape. To execute the shifting operation in this model the screwdriver of the system allows the knob (g) to get aligned with the framework (f). Now the manipulator of the framework grasps a knob and moves it to the right or left of the scanned tape. In this whole set, no springs are allowed so as to prevent friction in the system. Friction is not allowed in the system as it leads to thermodynamic dissipation.

In Bennett's model of Brownian computation, it is considered that each of the computational states will possess a unique antecedent state. A unique antecedent state is only possible when the computer executes reversible operations like the \textbf{NOT} gate. So, the antecedent state of the present state with memory 0 is 1, and vice versa. For a computer that executes irreversible operations, the antecedent state can be 0 or 1 when the memory cells have a 0 value.

To acquire a unique antecedent state for each state of the system, the interlocking of the device should be executed in such a way that the device has one degree of freedom, i.e., the computational process of this device meander within this single degree of freedom. This model has limitations like it requires a configuration space with a huge accessible portion.

Bennett in his work has presumed that the driving force to execute the computation requires some energy gradient but, in the work~\cite{norton2013brownian}, they have shown that one can sufficiently drive the computation process with the entropic force.

\section{Mathematical foundation of thermodynamic computation}\label{sec6}
The prime focus of this section is to explore the closed mathematical form of the thermodynamic cost of computing a parameter $\zeta$ from $\eta$. Since long ago various methods for the effective realization of reversible computation have been proposed~\cite{keyes1970minimal,likharev1982classical,bennett1982thermodynamics,bennett1989time,landauer1961irreversibility,feynman2018simulating,feynman1985quantum}. Here we are going to study the complete mathematical theory~\cite{li1992mathematical} for this process. The fundamental theorem 
\begin{equation}
E_{\alpha}(\eta, \zeta) \approx K_{\alpha}(\eta|\zeta) + K_{\alpha} (\zeta|\eta),
\end{equation} 
proposed in the work~\cite{li1992mathematical}, provides the key for the mathematical analysis of thermodynamic computation. Here $E_{\alpha}(\eta, \zeta)$ denotes the thermodynamic cost for computing $\zeta$ from $\eta$, and $K_{\alpha}(\eta|\zeta)$ describes the shortest plausible description in the form of length for $\eta$ given $\zeta$.

So far many physical realizations have been proposed for reversible computation. The accepted viewpoint regarding the thermodynamic cost of a computation process, where one replaces input $\eta$ by the output data $\zeta$, is equated with the number of irreversible bits that are erased during the process. The cost of this each unit of the erased bit is $kT\, \ln \,2$. To analyze the thermodynamics of computation we will consider the following axioms.

\vspace{.1in}
\textbf{Axiom 1:} No thermodynamic cost for a reversible computation process. 

\vspace{.1in}
\textbf{Axiom 2:} Any irreversible process that occurs in a computation process has a thermodynamic cost.

\vspace{.1in}
\textbf{Axiom 3:} For a reversible computational process where one replace the input $\eta$ by the output $\zeta$, no reversible process occurs.

\vspace{.1in}
\textbf{Axiom 4:} All computations are effectively physical in nature.

\vspace{.1in}
The first three axioms defined here are based on the principles of physics whereas, the fourth axiom is a form of the \textit{Church's thesis}. The notion of the fourth axiom is that every physical computation is equivalent to the notion of computation executed in a Turing machine.

Before we move on to analyze the thermodynamic cost of computation we will consider some definitions that we need for the analysis.

\textbf{Definition 6}  The Kolmogorov complexity of $\eta$ given $\zeta$ is

$K_{\alpha}(\eta|\zeta) = min\{|p_c| +\, |W_i|  : \xi_i(<p_c, \zeta>) = \eta, \, p_c \in \{0,1\}^\star, \, i \in \mathbb{N} \}$.

Here $\eta,\, \zeta \, \in \mathbb{N}$, where $\mathbb{N}$ represents the natural number system and $|\bullet|$ describes the cardinality of the bit string. The enumeration of the Turing machine is represented by $W_i$ and the parameter $\xi_i$ describes the enumeration of the partial recursive function. So one can say $W_i$ commutes $\xi_i$. Here $\xi_i(<p_c, \zeta>) = \eta$ is an effective bijection from $N \times N$ to $N$. $\{0,1\}^\star$ corresponds to $0,1,00,01,\cdots$.

Now let us analyze the thermodynamic cost for the computation process where we compute $\zeta$ from $\eta$. We want to calculate the minimum cost for the computation process. 

\textbf{Theorem 1:} $E_{\alpha}(\eta, \zeta) = K_{\alpha}(\eta|\zeta) + K_{\alpha} (\zeta|\eta)$ upto logarithmic term. 

\vspace{.2in} 
\textbf{Proof 1:} We are going to establish the upper and the lower bound for the thermodynamic cost. 

\textbf{\textit{Claim:}} $E_{\alpha}(\eta, \zeta)\leq K_{\alpha}(\eta|\zeta)+ K_{\alpha} (\zeta|\eta) + 2[K_{\alpha}( K_{\alpha} (\zeta|\eta)|\zeta) + K_{\alpha}(K_{\alpha}(\eta|\zeta)|\eta)]$.

\vspace{.2in}
\textbf{\textit{Proof}} We divide the computation in three parts. In the first part of the program it computes $\zeta$ from $\eta$, and is depicted by $p_c$ and $|p_c| = K_{\alpha} (\zeta|\eta)$. In the second part of the program $q_c$ computes $ K_{\alpha}(\eta|\zeta)$ from $\eta$ and $|q_c| = K_{\alpha}(K_{\alpha}(\eta|\zeta)|\eta)$, and in the final  part of the program $r_c$ computes $K_{\alpha} (\zeta|\eta)$ from $\zeta$, and $|r_c| = K_{\alpha}( K_{\alpha} (\zeta|\eta)|\zeta)$. Let's now analyze the computation process step by step. 

\begin{itemize}
\item In the first step  of the computation process $p_c$ computes $\zeta$ from $\eta$ and leaves behind garbage bits $g_a(\eta,\zeta)$.

\item We copy $\zeta$, and then use one of its copy and the garbage bits to reverse the computation process to get $\eta$ and $p_c$. 

\item We copy $\eta$ and then use one of its copy along with $q_c$ to compute $K_{\alpha} (\eta|\zeta)$ along with the garbage bits. 

\item We execute a shortest program which we depict as $s_c$ to compute $\eta$ from $\zeta$ with the help  of $\eta$, $\zeta$, $K_{\alpha} (\eta, \zeta)$. In this process, extra garbage bits are produced. 

\item We copy $s_c$ and repeat the process shown in the third and fourth bullet. This helps to cancel the extra garbage bits. So we have $p_c$, $q_c$, $r_c$, $s_c$, $\eta$, $\zeta$. 

\item We copy $\zeta$ again and similarly use one of its copy to compute $K_{\alpha} (\zeta|\eta)$. It again results in garbage bits.  

\item We execute a shortest program $p_c$ to compute $\zeta$ from $\eta$ with the help  of $\eta$, $\zeta$, $K_{\alpha} (\zeta, \eta)$. In this process, some extra garbage bits are produced.

\item We delete a copy of $p_c$ and repeat the process shown in the sixth and seventh bullet. This helps to cancel the extra garbage bits. So we have $\zeta$, $r_c$, $s_c$, $q_c$. 

\item We now compute $\eta$ from $s_c$ and $\zeta$ and then reduce a copy of $\eta$ by canceling it. Now we are left with $\zeta$, $r_c$, $s_c$, $q_c$. 

\item In the final step we erase $s_c$, $r_c$, $q_c$. 
\end{itemize}

We thermodynamically erased $s_c$, $q_c$, and $r_c$ in this computational process leaving behind the output $\zeta$. This provides the proof for our claim. Now we move on to the second claim, i.e., the upper bound of the thermodynamic cost of the computational process where $\zeta$ is computed from $\eta$. 

\textit{\textbf{Claim:}} $E_{\alpha} (\eta, \zeta) \geq K_{\alpha}(\zeta|\eta) + K_{\alpha}(\eta|\zeta)$.

\vspace{.2in}

\textit{\textbf{Proof:}} The length of the shortest program to compute $\zeta$ from $\eta$ is defined as $K_{\alpha}(\zeta|\eta)$. During the computation process, we produce garbage bits $g_c(\eta,\zeta)$. By definition~\cite{zurek1989thermodynamic}, we know that the cardinality of the garbage bits is greater or equal to the shortest program. So, to compute $\zeta$ from $\eta$, we have to erase the garbage bits which is equivalent to at least $K_{\alpha} (\eta|\zeta)$ bits. This proves our second claim. 

So both claims prove theorem 1.

\vspace{.2in}
\begin{center}
\textbf{Cognitive distance}
\end{center}

\vspace{.15in}
 Here we are going to explore a different aspect of computation. We will consider the analysis of pictures with binary string bits. Various distance measure is proposed for the measure of binary strings like Hamming distance, and Euclidean distance. In the case of pictures, if one changes some of the bits of the picture, then the distance measures like the Hamming and Euclidean ones, show small changes for the changed and unchanged picture but they look similar. Various approaches to studying the similarities in the picture have been proposed in pattern recognition. To date, the closed form of it is yet to be explored.

So far we have studied in this section, that the work done for the computational process where $\zeta$ is computed from $\eta$ is equivalent to the thermodynamic cost $E_{\alpha}(\eta,\zeta)$. This is the minimum amount of thermodynamic cost required for the execution of a computational process. To analyze the similarities of the pictures, we will consider our approach to be equivalent to human cognition, i.e., the human brain and eye work like a computer.

A distance measure when defined should be non-negative, and symmetric in nature. It must also satisfy the condition of triangular inequality.  It is not just these conditions that suffice the distance measure. To make it more concrete, we should be able to discriminate the strings that are similar and different from the given string $\eta$ with precision.  The distance measure considered in the work~\cite{li1992mathematical} is suitable only when the distance function is computable. So the cognitive distance is defined as

\textbf{Definition 7}  A total function which describes cognitive distance $C_D$ is a bijection from $N \times N$ to $N$. So,
\begin{itemize}
\item $\forall \,\, \eta$, $\zeta$, $C_D(\eta,\zeta) \geq 0$.The equality sign holds when $\eta = \zeta$.

\item The distance measure is symmetric, i.e.,  $\forall \,\, \eta$, $\zeta$, $C_D(\eta,\zeta) = C_D(\zeta,\eta)$.

\item $\forall \,\, \eta$, $\zeta$, and $\Xi$ we have $C_D (\eta, \zeta) \leq C_D(\eta, \Xi) + C_D(\Xi,\zeta)$. This describes the triangular inequality. 

\item For each values of $\eta$, one can recursively enumerate the set $\{\zeta : C_D(\eta,\zeta)\leq d_c\}$.

\item For each values of $\eta$, $\mid \{\zeta : C_D(\eta,\zeta)\leq d_c\} \mid \,\,\, \leq 2^{d_c}$.
\end{itemize}

Here $d_c$ represents the thermodynamic distance. Now we want to see whether the thermodynamic measure satisfies all the conditions of cognitive distance measure. Before we move on to analyze this, we describe a theorem that we require for the analysis. 

\vspace{0.1in}
\textbf{Theorem 2:} $\forall \,\, \eta$, the $\zeta$'s that satisfies the condition $E_{\alpha}(\eta,\zeta)\leq d_c$ is at most $2^{d_c}$ and at least $\Omega(2^{d_c})$.

\vspace{0.2in}

\textbf{Proof 2:} Let us assume the converse of the theorem, i.e., we consider for $\eta$ and $d_c$ there exists $\zeta$'s for which $E_{\alpha}(\eta,\zeta)\leq d_c$ will exceed $2^{d_c}$. So we can say that there exists some $\zeta_0$ for which $K_{\alpha}(\zeta_0,\eta)\geq d_c$. Now it is well known from the analysis that $E_{\alpha}(\eta, \zeta_0) \geq K_{\alpha}(\eta|\zeta_0) + K_{\alpha} (\zeta_0|\eta)$ and $ K_{\alpha}(\eta|\zeta_0) > 0$. So we have a contradiction.  

Let us now consider $\zeta$'s as $\eta p_c^R$, where $p_c^R$ describes the reversal of the program $p_c$. So we have $K_{\alpha}(\eta|\zeta) = O(1)$, and $K_{\alpha} (\zeta|\eta) = K_{\alpha}(p_c|\eta) + O(1)$. So, 

\vspace{0.1in}
$d_c\geq \, E_{\alpha}(\eta,\zeta)\geq K_{\alpha}(\eta|\zeta) + K_{\alpha} (\zeta|\eta)= K_{\alpha}(p_c|\eta) + O(1)$. 

\vspace{0.1in}
Therefore for every element of $\eta$, we have at least $\Omega(2^{d_c})$ elements of $p_c$'s which is equivalent to that many numbers of $\zeta$'s.

Now we move on to describe the thermodynamic cost function as a cognitive measure. 

\vspace{0.1in}
\textbf{Theorem 3:} The thermodynamic cost function $E_{\alpha}(\eta,\zeta)$ describes the cognitive distance function.  

\vspace{0.2in}

\textbf{Proof 3:} To be a cognitive measure function the thermodynamic cost function has to satisfy the condition given in \textbf{Definition 7}. The first three conditions get trivially satisfied as the thermodynamic measure function $E_{\alpha}(\eta,\zeta)$ represents a distance function. We can enumerate $E_{\alpha}(\eta,\zeta)$ so, the fourth condition gets satisfied. \textbf{Theorem 2} provides sufficient information to satisfy the fifth condition. So, the thermodynamic cost function can describe the cognitive distance.

This helps us to infer that the thermodynamic cost function provides the necessary information not only about the computational process but also about the distance measure required in pattern recognition theory. The thermodynamic cost function is the optimal cognitive distance measure function.

\section{Thermodynamics $\longleftrightarrow$ arrow of time for computer}\label{sec12}
Here we are going to explore a different viewpoint, where we will see the link between the thermodynamic arrow and the arrow of time for computers. To understand this we need to have a clear view of the notion of the thermodynamic arrow and that of the computation arrow. We are going to first analyze, what is meant by the thermodynamic arrow of time, and also about the `entropy increasing universe' along with the `entropy decreasing universe'. Then we will move on to analyze the computational arrow of time where we will explore how a physical process incorporates a computation.

\subsection{Thermodynamic Arrow}
The thermodynamic arrow of time can be expressed as a sort of causality. Let us consider a dynamic time parameter $t$, which has no prior thermodynamic direction. We consider macroscopic systems that are identical to each other but are separated by their time intervals $[0,t_1]$ and $[t_1,t_2]$.  The behavior of the system will be identical within the time period $[0,t_1]$, but different in the case of $[t_1,t_2]$ for the thermodynamic arrow. This is equivalent to the entropy increase of the system. So to understand the thermodynamic arrow of time we need to have a clear idea about entropy increasing universe as well as entropy decreasing universe.

For the universe, we will consider the state space to be formed of many smaller subsystems $\varkappa = \prod_j \otimes \varkappa_j$. Let us consider a measure for this state space as $\nu_a$. All these subsystems are categorized with a single environment $env$. So, the state space can be described as 
\begin{equation}\label{TAR1}
\varkappa =\varkappa_{env} \prod_j \otimes \varkappa_j,
\end{equation}
where $\varkappa_{env} = \prod_{j\in env} \otimes \varkappa_j$.

The dynamics of the state space are depicted by $\mathbb{M}^t$. It is an invertible function.  So there exist a map $\{\mathbb{M}^{-t}: \mathbb{M}^{-t} \circ \mathbb{M}^{t} = \mathbb{M}^{t} \circ \mathbb{M}^{-t} = \varphi\}$, where $\varphi \subset	 \varkappa$ and $\nu_a (\mathbb{M}^{t}(\varphi))= \nu_a (\varphi)$. Now having the dynamic map we will explore the change in the entropy.

\vspace{0.2in}
\begin{center}
\textit{Universe with increasing and decreasing entropy}
\end{center}

\vspace{.15in}

First, we will study what it's mean by the universe with increasing entropy. Let us consider a special initial microstate for this system $\varrho_0 \subseteq \varkappa$. The dynamics of this microstate are such that it spreads over the whole state space over time, and the measure $\nu_a$ is preserved throughout.  The initial microstate can be described as $\varrho_0 = \prod_j \otimes  \varrho_j$, where $\varrho_j$ represent the subsystems. Due to the microscopic correlation, the state after the evolution can not be represented by the product state of the subsystem. This correlation between these microstates is the reason behind the entropy increase.

So the dynamics of the state can be described as 
\begin{equation}\label{TAR2}
\mathbb{M}^t (\varrho_0)= \cup_n \varrho_n \otimes \varrho_{env},
\end{equation}
where  $\varrho_{env} \subseteq \varkappa_{env}$.  The initial boundary condition for the system is that it should evolve accordingly for all maps $\mathbb{M}^t$. We can  consider the state of the system where:

1. The macro subsystem has no microsystem correlation.

2. The thermal state of the system can be described using the Gibbs distribution.

3. The entropy of the macrosystem increases.

4. Miscosystem correlation flourishes among the subsystems.

when the system is analyzed after a long time from the initial condition but before thermalization is acquired.

Now we move on to analyze what it's mean by the universe with decreasing entropy. Broadly speaking, it can be considered as the time reversal of the universe with increasing entropy. For the universe of decreasing entropy, we have to define the future boundary condition. The condition is that at some time $\tau_a$ the state will be described by $\varrho_0^{T_a}$ with appropriate dynamics. The time-reversal is described by a map $\varrho^{T_a} = T_a (\varrho) \subset \varkappa$, where the measure satisfies the condition $\nu_a(\varrho^{T_a}) = \nu_a (\varrho)$. The way to analyze this is to consider the time reversal at some time $\tau_a/2$ of the universe with increasing entropy. The time reversal dynamics for this system is 
\begin{equation}\label{TAR3}
\varrho_{T_a t} = \mathbb{M}_{T_a \tau_a/2}^t (\varrho_0^{T_a}) = T_a \circ \mathbb{M}^{\tau_a- t} \circ \mathbb{M}^{-\tau_a} \circ T_a(\varrho_0^{T_a}).
\end{equation} 

So we can infer that if one takes the time-reversal of the terminating condition of the universe with increasing entropy, it will result in the initial condition of the universe with decreasing entropy. So at $\tau_a/2$, we get $\mathbb{M}_{T_a\tau_a/2}^t = \mathbb{M}^t$, and $\varrho_{T_a \tau_a} = \mathbb{M}^t \circ T_a \circ \mathbb{M}^{\tau_a} (\varrho_0)$. Similar to the universe with increasing entropy, one can describe the dynamics of the state with the reverse sets of conditions as defined for the increasing entropy universe. The conditions can be represented as 

1. In the macrosystem there exists a microsystem correlation.

2. The interaction causes the disappearance of the microsystem correlation. 

3. The Gibbs entropy of the macrosystem is decreasing.

4. At the end of each interaction the thermal state of the system is described by the Gibbs distribution.
 
It should be kept in mind that the interaction is analyzed very long before the future boundary condition and also when the system comes out of thermalization.

Schulman in his work~\cite{schulman2005computer}, has considered the computer as an open system and has argued from Landauer's Principle that one can align the computational arrow and the thermodynamic arrow of time. Now we will analyze the computation arrow for the universe with increasing as well as decreasing entropy. Though it seems intuitive that it violates Landauer's principle it will be shown that the approach taken in the formalism doesn't violate Landauer's Principle, and there is no such alignment between the computational arrow of time and the thermodynamic arrow of time.

\subsection{Computational Arrow of time}
Every physically acceptable computation can be considered as a combination of logical operations. Logical operations are nothing but a mathematical operation that maps input states to the output states. If the output state has a unique input state then it is defined as a logically reversible process, else it is called a logically irreversible process. Two basic logical operation are $NOT$ and $RESET \,\, TO \,\, ZERO$ $(RTZ)$ though the familiarity of $RTZ$ is less than that of $AND$, $OR$ logic operations. These two basic logic operations have the power to build all suitable combinations and it is the main center point of logical operation from a thermodynamic viewpoint. $RTZ$ reset every input to a zero output.

To study the thermodynamics of the logical operation let us consider a state space $\varkappa =\varkappa_S \otimes \varkappa_{env}$, where $\varkappa_S$ describes the logical processing state space and the environmental state space is depicted by $\varkappa_{env}$. For the universe with increasing entropy, the environment is a subset of $\varkappa_{env}$ and it has no correlation with the logical state space. A state space region is assigned for each input $\alpha_{inp}$ of the system where $X_{\alpha_{inp}} \subset \varkappa_S$ applied to the condition that $X_{\alpha_{inp}} \cap X_{\alpha_{inp}'} = \emptyset$ for $\alpha_{inp} \neq \alpha_{inp}'$. Similarly for the output $\beta_{out}$ we have  $X_{\beta_{out}} \subset \varkappa_S$ applied to the condition that $X_{\beta_{out}} \cap X_{\beta_{out}'} = \emptyset$ for $\beta_{out} \neq \beta_{out}'$. The input and output state is subjected to the condition that is invariant to the time-reversal symmetry, i.e., $X_{\alpha}^{T_a} = X_{\alpha}$ and $X_{\beta}^{T_a} = X_{\beta}$. The dynamics of the system $\mathbb{M}_{L_{m}}$, when a logical operation $L_m$ maps the input state to the output state for all inputs that are mapped to the output is 
\begin{equation}\label{TAR4}
\mathbb{M}_{L_{m}} (X_{\alpha} \otimes \varkappa_E) \subseteq X_{\beta} \otimes \varkappa_{env},
\end{equation}
where $\varkappa_E \subset \varkappa_{env}$.

In the case of reversible logical operation, we will be able to execute a time-reversal operation without any asymmetry in them as the output state has a unique input state for it. Whereas, for irreversible logical operations we observe a time-reversal asymmetry in them as the output state can be mapped to a finite number of input states instead of a unique one. To get a clear picture of the time-reversal for the irreversible process, we have to also include the indeterministic operation\cite{maroney2005absence} in our basic set of operations. $NOT$ follows the time-reversal symmetry where $RTZ$ being logically irreversible, results to some indeterministic operations $UNSET \,\, From \,\, Zero$ $(UFZ)$. So, we can infer that the time-reversal for $RTZ$ is not a reversible simulation, but for $UFZ$ the simulation is deterministic in nature. So if one constructs a Turing machine using logically reversible operations and simulates irreversible operations then the time-reversal will account for the contribution of the simulation of the indeterministic operation. Therefore this will not represent a Universal Turing Machine. So one can infer that a computation process has an arrow of time.

Now we will describe a new concept (Keeping in mind the indeterministic operation), i.e., reversibility of logical operation with the inclusion of in-determinism. For a defined logical operation $L_m$, we will have a reversal operation $RL_m$. This reversal operation maps similar to the time-reversal operations on a computational process.  The reverse Logical operation $RL_m$ considers the output of the logical operation $L_m$ as the input state and maps it to the input state of the logical operation $L_m$ as the output state. If $L_m$ depicts an irreversible operation then $RL_m$ depicts an indeterministic operation. Similarly, one will define $RL_m$ as a logically reversible operation, if $L_m$ represents an indeterministic operation.

Now to construct reversal operation we consider that the system starts from $\varrho_0 = \cup_\alpha X_{\alpha} \otimes \varkappa_E$ and ends in $\varrho_{t_{L_m}} =  \mathbb{M}_{L_m}^{t_{L_m}} (\cup_\alpha X_{\alpha} \otimes \varkappa_E)$. The construction of the $RL_m$ is depicted as follows

1. The start of  the reversal operation in logical state $\alpha_{inp}$ is
\begin{equation}\label{TAR5}
\mathbb{W}_{L_m} (\alpha_{inp}) = \frac{\nu_a ((X_{\alpha} \otimes \varkappa_{env})\cap \varrho_0)}{\nu_a(\varrho_0)}.
\end{equation} 

2. The halting state i.e., the output state $\beta_{out}$ conditioned that the initial state was $\alpha_{inp}$ is 
\begin{equation}\label{TAR6}
\mathbb{W}_{L_m}(\beta_{out}|\alpha_{inp}) = \frac{\nu_a\, \Big( (X_{\alpha}\otimes \varkappa_{env}) \cap \mathbb{P} \cap \varrho_{t_{L_m}} \Big)}{\nu_a ((X_{\alpha} \otimes \varkappa_{env})\cap \varrho_0)},
\end{equation}
where $\mathbb{P} = \mathbb{M}_{L_m}^{t_{L_m}} (X_{\alpha} \otimes \varkappa_E)$.

3. The system  will start from the initial state $\alpha_{inp}$ and it will end in the final state $\beta_{out}$ is
\begin{equation}\label{TAR7}
\mathbb{W}_{L_m} (\alpha_{inp}, \beta_{out}) = \frac{\nu_a\, \Big( (X_{\alpha}\otimes \varkappa_{env}) \cap \mathbb{P} \cap \varrho_{t_{L_m}} \Big)}{\nu_a(\varrho_0)}.
\end{equation}

4. The end of the reversal operation in the logical state $\beta_{out}$ is 
\begin{equation}\label{TAR8}
\mathbb{W}_{L_m} (\beta_{out}) = \frac{\nu_a ((X_{\beta} \otimes \varkappa_{env})\cap \varrho_{t_{L_m}})}{\nu_a(\varrho_0)}.
\end{equation}

5. The inverse of the second construction, i.e., the system starts from the initial state $\alpha_{inp}$ conditioned  that it will reach end logical state $\beta_{out}$ is 
\begin{equation}\label{TAR9}
\mathbb{W}_{L_m}(\alpha_{inp}|\beta_{out}) = \frac{\nu_a\, \Big( (X_{\alpha}\otimes \varkappa_{env}) \cap \mathbb{P} \cap \varrho_{t_{L_m}} \Big)}{\nu_a ((X_{\beta} \otimes \varkappa_{env})\cap \varrho_{t_{L_m}})},
\end{equation}

where it is considered that $\mathbb{W}_{L_m} (\alpha_{inp}) = \mathbb{W}_{L_m} (\beta_{out}) \neq 0$ and $\mathbb{W}_{L_m}(\beta_{out}|\alpha_{inp})$, $\mathbb{W}_{L_m}(\beta_{out}|\alpha_{inp})$ $\in \{0,1\}$. Now we will calculate the time reversal  of this reversal operation with input state $\alpha_{inp}$, and output state $\beta_{out}$. The time reversal results to 
\begin{eqnarray}\nonumber
\mathbb{W}_{T_a L_m} (\alpha_{inp}) = \frac{\nu_a ((X_{\alpha} \otimes \varkappa_{env})\cap \varrho_0)}{\nu_a( \varrho_{t_{L_m}})},\\ \nonumber
\mathbb{W}_{T_a L_m} (\beta_{out}) = \frac{\nu_a ((X_{\beta} \otimes \varkappa_{env})\cap \varrho_{t_{L_m}})}{\nu_a(\varrho_{t_{L_m}})}, \\ \nonumber
\mathbb{W}_{T_a L_m} (\alpha_{inp}, \beta_{out}) = \frac{\nu_a\, \Big( \mathbb{P'} \cap (X_{\alpha} \otimes \varkappa_E) \cap \varrho_0 \Big)}{\nu_a(\varrho_{t_{L_m}})}, \\ \nonumber
\mathbb{W}_{T_a L_m}(\alpha_{inp}|\beta_{out}) = \frac{\nu_a\, \Big( \mathbb{P'} \cap (X_{\alpha} \otimes \varkappa_E) \cap \varrho_0 \Big)} {\nu_a ((X_{\beta} \otimes \varkappa_{env})\cap \varrho_{t_{L_m}})}, \\ \nonumber
\mathbb{W}_{T_a L_m}(\beta_{out}|\alpha_{inp}) = \frac{\nu_a\, \Big( \mathbb{P'} \cap (X_{\alpha} \otimes \varkappa_E) \cap \varrho_0 \Big)}{\nu_a ((X_{\alpha} \otimes \varkappa_{env})\cap \varrho_0)},
\end{eqnarray} 
where $\mathbb{P'} = \mathbb{M}_{L_m}^{t_{L_m}} (X_{\alpha}\otimes \varkappa_{env})$. One can easily verify that 
\begin{eqnarray}\nonumber
\mathbb{W}_{T_a L_m} (\alpha_{inp}) = \mathbb{W}_{L_m} (\alpha_{inp}), \\ \nonumber
\mathbb{W}_{T_a L_m} (\beta_{out}) = \mathbb{W}_{L_m}(\alpha_{inp}|\beta_{out}),\\ \nonumber
\mathbb{W}_{T_a L_m}(\beta_{out}|\alpha_{inp}) = \mathbb{W}_{L_m}(\beta_{out}|\alpha_{inp}),\\ \nonumber
 \mathbb{W}_{T_a L_m}(\alpha_{inp}|\beta_{out}) = \mathbb{W}_{L_m}(\alpha_{inp}|\beta_{out}).
\end{eqnarray}

Similarly, one can also check that 
\begin{eqnarray}\nonumber
\mathbb{W}_{R L_m} (\alpha_{inp}) = \mathbb{W}_{L_m} (\alpha_{inp}), \\ \nonumber
\mathbb{W}_{R L_m} (\beta_{out}) = \mathbb{W}_{L_m}(\alpha_{inp}|\beta_{out}),\\ \nonumber
\mathbb{W}_{R L_m}(\beta_{out}|\alpha_{inp}) = \mathbb{W}_{L_m}(\beta_{out}|\alpha_{inp}),\\ \nonumber
 \mathbb{W}_{R L_m}(\alpha_{inp}|\beta_{out}) = \mathbb{W}_{L_m}(\alpha_{inp}|\beta_{out}).
\end{eqnarray}

Now we will analyze the computational reversal, where we will consider a series of operations in the universe with increasing entropy. Let us consider that the system starts at some initial state $\alpha_{inp}$. The logical map $L_{m_0}$ transits it to an output state $\alpha_{1}$. The output state $\alpha_1$ now acts as an input state for the next logical operation $L_{m_1}$ and maps it to the state $\alpha_2$ and so on. It is assumed here that the microsystem correlation between the system and the environment has no effect on the evolution of the system. Mathematically, it can be represented as 
\begin{equation}\label{TAR10}
\{\alpha_{inp}\} \xrightarrow{L_{m_0}} \{\alpha_{1}\} \xrightarrow{L_{m_1}} \{\alpha_{2}\} \xrightarrow{L_{m_2}} \cdots  \xrightarrow{L_{m_{f}}} \{\alpha_{final}\}.   
\end{equation}

For a universe with decreasing entropy, one obtains the results due to the time-reversal of the system. Mathematically it is depicted as 
\begin{eqnarray}\nonumber
\{\alpha_{final}\}  \xrightarrow{T_a L_{m_f}} \{\alpha_{f-1}\} \xrightarrow{T_a L_{m_{f-1}}} \{\alpha_{f-2}\} \xrightarrow{T_a L_{m_{f-2}}} \cdots \\ \nonumber
  \xrightarrow{T_a L_{m_{1}}}  \{\alpha_{inp}\}. 
\end{eqnarray}

As discussed earlier, we know that the series of time-reversal operations will not depict the same computation as the logical operation. So we define a series of reverse logical operations for the universe with increasing entropy. Mathematically, it is depicted as 
\begin{eqnarray}\nonumber
\{\alpha_{final}\}  \xrightarrow{R L_{m_f}} \{\alpha_{f-1}\} \xrightarrow{R L_{m_{f-1}}} \{\alpha_{f-2}\} \xrightarrow{R L_{m_{f-2}}} \cdots \\ \nonumber
  \xrightarrow{R L_{m_{1}}}  \{\alpha_{inp}\}. 
\end{eqnarray}

Now if we execute the time-reversal of the reverse logical operations, we boil down to the logical operation defined in Eq.~\eqref{TAR10}. Time reversal of the reverse logical operation takes place in a universe with decreasing entropy. So one can infer that logical operation in a universe with increasing entropy is equivalent to the computation of time reversal of reverse logical operation in the universe with decreasing entropy. So the notion that the computation arrow is aligned with the thermodynamic arrow is at stake in this analysis.

Here, in this analysis, one encounters a disagreement over the Schulman notion of equivalence of the thermodynamic and computational arrow of time. Schulman's notion was based on Landauer's principle. Landauer's principle stands as the basis for all computational processes~\cite {bub2001maxwell,caves1990quantitative,caves1993information,piechocinska2000information} from a thermodynamic perspective. So the question arises as to how the same information operation is possible for the universe with increasing as well as decreasing entropy. The answer to this question is that in Landauer's principle, the computation process is assumed for the universe with increasing entropy. So we will now explore how Landauer's principle turns out in the universe with decreasing entropy.

To understand Landauer's principle in the universe with decreasing entropy, we will first go through Landauer's principle for the universe with increasing entropy. For the input logical state $\alpha_{inp}$, the state for the physical system is described by density matrix as $\rho_{\alpha_{inp}}$ whereas, for output logical state $\beta_{out}$ the state is described by density matrix as $\rho_{\beta_{out}}$. The entropy and the mean energy for the states $\forall \, \alpha_{inp}, \, \beta_{out}$ is 
\begin{equation}\label{TAR11}
S= -k_B T\, \ln[\rho_{\alpha_{inp}} \, \ln\, \rho_{\alpha_{inp}}] = -k_BT\, ln[\rho_{\beta_{out}} \, \ln\, \rho_{\beta_{out}}],
\end{equation}
and 

\begin{equation}\label{TAR12}
U = Tr [H_S \rho_{\alpha_{inp}}] = Tr [H_S  \rho_{\beta_{out}}].
\end{equation}
Here $H_S$ describes the Hamiltonian of the system. In the universe, with increasing entropy, we follow certain assumptions. They are:

1. The total Hamiltonian $H$ of the system describes the evolution of the system. The total Hamiltonian of the system is composed of the energy of the system $H_S$, the energy of the environment $H_{env}$, and their interaction $V_{Senv}$. Mathematically it is described as 
\begin{equation}\nonumber
H = H_S  + H_{env} + V_{Senv}.
\end{equation}

2. It is considered that the environment can be described by Gibbs's canonical state for temperature $T$. It is also assumed that the system and the environment have no initial correlation. Thus, the density matrix of the environment can be expressed as 
\begin{equation}\nonumber
\rho_{env} (T) = \frac{e^ {-H_{env} /k_BT}}{Tr [e^ {-H_{env} /k_BT}]}.
\end{equation} 

3. The interaction energy before and after the interaction can be described as 
\begin{equation}\label{TAR13}
Tr [V_{Senv} \, \rho_0] \approx 0,
\end{equation}
and
\begin{equation}\label{TAR14}
Tr [V_{Senv} e^{-iHt} \rho_0 e^{iHt}] \approx 0,
\end{equation}
respectively. Using the proposed theory in~\citep{maroney2009generalizing,gibbs1902elementary,tolman1938principles,partovi1989irreversibility}, and using 
\begin{eqnarray}\nonumber
\rho_{In} = \sum_{\alpha_{inp}} P(\alpha_{inp}) \rho_{\alpha_{inp}},\\ \nonumber
\rho_{It} = e^{-iHt} \rho_0 e^{iHt},\\ \nonumber
P(\beta_{out}) = \sum_{\alpha_{inp}} P(\beta_{out}|\alpha_{inp}) P(\alpha_{inp}), \\ \nonumber
\rho_{f} = Tr_{env} [\rho_{It}], \quad \rho'_{env} = Tr_S [\rho_{It}],
\end{eqnarray}
we have 
\begin{eqnarray}\label{TAR15} \nonumber
\sum_{\alpha_{inp}} P(\alpha_{inp}) \, \ln [P(\alpha_{inp})] - \sum_{\beta_{out}} P(\beta_{out}) \, \ln [P(\beta_{out})]  \\ 
\geq \frac{Tr[H_{env} \rho_{env} (T)]}{k_BT} - \frac{Tr[H_{env} \rho'_{env} (T)]}{k_BT}, \, \, \, \, \,
\end{eqnarray}
where $P(\alpha_{inp})$ describes the probability of occurance of the logical state $\alpha_{inp}$. So one can derive the form of Landauer's Principle as $\Delta Q \geq -\Delta H k_BT\, \ln 2$, where $\Delta Q$ describes the expectation value for the heat generation, and $\Delta H$ describes the change in the Shannon information. Now we will analyze Landauer's Principle for the universe with decreasing entropy. For the entropy-decreasing universe, we consider the same logical states as defined in an entropy-increasing universe. We also continue with the same assumption for the dynamics of the system and for the interaction as in the case of the universe with increasing entropy. To incorporate the future boundary condition we consider that the environment is described by Gibbs's canonical state even after the operation and there exists no correlation between the system and the environment. The density matrix of the evolved system is described as $\rho_{It} = \sum_{\alpha_{inp},\beta_{out}} P(\beta_{out}|\alpha_{inp}) P(\alpha_{inp}) \, \rho_{\beta_{out}} \otimes \rho_{env}(T)$. Similar to the analysis done for the universe with increasing entropy, one can calculate and have 
 
\begin{eqnarray}\label{TAR16} \nonumber
-\sum_{\alpha_{inp}} P(\alpha_{inp}) \, \ln [P(\alpha_{inp})] + \sum_{\beta_{out}} P(\beta_{out}) \, \ln [P(\beta_{out})]  \\ 
\geq \frac{Tr[H'_{env} \rho_{env} (T)]}{k_BT} - \frac{Tr[H'_{env} \rho'_{env} (T)]}{k_BT}, \, \, \, \, \,
\end{eqnarray}
where $H'_{env}$ is the Hamiltonian for the environment for the entropy-decreasing universe. So Landauer's Principle for this universe is $\Delta Q \leq -\Delta H k_BT\, \ln 2$. This is what is expected in the entropy-decreasing universe. So the result of contradicting the notion of alignment of the thermodynamics and the computational arrow of time is valid.

\section{Finite state machine: thermodynamic interpretation}\label{sec14}
\begin{figure}[h]
  \includegraphics[width=1.0\columnwidth]{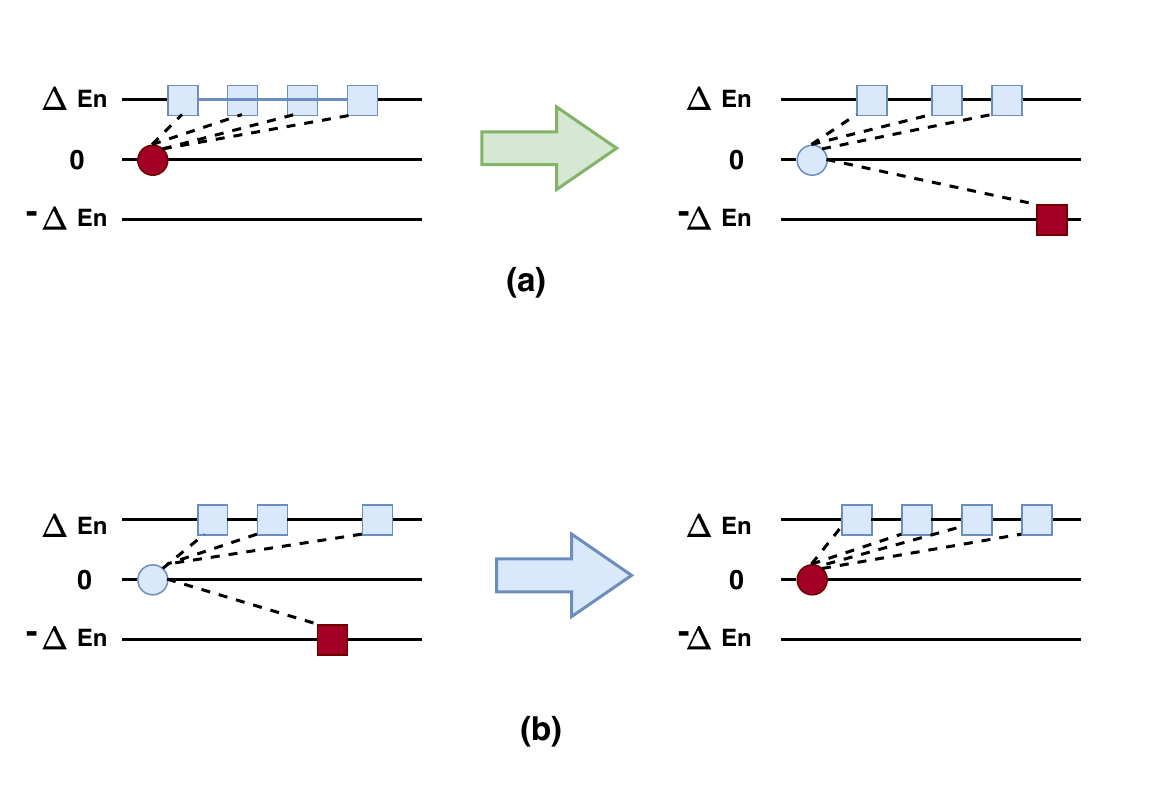}
  \caption{Schematic diagram of the $N-it$ setter and $N-it$ flip is shown in (a) and (b) respectively for $N= 4$. The label states are depicted by circles, and the dynamic states are described by squares. The favorable state is shown in brown color. For $N-it$, setter the final target state is transferred from the high energy level to the lower one whereas, in the case of $N-it$ flip, the reverse procedure is followed. }
  \label{fig9}
  \end{figure}

We are going to explore finite state machines (FSM) from a thermodynamics point of view~\cite{chu2018thermodynamically} in this section. For the analysis, we will consider two processes, which will be the fundamental blocks that will guide us in building physical models of FSM. These models help us to describe FSM from a thermodynamic viewpoint. We will describe these two processes as $N-it$ setters, and $N-it$ flips. These two processes are the generalization of bit-set and bit-flip operation. Here, we will consider three energy level systems with multiple numbers of states. The schematic representation of these two processes is shown in Fig.~\ref{fig9}.

We will consider a two-level system $\{c_1, c_2\}$ to conceive the idea of bit-flip and bit-set. Let us first explore the bit-flip process. For this process, we consider the state of the process to be correlated with energy level $\{-\Delta En, 0\}$. The system for this model will be kept in contact with a heat bath at some constant temperature $T$. The probability that the system in the state $c_1$ after a long time is $p(C_1) = e^{\Delta En/ kT}/ (1+ e^{\Delta En/ kT})$. For the allowed transitions between the states of the system, the energy levels are depicted as $r_{c_1\rightarrow c_2}$ and $r_{c_2\rightarrow c_1}$ where $r_{c_1\rightarrow c_2}= r^-$ and $r_{c_2\rightarrow c_1} = r^+$ initially. According to the proposed postulates in~\cite{crooks1998nonequilibrium}, we know that the system should be balanced, i.e., $r^+/ r^- = e^{\Delta En/k T}$. To execute the bit-flip operation the energy level is raised to $En_1 = \Delta En$. So the transition states in this condition is $r_{c_1\rightarrow c_2}= r^+$ and $r_{c_2\rightarrow c_1} = r^-$. The work done to execute this whole process is 
\begin{equation}\label{FSMT1}
\langle W_a\rangle =  \frac{2r^+}{r^+ + r^-} \,\, \ln \Big(\frac{r^+}{r^-} \Big),
\end{equation}
and the entropy  of the system is 
\begin{equation}\label{FSMT2}
S_{total} = \frac{r^+ - r^-}{r^+ + r^-}\,\, \ln \Big(\frac{r^+}{r^-} \Big).
\end{equation}
Similarly, one can get the Gibbs free energy for the system from the entropy. Whereas, for the bit-set operation the work done for the execution of the process can be described as 
\begin{equation}\label{FSMT3}
\langle W_a\rangle = -\frac{1}{2}\, \ln \Big(\frac{r^+}{r^-} \Big),
\end{equation}
and the entropy  of the system is 
\begin{equation}\label{FSMT4}
S_{total} =  \ln \Big(\frac{r^+ +r^-}{2 \sqrt{r^+ r^-}} \Big).
\end{equation}

Now we will explore the process described as  $N-it$ setter and $N-it$ flips from this simple model analysis. First, we will explore $N-it$ setter, then we will move on to explore  $N-it$ flips. Similar to the bit-flip and bit-set operation, here also we will alter the energy levels of the systems. For these multiple-level systems with $N+1$ states, the operation is more complicated than that of the bit-flip operations. The multi-level states are denoted as $\{V_0, V_1, V_2,\dots, V_N\}$ with energy levels $\{0, \Delta En, \Delta En, \dots, \Delta En\}$  at the initial stage. After the execution of the process, the energy level at the final stage is described as $\{0, \Delta En, \Delta En, \dots, -\Delta En, \dots, \Delta En\}$ for the states $\{V_0, V_1, V_2, \dots, V_i, \dots, V_N\}$.  One can compute the probability of the state $V_i$ for the initial stage analogous to the  bit-flip method as $p(V_i) = \frac{e^{-\Delta En}}{1 + Ne^{-\Delta En}}$, and the probability of the state $V_i$ for the final stage is $p(V_i) = \frac{(r^+)^2}{(r^+)^2 + r^+ r^- + (N-1) (r^-)^2}$. So the work done for the execution of the process is $W_a= \frac{2r^-}{r^+ + N r^-} \, \ln \Big(\frac{r^+}{r^-}\Big)$. Now one can compute the entropy of the system as $S_{total} = \frac{2r^-}{r^+ + N r^-} \, \ln \Big(\frac{r^+}{r^-}\Big) + log \left(1 \frac{r^+}{r^-} + \frac{r^-(N-1)}{r^+} \right) - log \left(1+ \frac{Nr^-}{r^+} \right) $.

\vspace{.1in}
We will now explore the second operation, i.e., $N-it$ flip. Similar to the bit-flip and bit-set operation, here also we will alter the energy levels of the systems. Similarly to the $N-it$ setter the states are described by $\{V_0, V_1, V_2,\dots, V_N\}$. One can compute the probability of the state $V_i$ for the initial stage analogous to the $N-it$ setter as  $p(V_i) = \frac{e^{\Delta En}}{e^{\Delta En} + 1 + Ne^{-\Delta En}}$, and the probability of the state $V_i$ for the final stage is $p(V_i) = \frac{r^+}{r^+ + Nr^-}$. So the work done for the execution of the process is $W_a =  \frac{(r^+)^2}{(r^+)^2 + r^+ r^- + (N-1) (r^-)^2}\, \ln \left(\frac{r^+}{r^-}\right)$. So, one can compute the entropy of the system as $S_{total} =  \frac{(r^+)^2}{(r^+)^2 + r^+ r^- + (N-1) (r^-)^2}\, \ln \left(\frac{r^+}{r^-}\right) + log \left(1+ \frac{Nr^-}{r^+} \right) - log \left(1 \frac{r^+}{r^-} + \frac{r^-(N-1)}{r^+} \right)$.

\vspace{.2in}
\begin{center}
\textbf{Physical model for FSM}
\end{center}

\vspace{.15in}

To analyze the thermodynamics of FSM, we have to construct a model for FSM which is thermodynamically consistent. FSM is equivalent to an inhomogeneous Markov chain, where the transitions depend on the tape symbol. To design a model for the physical implementation of FSM, one has to keep in mind the transition based on the tape symbol. First, we will consider a Markov chain model $M_c$, which is nothing but a naive translation of FSM.  However, the naive translation of FSM will not provide a sufficient mechanism to implement the physical system. The naive transformation is done as follows:

1. The states of $M_c$ are described as $\vartheta_i$. These are equivalent to the states defined for FSM. 

2. The transition $ tran_{ij}$ from a state to another state, i.e, from $\vartheta_i$ to some $\vartheta_j$ conditioned that it is associated with an input symbol $a$.

3. When an internal transition takes place, the external tape executes a movement in the forward direction, and then it takes the machine to the next tape element which has the symbol $a$.  During this transition, the transition rate $r_{ij} =1$ if and only if the machine is in the tape with some input symbol.

Now if we try to implement this model, we encounter some difficulties, and we will need some helper states which are described as dynamic states~\cite{wolpert2019space}. The difficulties come from the transition process. To see this, let us consider a transition from $\vartheta_i$ to $\vartheta_j$, where it is assumed that this transition occurs with a low transition rate $r^-$ with an input symbol $a$. So, the different transition form $\vartheta_i$ to $\vartheta_m$ is associated with the symbol $a$. For the execution of this transition, the state $\vartheta_j$ should be at a high energy level. Now, it might be possible that there exists some transition from $\vartheta_k \neq \vartheta_i$ to $\vartheta_j$ with a transition rate $r^+$, which demands $\vartheta_j$ to be at a lower energy level.  Just by adjusting the energy levels, we are not able to tackle this situation. To overcome this, helper states are required which is described as the `dynamic state'.

For each logical state, we will consider the $n$ number ($n$ is the number of symbols) of dynamic states which will keep track of all transitions even the self-transition. The dynamic states of the system for a input symbol $a$ is defined as $Ds_{a} = \{Ds_{a}^{1\rightarrow 1{a}}, \dots, Ds_{a}^{M\rightarrow M{a}} \}$, where $M$ denotes the number of internal states. Similarly, the dynamic states for the system correlated with the label states $\vartheta_i$ is $Ds^i = \{Ds_a^{1\rightarrow 1(a)}, Ds_b^{1\rightarrow 1(b)}, \dots \}$. The changes in the modified form of the Markov model from the pre-defined form is 

1. Now for every label state, we have $n$ number of dynamic states with tape symbols. 

2. All the forms of transition are taken into account by this dynamic set of states. 

\begin{figure}[h]
  \includegraphics[width=1.0\columnwidth]{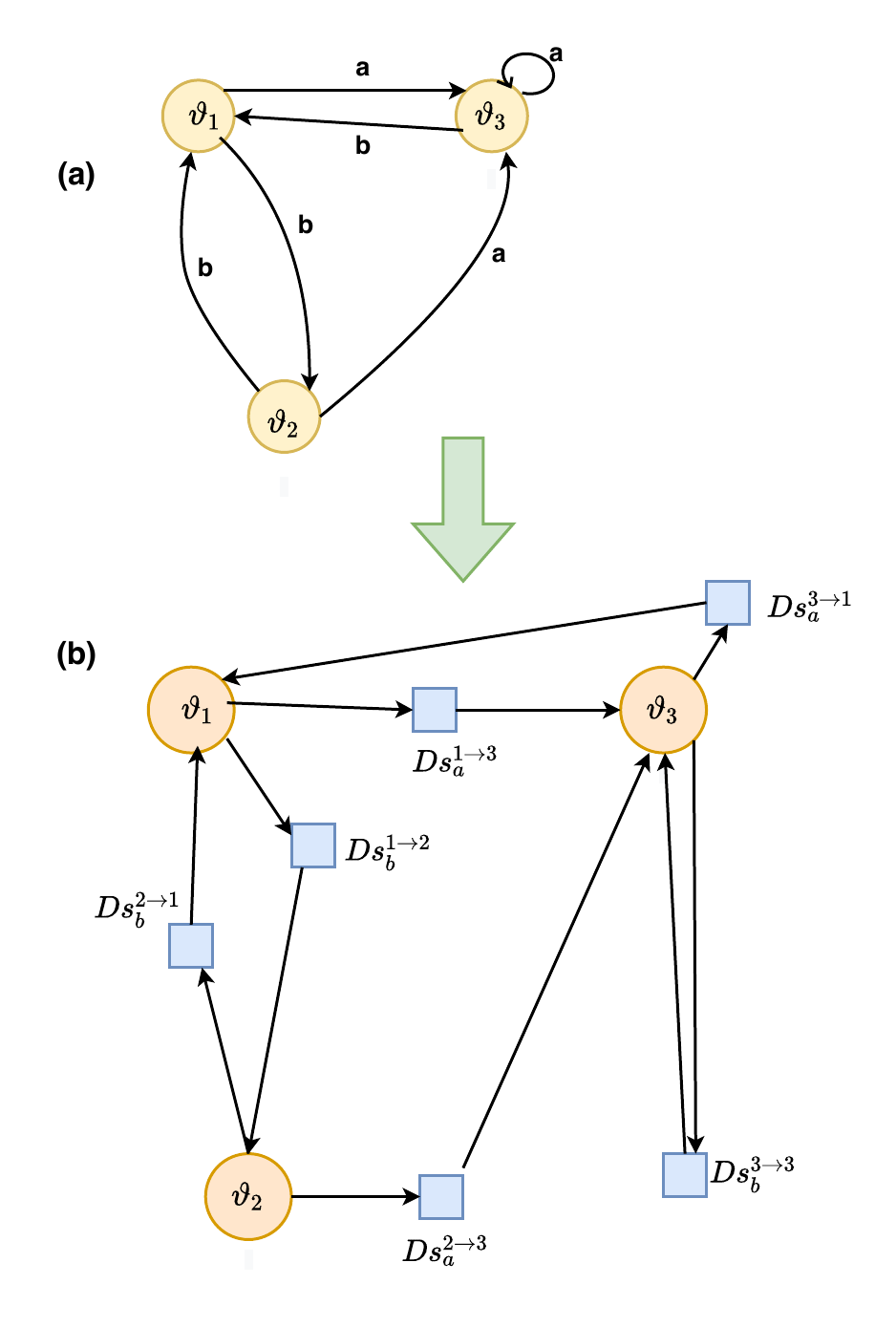}
  \caption{A state diagram of an FSM with three states where the accepting state is represented by $\vartheta_3$  is shown in (a). The details to develop a state diagram for a computational process have been described in detail with an example while we were exploring finite automata in section~\ref{sec3}. The transition diagram of the Markov model designed for the same FSM is shown in (b). The dynamic states are described by the square blocks and the label state by the circles. The direction of the computation is shown by the arrows.}
  \label{fig10}
  \end{figure}

The schematic of an FSM model using the Markov model is shown in Fig.~\ref{fig10}.   

\begin{figure}[h]
  \includegraphics[width=1.0\columnwidth]{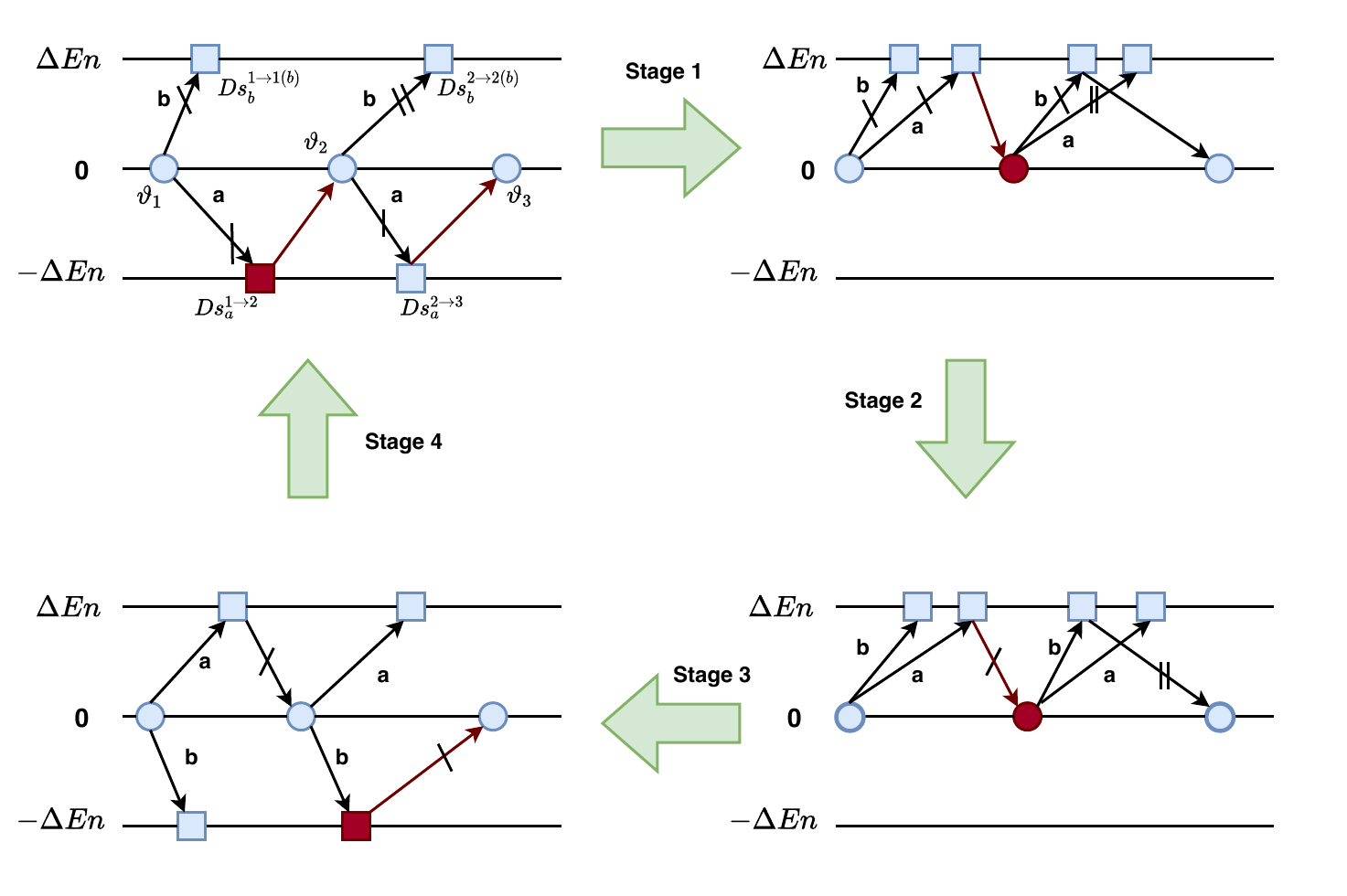}
  \caption{A schematic representation of the cyclic process for the computation is shown. The bluish circle and the squares describe the label states and dynamic states. The input tape is marked with brown color having the current symbol and also the current state is highlighted by brown color. }
  \label{fig11}
  \end{figure}

Now we will use this proposed Markov model to simulate any FSM. To do that one has to consider three energy levels $\{0, \pm \Delta En\}$, where the label states are at $0$ energy level and the dynamic states at $\pm \Delta En$. The schematic representation of the cycle to execute an FSM is shown in Fig.~\ref{fig11}. This protocol will be analyzed for an FSM. The protocol is independent of the prior knowledge of the states. We will now describe the protocol step by step. 

We consider an arbitrary stage for the beginning stage of the cycle of operation, where the label states $\vartheta_i$ have energy $0$. The corresponding dynamic state $Ds_{a}^{i\rightarrow i(a)}$ of the system for the specific label state $\vartheta_i$ have energy  $Enr = -\Delta En$, and the other dynamic state for symbol other than $a$ have energy $Enr= \Delta En$. We may visualize errors in the computation as one will find the machine in different initial states. Now we will describe the four stages of the cycle. 

\textit{Stage 1:} In this stage of the cycle, the dynamic states are transmitted to the energy level $\Delta En$. In this stage of the cycle, the bit flip operation is performed. The system in this stage of the cycle gets relaxed at the label state $\vartheta_j$ after a certain period of time. Before the transition of the cycle from stage 1 to stage 2, the system should be provided with some relaxation time.

\textit{Stage 2:} In this phase of the cycle, the energy barriers are removed. For the new configurations, new energy barriers are created. Such type of energy barriers helps us to reduce the energetic cost for the system and produce results with less error probability. This also helps in coarse-graining label states, which are required for the execution of the processes.

\textit{Stage 3:} In this phase of the cycle, the energy levels of the dynamic states are set based on the tape symbol of the system. Before the transition of the system to the next phase of the cycle, the system is provided some time for relaxation. In this stage of the cycle, we implement the $N-it$ setter. So the system enters the dynamic state $Ds_{a}^{i \rightarrow i(a)}$ for the label state $\vartheta_i$ with tape symbol $a$. Here the dynamic states $Ds_{a}$ are transmitted from the higher energy to the lower energy, whereas the other dynamical states remain at the high energy level.

\textit{Stage 4:} In this last phase of the cycle, the energy barrier is again removed and new are created for the outgoing configuration. So this stage brings back the system to its initial stage for the further execution of the process.

In this whole computational process, work is done only in two stages of the cycle, stage 1 and stage 3. In stage 1, work is done to raise the energy levels, and in stage 3, work is extracted to lower the dynamic state for the execution. So the work done in stage 1 can be expressed as 
\begin{equation}\nonumber
W_1= \frac{2 (r^+)^2}{(r^+)^2 + r^+ r^- + (N-1) (r^-)^2} \, \ln \left( \frac{r^+}{r^-} \right).
\end{equation}
And the work done in stage 3 can be expressed as 
\begin{equation}\nonumber
W_3 = -\frac{2 r^-}{r^+ +n r^-} \ln \left( \frac{r^+}{r^-} \right).
\end{equation}

So the total work done for the execution of the process is 
\begin{eqnarray}\nonumber
W_{total} & =  & \left(  \frac{2 (r^+)^2}{(r^+)^2 + r^+ r^- + (N-1) (r^-)^2} -  \frac{2 r^-}{r^+ +n r^-} \right)\\ \nonumber
 &\times & \ln \left( \frac{r^+}{r^-} \right),\\ \nonumber
 & = & 2 \ln \left( \frac{r^+}{r^-} \right) + \mathcal{O} \left(  \frac{r^-}{r^+}   \ln \left( \frac{r^+}{r^-} \right) \right).
\end{eqnarray}

For $r^+ >> r^-$ the results show no dependence on the symbol count and also on the size of the machine. Along with the energy cost for the execution of the cyclic process for the computation, there is also some additional energy cost. Some amount of energy is required to reset the machine to its initial state. So one can infer the thermodynamic cost is directly proportional to the number of states of an FSM but has no dependence on the number of computational steps that are executed in the process.

\section{Turing Machine (TM): Thermodynamic Interpretation}\label{sec16}

In this section, we are going to analyze the thermodynamic cost for the different physical models that are considered for Turing machines. First, we will discuss the entropic one-way-ness which confirms the existence of one-way computation. Secondly, we will analyze the reversible Turing machine using different physical models and evaluate its thermodynamic costs. Stochastic thermodynamics is considered for the analysis of the dynamics of this model of TM. From the analysis, we will be able to infer that the thermodynamic complexity is bounded, whereas the Kolmogorov complexity happens to be unbounded.

\subsection{One-way-ness}
For the analysis of one-way-ness in the input machine (Turing machine), we will consider Bennett's Turing machine model~\cite{bennett1989time,bennett1985fundamental}. Binary memory and the measured system are the logical structures of this model, and the logical states are controlled by the controlled-NOT gate (C-NOT).

\begin{figure}[h]
  \includegraphics[width=1.0\columnwidth]{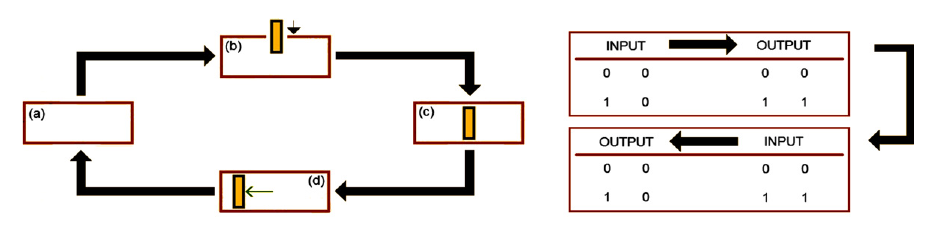}
  \caption{A pictorial representation of Bennett's algorithm for the Szilard one-molecule engine is shown. There exists a one-to-one correspondence between the initial and the final state for these reversible Turing machines. The thermodynamic analysis for this model comprises the transition that takes place in this cycle where non-random data is changed to random and vice versa in the cycle. [Adapted from Ref.~\cite{de2014one}. Copyright 2014 Elsevier ] }
  \label{fig12}
  \end{figure}

In Bennett's algorithm, two non-commutative stages are utilized for the restoration and measure of the states of the memory. To execute the process, a partition is inserted in the adiabatic box that divides it into two equal halves. It is assumed that no thermodynamic work is done during this insertion. Due to this insertion, the memory split up into two states. Before the execution of this process, the memory state is stored in the target bit by applying the C-NOT gate. Due to the insertion of the partition, the logical state of the measuring and the measured system gets correlated. The pictorial representation of the process is shown in Fig.~\ref{fig12}. The operation executed in this process randomizes the memory state.

Now we are left with the merging of the two states of the system. Due to the merging of the memory state, we are going to observe certain changes in some of the macroscopic variables. Thermodynamically this operation is represented by the isothermal expansion of the system, i.e., the adiabatic box.  This is equivalent to the process of compressing the piston in a Szilard engine model~\cite{szilard1929entropieverminderung}. To execute the merging of the states, the partition is moved toward the left side of the chamber. Due to this operation, the measured and the measuring system get detached from each other. This is represented by the fourth stage of the cycle shown in Fig.~\ref{fig12}. The system is reverted back to the initial stage, and the memory is restored to its initial state. This process is equivalent to the information erasure process. Due to this erasure, we can infer from Landauer's principle, that heat will be generated in the adiabatic chamber. This will cause an increase in the entropy by an amount of $\Delta S \geq k_B\, \ln\, (2)$. The thermodynamic reversibility is maintained by the equivalent amount of decrease in the entropy of the random data.

\begin{figure}[h]
  \includegraphics[width=1.0\columnwidth]{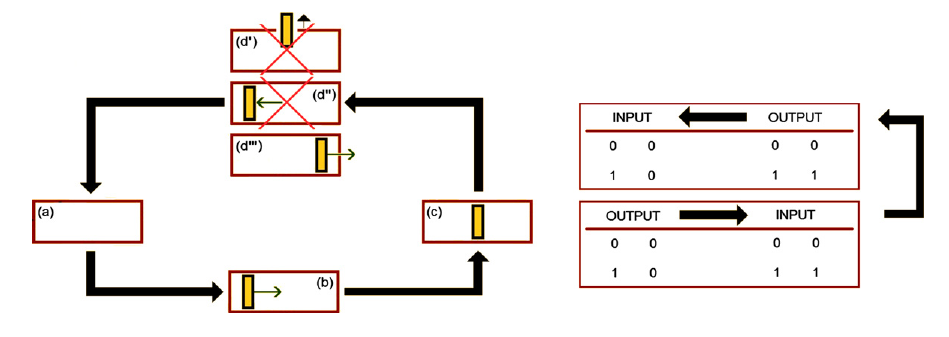}
  \caption{The reverse cycle for the two non-commutative stages is shown. Three approaches for the completion of the reverse cycle are considered, out of which one satisfies all the conditions. The other two processes are discarded as shown by the crossover on the box. The other two approaches were removing the partition wall without dissipation, and the other approach was to move the wall towards the extreme left of the chamber. In the adopted approach, the wall is moved towards the right of the chamber. [Adapted from Ref.~\cite{de2014one}. Copyright 2014 Elsevier]}
  \label{fig13}
  \end{figure}

Due to the thermodynamic reversibility of the process, one can reverse and rebuild the memory. Thermodynamically this operation can be described by the isothermal compression of the measured state as shown in Fig.~\ref{fig13}. The process is executed by moving the partition wall from the left side of the chamber with volume ($\mathcal{V}$), to the center of the chamber ($\frac{\mathcal{V}}{2}$) to distribute it into two equal half. During the execution of the process, heat is extracted, and it gets converted to thermodynamic work. So in the whole process where erasure and its reversal process take place, we observe no net entropy change, as shown in Eq.~\eqref{OWNT1}.
\begin{equation}\label{OWNT1}
\int_{\mathcal{V} \rightarrow \frac{\mathcal{V}}{2}} k_B \frac{d\mathcal{V}}{\mathcal{V}} + \int_{\mathcal{V} \leftarrow \frac{\mathcal{V}}{2}} k_B \frac{d\mathcal{V}}{\mathcal{V}} = 0.
\end{equation}

The problem that one encounters is the closing of the cycle. An approach to solve this problem has been proposed in~\cite{de2014one}. The approach one can think of is by moving the partition wall to the extreme right of the chamber as shown in Fig.~\ref{fig13}(d$^{\prime\prime\prime}$). The measured and the measuring system gets detached when the partition is moved to the extreme left of the chamber. Now we will discuss the method to execute this process.

The change that occur in the memory state $\mathcal{M_D}$ can be  expressed as 
\begin{equation}\label{OWNT2}\nonumber
\frac{d}{dt} \mathcal{M_D} = \mathcal{\phi_A} \langle \varsigma,  \mathcal{M_D} \rangle,
\end{equation}
where $\varsigma$ describes the outcome of the measurement, and $\mathcal{\phi_A}$ represents the matching metric that describes the correlation between the measuring and the measured system. If $\mathcal{\phi_A}=1$, then there exists a perfect correlation of the feedback control for Bennett's Turing machine. Whether one runs the forward cycle or the reverse cycle the dynamics can be expressed in terms of the correlation as 
\begin{equation}\label{OWNT3}
\frac{d}{dt} \mathcal{M_D} = \left\{
\begin{array}{lc}
   \varsigma  \mathcal{M_D}, &\  -ve\, feedback,\\
   -\varsigma  \mathcal{M_D}, & \  +ve\, feedback.
\end{array}\right.
\end{equation}

So, one can define the erasure in the reverse process from Eq.~\eqref{OWNT3} as
\begin{equation}\label{OWNT4} \nonumber
- \varsigma\Delta t \Big|_{+ve \, feedback}= \int_{\mathcal{M_D}_{a}}^{\mathcal{M_D}_{b}} = \frac{d\mathcal{M_D}}{\mathcal{M_D}}= \ln \left( 1+ \frac{\Delta \mathcal{M_D}}{\mathcal{M_D}_{a}} \right),
\end{equation}
where $\mathcal{M_D}_{a}$, $\mathcal{M_D}_{b}$ describes the initial and the final state. When $\frac{\Delta \mathcal{M_D}}{\mathcal{M_D}_{a}} = -\frac{1}{2}$, the information that is discarded in the process is $\varsigma = \ln 2$. This occurs when the two logical states get merged and prevent the loss of randomness. For the forward direction, the amount is $\varsigma = -\ln 2$. As the system is immersed in an isothermal heat reservoir, the erasure of information will have a thermal effect by a factor $k_B T$. So for the reverse cycle, the system prevents the entropy increase by an amount of $\Delta S \leq k_B \, \ln 2$, where the equality holds for the perfect correlation. For the forward cycle, we encounter the reverse process where the system prevents the decrease in the entropy of the system by an amount $\Delta S \geq k_B \, \ln(2)$, where the equality holds for the perfect correlation. So one can infer that the entropic one-way-ness guarantees the existence of a one-way permutation.

\subsection{Thermodynamics of reversible TM}
Thermodynamically, one can consider a machine to be coupled with a heat reservoir at temperature $T$ and a work reservoir. The primary motivation of the work reservoir is to drive the computational process in a definite direction. The Turing machine defined in section~\ref{sec3} is irreversible in its form. If in computation, we overwrite the input tape by the output one, then the computation will be an irreversible one. So to have a reversible TM, it demands two tapes one for the input and the other for the output. In a reversible TM, we will be able to retrace the computational path and trace back to the initial state of the TM.

A logically reversible TM for the computational process was proposed by Bennett. In his work, he has shown that an irreversible TM needs four times less number of steps for the execution of a computation, than that of the logically reversible TM.  Now in a recent work~\cite{strasberg2015thermodynamics}, they have modified Bennett's treatment. Bennett in his work considered a single computation, whereas in this model they develop a TM which has the power to process a continuous stream of input string like ($\dots$, $a$, $\mathcal{S}_{inp}^{\prime}$, $a$, $\dots$, $a$, $\mathcal{S}_{inp}$, $a$, $\dots$). The input tape is an infinite tape with an input string described by  $\mathcal{S}_{inp}^{\prime}$,  $\mathcal{S}_{inp}$. They will be separated from each other by blank symbols. The blank symbols will denote the beginning and the end of the input string. Following this, the output string can be described as  ($\dots$, $a$, $\mathcal{S}_{out}^{\prime}= \mathbb{U}(\mathcal{S}_{out}^{\prime})$, $a$, $\dots$, $a$, $\mathcal{S}_{out} = \mathbb{U}(\mathcal{S}_{out})$, $a$, $\dots$), where $\mathbb{U}: \mathcal{S}_{inp} \rightarrow \mathcal{S}_{out}$. This model is equivalent to the model studied in other works~\cite{mandal2012work,barato2013autonomous,mandal2013maxwell,deffner2013information,barato2014stochastic,strasberg2014second}, where one manipulates the external tape to extract work.

The new TM model consists of four tapes, out of which two are the input and the output tape, and the other two are working tape and history tape. The machine comprises the working and the history tape, and the other two tapes are provided externally. This model also consists of a computation cycle. It is a five-stage cycle. The stages of this cycle are described as follows:

\textit{Stage 1:} In the first stage of the cycle, the system receives an input.  This input is copied from the input tape to the working tape.

\textit{Stage 2:} In the second stage of the cycle, the system performs the computation process. So after the execution of the computation, the working tape has the output data. During the process, the history tape keeps the track of each step of the computational process to satisfy the reversibility of the model.

\textit{Stage 3:} In this phase of the cycle, the output data in the working tape is copied to the output tape, and the working tape is reset to zero.

\textit{Stage 4:} This stage of the cycle is just the reverse of the \textit{Stage 2}. The reverse computation is done in this stage of the cycle using the data of the history tape, and then the history tape is reset.

\textit{Stage 5:} In the last stage of the cycle, the working tape having the input data is copied to the input tape, and the working tape is reset to  zero. This process complete the computational cycle.

This TM can be modeled by a continuous Markov process. The Markovian master equation for this Markov process for the system $\mathcal{Y}$ with states $\mathbb{Y}$ is 
\begin{equation}\nonumber
\frac{d}{dt} p_{y} (t) = \sum_{y^{\prime}} \mathcal{W}_{y,y^{\prime}}  p_{y^{\prime}} (t),
\end{equation}
where $\mathcal{W}_{yy^{\prime}}$ represents the rate matrix, and $p_{y} (t)$ is the probability to find the system in the state $y \in \mathbb{Y}$. Now for the thermodynamic interpretation of this TM, we will consider that each state of the system will be associated with an energy $E_y$. Due to this, an additional property is attached to the rate matrix, which is defined as $\ln [\mathcal{W}_{y,y^{\prime}}|\mathcal{W}_{y^{\prime},y}] = - \beta (E_{y} - E{y}^{\prime})$. So, the Markovian master equation gets associated with the reversible TM by allowing the transition in the computation process for both forward as well as backward directions. In this model, it has been taken care that each state has at most two adjacent states the predecessor and the successor state. If there exist more branches then the system will have multiple predecessors and successors, which will cause the computational process to lose its reversibility. So the final form of the Markovian master equation is 

\begin{eqnarray} \nonumber \label{TRTM1}
\frac{d}{dt} p_{\eta} (t) & = & - \left( \mathcal{W}_{\eta+1,\eta} - c_{\eta-1,\eta} \right) p_{\eta}(t)\\ 
& +  &  \mathcal{W}_{\eta,\eta+1} p_{\eta+1}(t) + \mathcal{W}_{\eta,\eta-1} p_{\eta-1}(t),
\end{eqnarray}
where $\eta \in \mathbb{Z}$, and $\mathcal{W}_{\eta,\eta+1}$, $\mathcal{W}_{\eta,\eta-1}$ describes the forward and the reverse rate respectively. The rate matrix $\mathcal{W}$ will be decomposed into blocks for each input during the computation process in stage 2 and stage 4. The transition between the different blocks of the rate matrix is prohibited during the computation process.

For the thermodynamic analysis of this model, we will couple the system with an energy landscape, which happens to be linear in its form for the computational path, i.e., the logical and the successor state are separated from each other by a constant amount $\epsilon_c$. Due to this addition of energy, the forward and the backward transition rate should obey the local detail balance. This in turn fixes the temperature for the environment. Now we will move on to the thermodynamic interpretation of the computation process using this TM model.

We will analyze the thermodynamics of this model at a coarse-grained level, subject to the condition that the system (computer) was active for quite some time, so that the variance ($ \langle \eta^2 \rangle - \langle \eta \rangle^2$) is large than unity.  The linearity of the energy is necessary as we are interested in the steady state regime for the analysis. So the rate of transition for this model can be described as
\begin{equation}\nonumber
\mathcal{W}_{\eta,\eta+1}= \Gamma_a e^{-\beta \epsilon_c/2} \quad \mathcal{W}_{\eta+1,\eta} = \Gamma_a e^{\beta \epsilon_c/2},
\end{equation}
where $\Gamma_a$ represents the overall time scale. The Fokker-Planck equation (FPE) using Eq.~\eqref{TRTM1} for this model can be expressed as 
\begin{equation}\nonumber
\frac{1}{\Gamma_c} \frac{\partial}{\partial t} p_{\eta}(t) = \frac{\partial}{\partial \eta} \left(-2 \sinh (\frac{\beta \epsilon_c}{2})  + \cosh (\frac{\beta \epsilon_c}{2} ) \frac{\partial}{\partial \eta}  \right) \, p_{\eta}(t).
\end{equation}

The expectation value of the parameter $\eta$, which describes the count of the computational steps, and the speed of the computational process is described as
\begin{eqnarray}\nonumber \label{TRTM2}
\langle \eta \rangle (t) = 2 \Gamma_c \,t \, \sinh \left(\frac{\beta \epsilon_c}{2} \right), \\ 
vel \equiv \frac{d}{dt} \langle \eta \rangle (t) = 2 \Gamma_c \, \sinh \left(\frac{\beta \epsilon_c}{2} \right),
\end{eqnarray}

where $vel$ depicts the speed of the computational process. So the variance over the parameter which counts the number of computational steps can be expressed as 
\begin{equation}\nonumber
\langle \eta^2 \rangle (t)- \langle \eta \rangle^2 (t) =  2 \Gamma_c \,t \, \cosh \left(\frac{\beta \epsilon_c}{2} \right). 
\end{equation}

The entropy cost for the execution of the process for this model which one can describe by the Shannon entropy is evaluated as 
\begin{equation}\nonumber
S(t) = -\int d\eta\, p_{\eta} (t)\, \ln p_{\eta} (t) =  \frac{1}{2} \, \ln \left(4 \pi  \Gamma_c \,t \, \cosh \left(\frac{\beta \epsilon_c}{2} \right) \right).
\end{equation}

Using Eq.~\eqref{TRTM2}, we can  transform  the Shannon entropy in terms of the parameter $\eta$ as 
\begin{equation}\nonumber
S(t) = \ln \left(2 \pi  \Gamma_c \,t \, \cosh \left(\frac{\beta \epsilon_c}{2} \right) \langle \eta \rangle (t) \right).
\end{equation}

So the entropy production rate, which in other terms is described as the rate of change of the Shannon entropy for the computational  process can be expressed as 
\begin{eqnarray}\label{TRTM3} \nonumber
\dot{S}(t) & = & \frac{d}{dt} S(t) + \beta\, \epsilon_c \, vel, \\ 
& =  & \frac{1}{2t} + 2 \Gamma_c \, \beta \epsilon_c \, \sinh \left(\frac{\beta \epsilon_c}{2} \right) \geq 0.
\end{eqnarray}

Therefore the defined model works in a thermodynamically reversible manner without any dissipation. The rate of entropy production becomes small for the condition where $\eta \rightarrow 0$ doesn't imply, and the overall production rate of the entropy throughout the process will be zero. This model also satisfies Norton's notion regarding this matter~\cite{norton2013brownian,norton2014brownian}.

We will now explore a very recent work~\cite{kolchinsky2020thermodynamic} in this direction. They have considered that their physical system is associated with heat reservoirs, and its dynamics are influenced by the driving schemes~\cite{parrondo2015thermodynamics,van2013stochastic,esposito2010three}. They have considered stochastic thermodynamics for the analysis of the dynamics of these physical processes. Interested readers can go through the review article~\cite{arrighi2019overview} which provides a detailed analysis of the stochastic thermodynamics in different aspects of computation.  The state of this physical system can be equated to some logical state of a Turing machine.  In this work, the authors have considered two physical processes for TM, and have analyzed three thermodynamic quantities for each physical process. The three quantities that are analyzed are as follows:

(1) The heat generated during the execution of the realization of TM will be processed for each input $z$. It is denoted as $Q(z)$. 

(2) The heat generation for the entire computation that mapping the input $z$ to the output $y$. 

(3) The average heat generation $\langle Q\rangle$ that minimizes the entropy production in the physical process while evaluating the input.

The physical process that is considered for analysis is the `coin-flipping' process for the universal Turing machine (UTM). This physical model is a thermodynamically reversible model, where the input is samples of the `coin-flipping' distribution $p(z) \propto 2^{l(z)}$, where $l(z)$ depicts the string length. The heat generation for this process of UTM is described as
\begin{equation}\label{TRTM4}
Q_{cofp}(z) = l(z) + K(UTM(z)) + \mathcal{O}(1),
\end{equation} 
where $UTM(z)$ describes the output of UTM for the given input $z$, and $K(UTM(z))$ describes the \textit{Kolmogorov complexity} for the given input string $z$. From the definition given in Eq.~\eqref{TRTM4}, one can infer that the thermodynamic complexity is a bounded function.

Being motivated by the \textit{physical Church-Turing thesis}, the alternative physical process that the authors have considered in their work is a semi-computable process coined as \textit{domination realization}. Similarly to the first physical process, the heat generation or rather heat function of the process is described as 
\begin{equation}\label{TRTM5}
Q_{dore} (z) = K(z|TM(z)) + \mathcal{O} (1),
\end{equation}  
where $K(z|TM(z))$ represents the condition Kolmogorov complexity for the TM. This heating effect holds for this process even if $TM$ is not even a UTM. For a semicomputation process we have $Q_{dore} (z) \leq Q(z)  + \mathcal{O} (1)$. This conveys the fact that the heat generation in this process is less than the heat generation for any other semicomputable process of TM.

For a given universal Turing machine, one can define the Kolmogorov complexity for a given bit string $z \in \mathbb{S}^\star$ as 
\begin{equation}\nonumber
K_{UTM} (z) = \min_{UTM(z)=y} l(z). 
\end{equation}

The defined Kolmogorov complexity is an unbounded function. Now we will analyze the `coin-flipping' process in detail. The coin-flipping distribution is expressed as 
\begin{equation}\label{TRTM6}
CP(z) = 2^{-l(z)} \delta(f(z),y) \quad z\in dom \,\, UTM.
\end{equation}
where it is conveyed that the $UTM$ will not halt, if the input $z \notin dom \,\, UTM$. $CP(z)$ describes the probability of successive feeding of the bits in the UTM, where this UTM halts after receiving $z$ number of bits. We will now calculate the thermodynamic cost of this physical process. To do this, we first normalize the distribution shown in Eq.~\eqref{TRTM6} as 
\begin{equation}\nonumber
p_{z}^{cofp} = \frac{CP(z)}{\sum_{z\in dom \,\, UTM} CP(z)}.
\end{equation} 

Similar to this we can define the universal distribution $CP(y) = \sum_{UTM(z) =y} 2^{-l(z)}$ and its normalized distribution is $p (UTM(z)) =\frac{CP(UTM(z))}{\sum_{z\in dom \,\, UTM} CP(UTM(z))}$. The heat function for this coin-flipping distribution is expressed as 
\begin{eqnarray}\nonumber \label{TRTMa10}
Q_{cofp} (z)& = & l(z) + \ln CP(UTM(z)),\\ 
& = & l(z) - K(UTM(z)) + \mathcal{O}(1).
\end{eqnarray}

The minimum amount of heat required for the generation of the output for this physical process can be described as
\begin{equation}\label{TRTMa11}
\min_{UTM(z)=y} Q_{cofp}(z) = K(y) + \ln CP(UTM(z)).
\end{equation}
Using Eq.~\eqref{TRTMa10} in Eq.~\eqref{TRTMa11} we have 
\begin{equation}\nonumber
\min_{UTM(z)=y} Q_{cofp}(z) = \mathcal{O} (1).
\end{equation}

The average heat that is generated for a set of input strings generated from the distribution of  $p_{z}^{cofp}$ is 
\begin{equation}\nonumber
\langle Q \rangle_{p_{z}^{cofp}} =  \Big[ S(p_{z}^{cofp}) - S(p (UTM(z))) \Big] + \Upsilon (p_{z}^{cofp}),
\end{equation}
where $\Upsilon (p_{z}^{cofp})$ describes  the entropy production in a thermodynamic process.

Now, we will analyze the second physical process coined as \textit{dominating realization}. The heat function for this process has two properties, one is that it is semi-computable, and the second is that it is optimal for the physical process.  The associate function for a given TM, which is not universal is defined as 
\begin{equation}\nonumber
\mathcal{G}_b (z) = K(z|TM(z)).
\end{equation}
For a defined TM the heat function $Q_{dorl}$ happens to be an upper semicomputable function. To compute the heat function for this process, let us consider a TM that reads some long and incompressible data of $m_v$ bits for some input program $ip$. The heat function for this process can be evaluated as 
\begin{equation}\nonumber
Q_{dorl} (ip)= K(ip|TM(ip)) \approx m_v. 
\end{equation}

Now when we consider the TM to be a universal TM, it guarantees us that there exists some desired output for a well-defined program. This provides us with the required element of information for the analysis of the thermodynamic complexity of this physical process. So we can now compute the minimum amount of heat that is required for the computational process to execute its operation to give a desired output $y$. The amount is bounded by a constant and is expressed as 
\begin{equation}\nonumber
\min_{UTM(z) = y} Q_{dorl} (z) = \mathcal{O} (1). 
\end{equation}
Through a thorough analysis, it has been conveyed that this bound holds even if the TM is not an universal TM.

Finally, we can analyze the expected heat that is being generated while this physical process executes some computation. For the analysis, we will consider the input to be a random sample from the input distribution. For the comparison between the two processes, we will consider that the input distribution will result in minimum entropy production. The analysis shows that the expectation of heat that is being generated by this physical process during the execution of computation for a given input distribution is infinite.

To get a clear view of the two physical processes we will give a short comparison between the physical processes. 

(1) For both the physical process, the minimum amount of heat that is required for the computation to generate output $y$ is bounded by a constant. The constant defined for these two processes has no relation among them, but the thermodynamic cost for the dominating physical process is larger in principle than that of the coin-flipping.

(2)  For the coin-flipping process, one has to know the shortest route for the output $y$ to get the bounded form of the heat production, whereas, in the case of the dominating process the condition is quite simple and advantageous. The condition says that we can get the bounded form if the computation is fed by input which demands the print of the output $y$.

(3) The heat function happens to be an upper semicomputable function for the dominating physical process, whereas it is a lower semicomputable function for the coin-flipping process.  So $Q_{dorl} (z)- Q_{cofp} (z) > c_\gamma$ where $c_{\gamma}$ is a real number. The excess amount of heat that is generated in the dominating physical process is bounded. It is expressed as $Q_{dorl} (z)- Q_{cofp} (z)  \leq \ln K(UTM(z))$.

The above methods have provided insight into the physical realization of the TM from a thermodynamic viewpoint. Turing machine happens to be the center of attraction to both physics and computer science. So the physical realization of TM with a more feasible and realistic model needs further investigation.

\section{ Error Correction: Thermodynamic Interpretation}\label{sec18}
After the advent of quantum error correction theory~\cite{shor1995scheme,calderbank1996good,steane1996error}, it has experienced a rapid development. Some of the quantum codes were just mapped from the classical error correction codes mainly the CSS codes~\cite{steane1996error}. It was then further generalized to the stabilizer codes~\cite{calderbank1997quantum,calderbank1998quantum,gottesman1996class,gottesman1997stabilizer}. Firstly, we will discuss an unsophisticated model for the analysis of error correction as described in the work~\cite{vedral2000landauer}. Here in this work, the authors have considered a reversible cycle to explain the error correction process. This cycle is a modified version of Bennett's `Maxwell's demon'. We will explain the classical codes from a thermodynamics point of view, then we will explore the quantum error correction theory using this reversible cycle. From a thermodynamics viewpoint, quantum error correction is equivalent to a refrigerator process. In this process, one tries to sustain a steady entropy for the system even when it is subjected to environmental noises. Due to the decrease in entropy, one might intuitively think that the protocol defined for the quantum error correction violates the second law of thermodynamics. A careful analysis of the process will show that it does not violate the second law.

The author in their work~\cite{vedral2000landauer} has considered a singlet atom confined in a box to link the information with thermodynamics. The classical information can be modeled using this system, where the atom when happens to be on the `left-hand side' (LHS) is described as 0, and when on the `right-hand side' it will be described as 1. Now we allow the system to expand isothermally considering that the atom exists on one of the sides. Due to this expansion, one encounters an increase in the entropy by an amount of $\Delta S= k_B\, \ln 2$. The atom has an equally likely probability to jump the either side of the box from its initial state. This comprises the errors that can occur in the system. The protocol for the error correction is described below:

(1) We will consider that the atoms are in the LHS and RHS of the respective boxes. 

(2) Let us consider that some error occurs to the particle in the box $\mathcal{A}_\alpha$. The error is defined as the probability of the atom in the box $\mathcal{A}_\alpha$ to be in the LHS or RHS of the box. 

(3) Another system $\mathcal{B}_\alpha$ is kept to keep a track on the system $\mathcal{A}_\alpha$. The system $\mathcal{B}_\alpha$ correlates itself with the system $\mathcal{A}_\alpha$. 

(4) Based on the state of the system $\mathcal{B}_\alpha$ one will move the system $\mathcal{A}_\alpha$ to its respective side. This leaves the system $\mathcal{B}_\alpha$ in a randomized state.  The execution of this process needs no work.

(5) This is the last step of the protocol where the system $\mathcal{B}_\alpha$ is brought back to its initial state by isothermal compression.

\begin{figure}[h]
  \includegraphics[width=1.0\columnwidth]{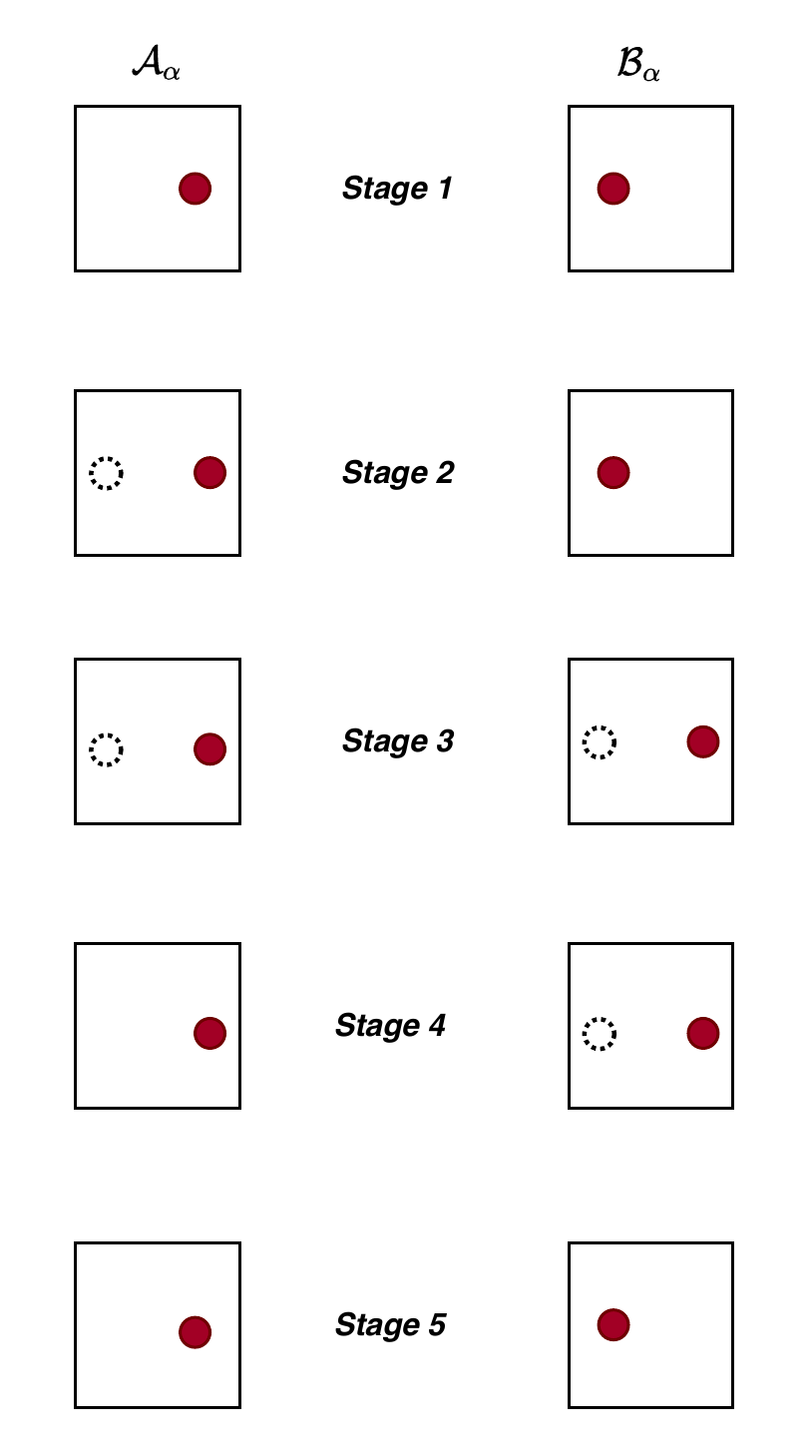}
  \caption{A pictorial representation of the reversible cycle which implements the classical error correction protocol is depicted. A detailed description of each stage of the cycle is given in the text.}
  \label{fig14}
  \end{figure}

The pictorial representation of the protocol is described in Fig.~\ref{fig14}. The energy of the atom in the system $\mathcal{A}_\alpha$ is $k_B T \, \ln 2$. Due to the error, the system undergoes a decrease in the energy by an amount $\Delta F = -k_BT\, \ln  2$, and the entropy of the system increases by an amount $S=k_B \ln 2$. In the third stage of the cycle, the system  $\mathcal{B}_\alpha$ has the information of the $\mathcal{A}_\alpha$ as they happen to be correlated. In the final stage of the cycle, the $k_B T \, \ln 2$ amount of work is done in the system $\mathcal{B}_\alpha$ to reset the system. The amount of entropy change during this stage of the process is $S= -k_B \,\ln 2$.   So the error that has been described in this model is nothing but the inability to do work. So to make the system free of this inability the system is correlated to another system. This makes the second system possess the same inability. In the reset, step entropy is wasted so the system regains its initial state. Now we will analyze the error correction in the quantum realm using this protocol.

The main motivation of the quantum error correction theory is to preserve the quantum state. The protocol is described below for the pure states. One can similarly describe the protocol for the mixed states. The protocol is as follows:

(1) The state of the system after the introduction of the error is described as $\sum_{j} Err_j |\psi_{cs}\rangle |m\rangle |env_{j}\rangle$, where $|\psi_{co}\rangle$ describes the encoded state, $|m\rangle$ represents the state of the measurement device, and the environmental state is described $|env\rangle_j$. The environmental states are orthogonal in nature. Here, it has been assumed that $\langle env_j|env_k\rangle = \delta(j,k)$.

(2) Now in the next phase of the protocol, the environmental effect is traced out. The state of the system after tracing out is $\sum_j Err_j |\psi_{co}\rangle \langle \psi_{co}| Err_j^{\dagger} \otimes |m\rangle \langle m |$.

(3) In this phase of the protocol, one observes the system, due to which correlation is generated between the measurement apparatus and the system. The state of the system is described as $\sum_j Err_j |\psi_{co}\rangle \langle \psi_{co}| Err_j^{\dagger} \otimes |m_{j}\rangle \langle m_{j} |$, where $\langle m_j | m_k \rangle = \delta(j, k)$. We have to keep in mind that if the observation is not perfect, then the protocol will not work properly, as we will not be able to get the orthogonal states of the system.

(4) During this phase of the cycle, the system gets uncorrelated. In this phase, the correction of the error is done. The state of the system after the execution of this step is $ |\psi_{co}\rangle \langle \psi_{co}|  \otimes \sum_j |m_{j}\rangle \langle m_{j} |$. This is not the exact state that we have started with. We have to reset the measurement state to its initial state.

(5) This is the last stage of the protocol, where we reset the measurement system to its initial state.  To do this, we include a garbage system to the system. Now, the state of the system is $ | \psi_{co}\rangle \langle \psi_{co}|  \otimes \sum_j |m_{j}\rangle \langle m_{j} | \otimes |m\rangle \langle m |$. The state of the system by swapping the garbage system and the measurement system can be expressed as $|\psi_{co}\rangle \langle \psi_{co}| \otimes |m\rangle \langle m | \otimes \sum_j |m_{j}\rangle \langle m_{j} |$. Now the system is reset for another cycle of quantum error correction.

Similar to the analysis done for the classical code, we will analyze the entropy change that takes place throughout the protocol. The entropy of the system after the first stage of the cycle is $\Delta S = S(\rho_{ab})$, where $\rho_{ab} = \sum_{j} Err_j |\psi_{co}\rangle \langle \psi_{co} | Err_j^{\dagger} $ is the density matrix. We don't encounter a change in the entropy for the other steps except for the reset step.  The entropy of the system during this reset step is $\Delta S = - S(\rho_{ab})$. The garbage system encounters a gain in the entropy of the system.  The information gained by the system during the correlation step is $S(\rho_{ab})$. So from the observation, we can infer that the gain in the entropy by the garbage system is more than the information gain. This confirms that the error correction protocol is successful from a thermodynamic viewpoint.

In the work~\cite{cafaro2014entropic}, the author has further extended the model described in the work~\cite{vedral2000landauer} to include the non-maximally mixed states for the analysis of quantum error correction. In their work, they have conveyed that their system exhibits a constant entropy even when subjected to the impact of environmental noises. In their model, they have introduced some ancilla qubits which keep track of the error. In the refrigeration process, this is equivalent to the transfer of information of the error from the data qubit to the ancillary qubit. This causes an entropy change and cools down the data qubits. They have also explored the changes that occur in the quantum error correction cycle when the measurement system is not perfect, and also the information gain is not optimal. Another important aspect, i.e., how the non-orthogonal states will have an impact on the thermodynamic analysis of the error cycle is also addressed in this work.

Now we will explore a modern formalism to describe quantum error-correcting conditions from a thermodynamic viewpoint as studied in the work~\cite{korepin2002thermodynamic}. Different interactions in a system have been modeled through various Hamiltonian, like the spin chain model. One has encountered applicability of the spin model in quantum information theory like in space free from decohorence~\cite{kempe2001theory}, entanglement~\cite{osterloh2002scaling}, gates~\cite{divincenzo2000universal}, and even in topological quantum computing~\cite{kitaev2003fault,freedman2003topological}. Here in this work, the authors have used spin chain formalism to develop the condition of the quantum codes.

In the seminal works~\cite{preskill1998reliable,gottesman1998theory,bennett1996mixed,knill1997theory}, they have proposed the necessary and sufficient condition for the quantum codes to correct the errors. The condition to correct the errors $err$ that appear during transmission, or during storage in quantum devices can be defined as $\langle \psi |Err_{a}^{\dagger}Err_b |\psi\rangle$, where $\psi$ is a vector $\in \mathbb{C}$, and $Err_{a}$, $Err_{b} \in err$. The errors are defined as a set of linear operators $err = \{Err_{a}: \mathbb{B} \rightarrow \mathbb{B}\}$, where $\mathbb{B}$ represents the encoding space. This encoding space has a set of orthonormal basics defined as $\{|\psi_i \rangle \}$. So, one can alternatively define the condition for quantum error correction as $\langle \psi_j |Err_{a}^{\dagger}Err_b |\psi_i\rangle =\varpi_{ab} \delta(i,j)$, where $\varpi_{ab}$ is a constant.

For the analysis, the authors have first considered the XXO model. The Hilbert space for this model is $\mathbb{B}\simeq (\mathbb{C}^2)^{\otimes n_c}$, where $n_c$ represents the size of the model or one can convey it as the length of the lattice. The Hamiltonian of the system is described as 
\begin{equation}\label{QECT1}
\mathcal{H} = -\sum_{i=1}^{n_c} \left(\sigma_{j}^{x} \sigma_{j+1}^{x} + \sigma_{j}^{y} \sigma_{j+1}^{y} + \iota \, \sigma_{j}^{z} \right),
\end{equation}
where $\sigma^a$ for $a=x,y,z$ describes the Pauli matrices and $\iota$  describes the magnetic field, and it is real in its form. This spin chain model is also called as ``isotropic XY model", and was proposed in the seminal work~\cite{lieb1961two}. For any thermodynamical analysis, we need the partition function $Z$ of the system, then we can extract the information of other thermodynamic variables. The partition function for this system for $T>0$ can be evaluated as
\begin{equation}\label{QECT2}
Z = tr (2^{-\mathcal{H}/T}) = \sum_{En} 2^{n_cS} 2^{-En/T},
\end{equation} 
where $2^{nS}$ describes the degeneracy that exists within the energy levels, and S depicts the entropy of the system. For this system,  we observe that the energy and the entropy are directly proportional to $n_c$. 

Now for this model, the partition function of the system can be evaluated as 
\begin{eqnarray}\nonumber
Z & = & tr_{\mathcal{C}1} (2^{-\mathcal{H/T}})\\\nonumber
& \sim & \frac{n}{\pi} \int_{-\pi}^{\pi} \ln (1+ 2^{-e(p)/T}) \, dp,   
\end{eqnarray}
where $e(p)= -4\cos\, p +2\iota$. Here, the trace is taken over the subspace $\mathcal{C}1$. So, the free energy for the system can be described as $F_{en} = \frac{1}{2\pi} \int_{-\pi}^{\pi} \ln (1+ 2^{-e(p)/T})$. So the entropy of the system can be described as 
\begin{eqnarray}\label{QECT3}\nonumber
S =  \int_{-\pi}^{\pi} \Big(\frac{1}{2\pi} \ln \left(\frac{1}{2\pi}\right) - \mathcal{J}(p) \, \ln [\mathcal{J}(p)]\\ 
 - \{\left(\frac{1}{2\pi} \right)- \mathcal{J}(p)\}\, \ln \Big[\left(\frac{1}{2\pi}\right)-\mathcal{J}(p)\Big] \Big), 
\end{eqnarray}
where $\mathcal{J}(p_i) = \frac{1}{n_c(p_{i+1}-p_i)}$. Here $[p]$ denotes the collection of the momentum of the particles. Now we define a correlation function $ \langle \mathbb{O}\rangle_T$, where $\mathbb{O}$ represents the product of Pauli operators in linear combination. The mathematical definition of this function is defined as 
\begin{equation}\nonumber
\langle \mathbb{O}\rangle_T  = tr \frac{2^{-\mathcal{H}/T} \mathbb{O}}{Z}.
\end{equation}
This is also called the thermodynamic correlation function. For this model, the correlation is evaluated as
\begin{equation}\nonumber
\langle \mathbb{O}\rangle_T = \lim_{n_c \rightarrow \infty}  \frac{\langle [p] | \mathbb{O}|[p]\rangle}{\langle [p]|[p]\rangle}.
\end{equation}

The eigenvector for this system happens to be in thermo-equilibrium space with the same $\mathcal{J}$. The thermodynamic correlation function is independent of the wave vectors $|\psi\rangle$.

In the field of statistical mechanics the probability that the system collapses in an eigenstate $|\phi_b\rangle \in \mathbb{B}$ with eigenvalue $E(|\phi_b\rangle \in \mathbb{B})$ is $\frac{2^{-\mathcal{H}/T}}{Z}$. So the subspace of this system is at the thermo-equilibrium state and one can compare this with Schumacher's subspace. This subspace asymptotically converges to the quantum error correction criteria for all types of errors. One can equivalently define the quantum codes to be an approximation of the `thermo-equilibrium space'.  To get a clear idea about this spin chain model in quantum error correction, practical implementation of this process using gates and measurement tools are required.

\section{Brownian computer: Thermodynamic interpretation}\label{sec20}

Here we will explore the thermodynamic properties of the Brownian computer~\cite{norton2013brownian}. To analyze, we will first study the expansion of a single molecule gas. Using this expansion model we will explore Brownian computers with different constraints. Finally, from all this analysis, we will be able to infer that Bennett's claim about the necessary condition of thermodynamic reversibility for the operation of Brownian computers is unsustainable.

\subsection{Single molecule gas expansion}

Let us consider an ideal singlet gas molecule at a particular temperature $T$. We will execute a process, where the molecule will have $m$-fold expansion. The thermodynamics of this process will be analyzed for this particular temperature $T$. A large chamber is divided into $m$ parts by partitions. At the initial stage, the gas molecule is at the first cell of volume $V$ of the $mV$ volume chamber. The partitions are removed and the single gas molecule expands into the large-volume chamber.

The system Hamiltonian is 
\begin{equation}\label{smge1}
H = \mathbb{L} (p),
\end{equation}
where $p$ is the momentum of the molecule. The Hamiltonian of a system is generally a function of spatial and momentum coordinates. Interestingly for this system, the Hamiltonian is independent of the spatial coordinate of the molecule. It has been assumed in this model that, the axis of the chamber and the spatial coordinate are aligned.  

The molecule at thermal equilibrium exhibits Boltzmann distribution over the considered phase space. Mathematically,
\begin{equation}\label{smge2}
\Pi_a (x,p) = \frac{exp(-H/k_BT)}{Z},
\end{equation}
where the chamber is considered to extend from $x=0$ to $x= L$. $Z$ represents the partition function of the system and $k$ is the Boltzmann constant. The partition function for this system is 
\begin{equation} \label{smge3}
Z = VL \, \int_{p} e^{\Big(-\mathbb{L} (p)/k_BT \Big)} dp.
\end{equation} 

The information on the partition function provides sufficient information for the analysis of the thermodynamic variables. With the help of Eq.~\eqref{smge3}, we can describe the thermodynamic entropy for the system as

\begin{equation}\label{smge4}
S = \frac{\partial}{\partial T} (k_BT\,\, \ln Z) = k_B \, \ln (VL) + C_p(T),
\end{equation}
where the contribution from the momentum perspective is included in the constant $C_p(T)$. The entropy exhibits logarithmic dependence over volume as the Hamiltonian of the system shown in Eq.~\eqref{smge1} has no dependence on position coordinate.

So, the entropy change in this n-folded expansion of the gas molecule is $S = k_B \, \ln (mV) - k_B\, \ln V = k_B \, \ln\, m$. There is no change in the mean energy while this process is executed. So, the free energy $F = E - TS$ (where $E$ is the mean energy) of the system is 
\begin{equation}\label{smge5}
F = - k_B T \, \ln \, m.
\end{equation}
This is equivalent to $F= -k_BT \, \ln Z$. So, we can infer that the system is not driven by any external force but by the entropic force.

\subsection{Undriven Brownian computer}
From the point of view of thermodynamics, one can consider a Brownian computer to be equivalent to the singlet molecule expansion. First, we will describe the undriven Brownian computer, i.e., the system is not driven by any external energy force. We have subdivided the analysis into two parts. One is without the trap and the other is with a trap. A trap is the energy gradient from the existing energy of the particle where the molecule gets trapped.

\vspace{0.2in}
\begin{center}
\textit{Without Trap}
\end{center}

\vspace{.15in}

Here we discuss the Brownian computer which is undriven by any energy force,  and also without any trap. Alike singlet molecule expansion, the Brownian particle has the freedom to expand irreversibly into its confined configuration space. Though it is an irreversible process, it will terminate the computational process of the Brownian computer. This model describes the minimum amount of entropy that is being created in all forms of Brownian computers.

\begin{figure}[h]
  \includegraphics[width=1.0\columnwidth]{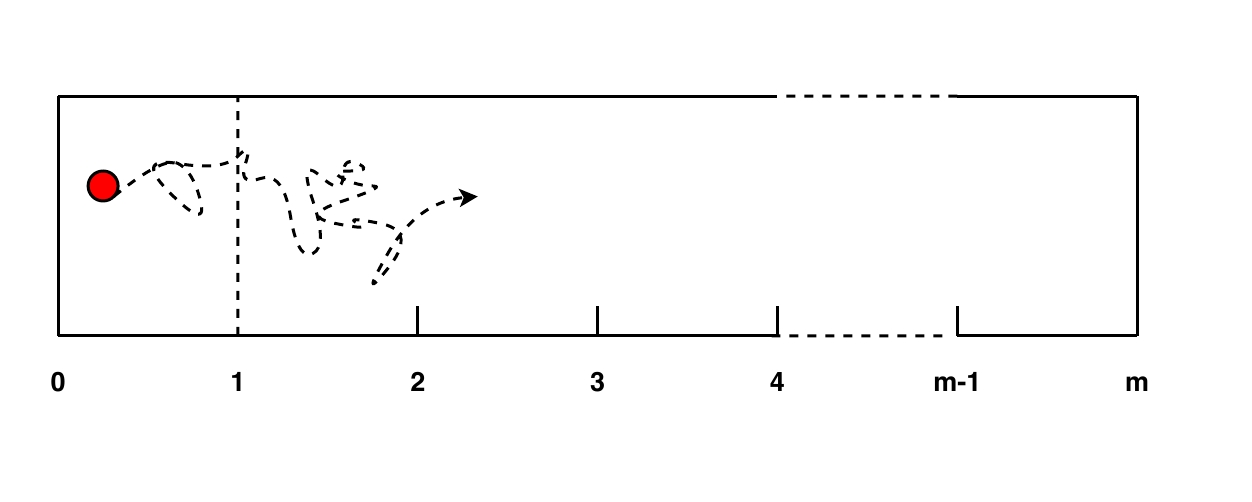}
  \caption{A schematic diagram of the Brownian computer without any energy trap is depicted.}
  \label{fig7}
  \end{figure}

In this model, we consider that the spatial configuration of the Brownian computer has the same energy throughout the channel of the phase space. No external energy force to drive the system is applied throughout the process. So the Hamiltonian of this model is equivalent to the Hamiltonian of the singlet molecule shown in Eq.~\eqref{smge1}. So, this model is independent of the spatial coordinate and has only one degree of freedom, i.e., the momentum of the Brownian particle. The analysis for this model is thus equivalent to that of the singlet molecule expansion. The pictorial representation of this model is shown in Fig.~\ref{fig7}.

For our analysis, similar to that of singlet molecule expansion, we distribute the configuration space of the computer into $m$ stages. Each of these states is equivalent to the square blocks of the Turing machine. We will consider that each state of the configuration space of the computer occupies volume $V$. We will set a counter $\lambda_a$ which will keep an update of the progress, and the states that have been accessed.

To execute the process, the initial state of the computer is defined by confining the Brownian particle to the first stage, i.e., $\lambda_a =0$ to $\lambda_a = 1$. The other stages of the computer are unlocked in the next phase of the process. This is equivalent to the case of singlet molecule gas expansion, where the partitions are removed so that the molecule gets access to the whole volume of the system. Similarly, the Brownian particle undergoes a random walk and explores the full volume $mV$ of the configuration space. The entropy of the system is equivalent to Eq.~\eqref{smge4}. So, the entropy change of the system is

\begin{equation}\label{Wt1}
S = k_B \, \ln (mV) - k_B\, \ln V = k_B \, \ln\, m.
\end{equation}

This is the minimum amount of thermodynamic entropy required for the operation of the Brownian computer. The free energy for this model of Brownian computer is equivalent to Eq.~\eqref{smge5}.

\vspace{0.2in}
\begin{center}
\textit{With Trap}
\end{center}

\vspace{.15in}

\begin{figure}[h]
  \includegraphics[width=1.0\columnwidth]{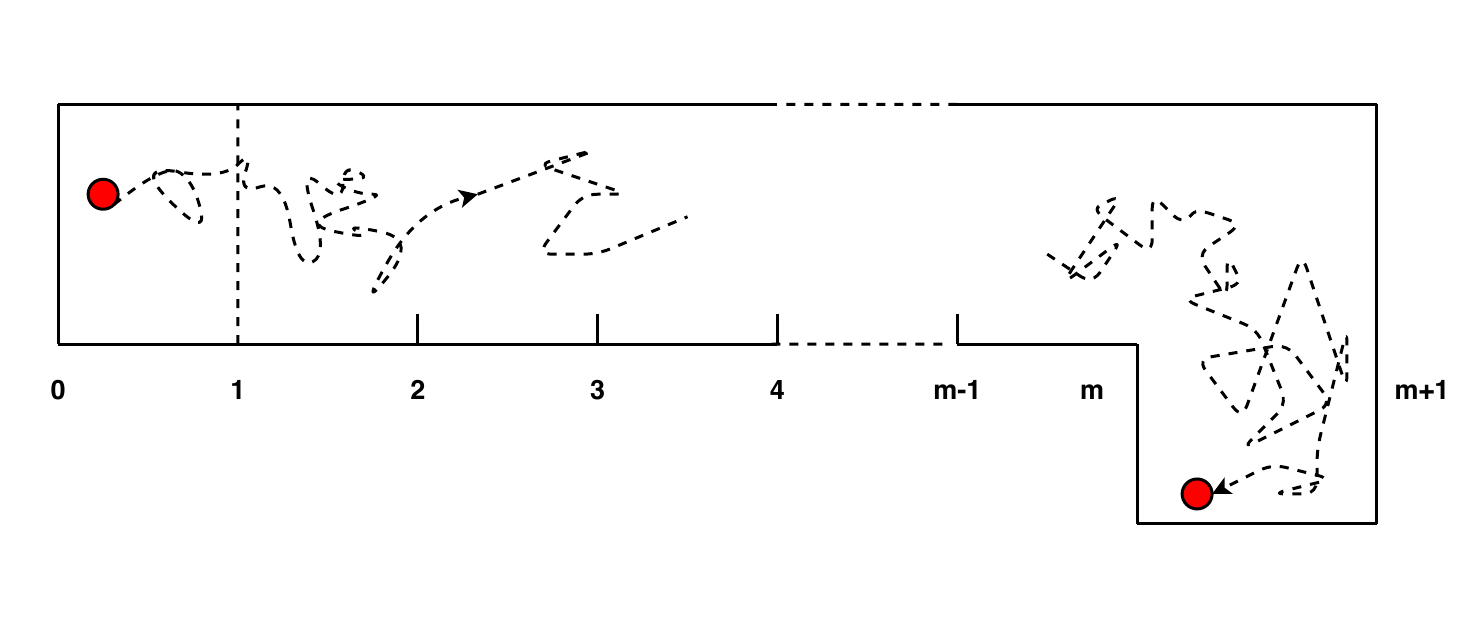}
  \caption{Schematic diagram of the Brownian computer with an energy trap.}
  \label{fig8}
  \end{figure}

In the model studied in the previous section the system at equilibrium is evenly distributed over the whole channel. This model is not computationally useful. The simple remedy to this model is to add an extra stage to the channel of the configuration space of the system, i.e., $\lambda_a = m$ to $\lambda_a = m+1$. This stage is what we call the trap stage for the system. The energy of the system during this stage will be $E_{trap}$. The energy $E_{trap}$ will be less than the position energy than that of the other stage of the system. So, unlike the case of the previous section, the energy of the system is now dependent on the spatial configuration. The schematic representation of this model is shown in Fig.~\ref{fig8}. So this stage represents the halting state of the computation in this model, i.e., when the computer reaches this state it defines the completion of the process of computation.

The Hamiltonian of the system thus changes from the previous model, and its form is described as 
\begin{equation}\label{WT2}
H = \mathbb{L} (p) + \Psi (\lambda_a),
\end{equation}
where
\begin{eqnarray} \nonumber
\Psi(\lambda_a)=\left\{
\begin{array}{lc}
  0, &\  0<\lambda_a < m ,\\
  - E_{trap}, & \  m < \lambda_a < m+1.
\end{array}\right.
\end{eqnarray} 

The Boltzmann distribution for this model similar to Eq.~\eqref{smge3} is expressed as 
\begin{equation}\label{WT3}
\Pi_a (\lambda_a,p) = \frac{exp(-H/k_BT)}{Z},
\end{equation}

where the partition function for this model is expressed as 

\begin{eqnarray}\label{WT4}\nonumber
Z & = & \int exp(-H/k_BT) dx dp, \\ \nonumber
& = & \int_{p} e^{(-\mathbb{L} (p)/k_BT)} dp \times \int_{\lambda_a= 0}^{m+1} e^{(-\Psi (\lambda_a)/k_BT)} V d\lambda_a, \\
& = & N_b V (m + e^{E_{trap}}/k_BT).
\end{eqnarray} 
Here the momentum contribution of the system is represented by $N_b = \int_{p} e^{(-\mathbb{L} (p)/k_BT)} dp$. The probability density that the state is in a specific stage  specified by the counter $\lambda_a$ can be described as

\begin{eqnarray} \label{WT5}
p(\lambda_a)=\left\{
\begin{array}{lc}
  \frac{1}{m + exp({E_{trap}}/k_BT)}, &\  0<\lambda_a < m ,\\
  \\
  \frac{ exp({E_{trap}}/k_BT)}{m + exp({E_{trap}}/k_BT)}, & \  m < \lambda_a < m+1.
\end{array}\right.
\end{eqnarray} 

The probability that the computer is at the trap stage is $P_a = \frac{1}{1 + m \, \, exp(-{E_{trap}}/k_BT)}$.

At equilibrium, state the entropy for this model can be calculated as
\begin{eqnarray}\label{WT6}\nonumber
S & = & \frac{\partial}{\partial T} (k_BT \, \ln Z),\\ \nonumber
& = & \frac{\partial}{\partial T} (k_BT \, \ln N_a) +\frac{\partial}{\partial T} (k_BT \, \ln [m + exp({E_{trap}}/k_BT)]) \\ \nonumber
& + & \frac{\partial}{\partial T} (k_BT \, \ln V), \\ \nonumber
& = & C_p (T) + k_B \,\ln V + k_B \,\ln [m + e^{({E_{trap}}/k_BT)}] \\
& - & \frac{P_a \,E_{trap}}{T}.
\end{eqnarray}

The momentum contribution gets absorbed in the first term of the Eq.~\eqref{WT6}. So one can calculate the entropy gain due to the inclusion of the trap to the system, and it is expressed as 
\begin{equation}\label{WT7} \nonumber
S_{add} = k_B \ln [m + e^{({E_{trap}}/k_BT)}] - \frac{P_a \,E_{trap}}{T}.
\end{equation}

In the case of the singlet molecule and the Brownian computer without a trap, we have encountered no change of entropy for the environment. But in this case, when the system reaches the halting state it releases energy to the environment. So the thermodynamic entropy of the environment gets increased by an amount of $S_{envr} = \frac{P_a \, E_{trap}}{T}$. So the total change in the entropy that occurs due to the process is 
\begin{equation}\label{WT7a} \nonumber
S_{total} = k_B \ln [m + e^{({E_{trap}}/k_BT)}].
\end{equation}

The free energy for this model of Brownian computer can be calculated as 
\begin{equation}\label{WT8} \nonumber
F = - k_BT \, \ln \, m - k_BT \, {({E_{trap}}/k_BT)}.
\end{equation}
The results are equivalent when derived using the definition $F= -k_BT\, \ln\, Z$. So the trap increases the amount of entropic force required to drive this model of the Brownian computer.

\subsection{Driven Brownian computer}

So far we have discussed about two models of Brownian computer. The model with a trap is sufficient enough to operate as a Brownian computer. But to speed up the process of computation we can add some external energy force to the system. Though it complicates the system, we get an advantage due to this driving energy. First, we will analyze the simpler version of this energy-driven model, i.e., we will consider the driven Brownian computer without an energy trap and then we will analyze the computer with an energy trap.

\vspace{0.2in}
\begin{center}
\textit{Without Trap}
\end{center}

\vspace{.15in}

In this section, we will explore the driven Brownian computer without an energy trap.  Similar to the previously studied models we will consider that the parameter $\lambda_a$ will keep track of the progress the system makes during the computation within the accessible channel of the confined phase space. For this model, we will consider the energy gradient to be a function of this spatial parameter $\lambda_a$. Let us assume that for each stage, the energy ramp is $\epsilon_{r}$. This can be pictorially represented similar to Fig.~\ref{fig7} with driven energy acting on the system. So the Hamiltonian of the models results to 
\begin{equation}\label{DWT1}
H =  \mathbb{L} (p) - \epsilon_r \, \lambda_a.
\end{equation}

The work of the energy ramp is to boost the computation and help to reach the halting state much earlier than previously studied models. 
The Boltzmann distribution for this model similar to Eq.~\eqref{smge3} is expressed as 
\begin{equation}\label{DWT2}
\Pi_a (\lambda_a,p) = \frac{exp(-H/k_BT)}{Z},
\end{equation}
where the partition function takes the form 
\begin{eqnarray}\label{DWT3}\nonumber
Z & = & \int exp(-H/k_BT) dx dp, \\ \nonumber
& = & \int_{p} e^{(-\mathbb{L} (p)/k_BT)} dp \times \int_{\lambda_a= 0}^{m} e^{(- \epsilon_r \,\lambda_a)/k_BT} V d\lambda_a, \\
& = & N_b\, V \,  (k_BT/\epsilon_r) \,\, (exp(\epsilon_r m/k_BT)-1).
\end{eqnarray} 

Similar to the previous models, we are not concerned about the momentum contribution which is depicted by $N_b$ in Eq.~\eqref{DWT3}. To calculate the probability density of the configuration space accessible to the system we integrate over the momentum contribution. It is expressed as 
\begin{equation}\label{DWT4}
p(\lambda_a) = (\epsilon_r/k_BT)  \,\, \times \frac{exp( \epsilon_r \lambda_a / k_BT )}{exp( \epsilon_r m / k_BT )-1}.
\end{equation}

With the help of the probability density defined in Eq.~\eqref{DWT4}, we can calculate the probability of the computer being in the halting state of the system 

\begin{eqnarray}\label{DWT5} \nonumber
P & = & \int_{m-1}^m p(\lambda_a) d\lambda_a, \\ \nonumber
& = & \frac{exp( \epsilon_r m / k_BT ) - exp( \epsilon_r (m-1) / k_BT )}{exp( \epsilon_r m / k_BT ) -1},\\ 
& \approx & 1- exp(-\epsilon_r /k_BT).
\end{eqnarray}
Here we have considered that $\epsilon_r m/ k_BT >> 1$. The energy ramp considered for this model can be expressed in terms of the success probability as $\epsilon_r = k_BT \, \ln \,(1/ (1-P))$. So if we want a high success probability $P$, we observe that the energy ramp results to a high value. The convention of the equilibrium condition of the Brownian computer conveys that the driving force $\epsilon_r$ should be less than $k_BT$. If the driving force is able to comply with this condition, then the Brownian computer spends most of its time in the final state (halting state). To satisfy this condition for this model we have to consider the energy ramp $\epsilon_r << k_BT$. So the probability $P$ takes the form $P = 1- exp(-\epsilon_r /k_BT) \approx \epsilon_r/k_BT$. 

The mean energy of the model which is associated with spatial configuration can be evaluated as 

\begin{eqnarray}\label{DWT6} \nonumber
E(m) & = & k_BT^2 \, \frac{\partial}{\partial T} \ln Z, \\ \nonumber
& = & E_{mom} (T) + k_BT - \frac{\epsilon_r m}{1- exp(- \epsilon_r m/k_BT)}, \\ 
\end{eqnarray}
where $E_{mom} (T)$ is the contribution that comes from the momentum. The thermodynamic entropy for this system for the halting state at equilibrium is expressed as 

\begin{eqnarray}\label{DWT7} \nonumber
S(m) & = & \frac{\partial}{\partial T} (k_BT \, \ln Z(m)),\\ \nonumber
& = & C_p(T) -\frac{ E_{mom}}{T} + k_B\, \ln\,V  + k_B\, \ln \Big(\frac{k_BT}{\epsilon_r}\Big)\\ \nonumber
& + & k_B\, \ln \,(exp( \epsilon_r m / k_BT ) - 1) + E(m)/T, \\ \nonumber
& = & k_B \ln \,V  + \frac{E(m)}{T} \\ \nonumber
& + &   k_B\, \ln \Big(\frac{k_BT}{\epsilon_r}\Big) + k_B\, \ln \,(exp( \epsilon_r m / k_BT ) - 1).\\
\end{eqnarray}
where we considered the momentum contribution $C_p (T)$ to be equivalent to $\frac{ E_{mom}}{T}$. Entropy for the initial state can be calculated by substituting $m=1$ in Eq.~\eqref{DWT7}. It results to 
\begin{eqnarray}\label{DWT8} \nonumber
S(1)& = & k_B \ln\,V  + k_B\, \ln \Big(\frac{k_BT}{\epsilon_r}\Big) + k_B\, \ln \,(exp( \epsilon_r / k_BT ) - 1) \\ \nonumber
& + & \frac{E(1)}{T}.
\end{eqnarray} 

So, the increase in the entropy of the computer when it transits from the initial stage of the process to the final stage of the computational process can be calculated as
 
\begin{eqnarray}\nonumber
S_{add} & = & k_B\, \ln \,(exp( \epsilon_r m / k_BT)-1) \\ \nonumber
& + & \frac{E(m)- E(1)}{T} - k_B\, \ln \,(exp( \epsilon_r / k_BT) -1).
\end{eqnarray}

Unlike the case of the Brownian computer without a trap, this model will transfer heat to the environment while the computation is executed. The entropy change that occurs in the environment can be described as 
\begin{equation}\nonumber
S_{envr} = -\frac{E(m)- E(1)}{T}.
\end{equation}

So, the total  change in the thermodynamic entropy of the system can be expressed as 
\begin{eqnarray}\label{DWT9} \nonumber
S_{total} & = & S_{add} + S_{envr},\\ 
& = & k_B \, \ln \left( \frac{exp( \epsilon_r m/ k_BT) -1}{exp( \epsilon_r / k_BT) -1} \right).
\end{eqnarray}

Now we describe the free energy of the model under analysis. The free energy can be evaluated as 
\begin{eqnarray}\label{DWT10}
F = - k_B T\, \ln \left( \frac{exp( \epsilon_r m/ k_BT) -1}{exp( \epsilon_r / k_BT) -1} \right).
\end{eqnarray}

The thermodynamic cost for the execution of the process is high as one requires the driven energy to be on the higher side for the higher probability of the computation to halt at its final state.

\vspace{0.2in}
\begin{center}
\textit{With Trap}
\end{center}

\vspace{.15in}

Now we will analyze the model where the Brownian computer will be driven by an external energy force along with an energy trap. So this model is the most complicated version of the models studied so far. The pictorial representation of this model is equivalent to Fig.~\ref{fig8} with the external force that drives the system for fast computation. The Hamiltonian of this model can be expressed as 
\begin{equation}\label{DWT11}
H = \mathbb{L} (p) + \Psi (\lambda_a),
\end{equation}
where
\begin{eqnarray} \nonumber
\Psi(\lambda_a)=\left\{
\begin{array}{lc}
  \epsilon_r \lambda_a, &\  0<\lambda_a < m ,\\
  -\epsilon_r m - E_{trap}, & \  m < \lambda_a < m+1.
\end{array}\right.
\end{eqnarray}

Following the procedures used to derive the partition function for the model under analysis, we can describe the partition function for the system is 
\begin{eqnarray}\label{DWT12} \nonumber
Z (m) & = & N_b\, V \,  (k_BT/\epsilon_r) \,\, (exp(\epsilon_r m/k_BT)-1),
\end{eqnarray}
and
\begin{equation}\nonumber
Z(m+1) = N_b V (m + e^{E_{trap}}/k_BT).
\end{equation}

So the partition function for the system is equivalent to the partition functions of the driven Brownian computer and with the model undriven Brownian computer with an energy trap. So the analysis of the thermodynamic entropy for the system can be equivalently executed from the previous two models. So we skip the detailed analysis for this model. The thermodynamic entropy to drive this model of the Brownian computer is the most out of all the models analyzed so far.

If we summarize the results of the thermodynamical analysis of the Brownian computer we can infer that the $m$ stage Brownian computer is actually a thermodynamically irreversible process with a minimum entropy amount $k \,\ln \,m$. With the addition of different parameters (like energy trap, and external energy force) to the systems, we observe additional thermodynamic entropy change in the system. According to Bennett, the Brownian computer is a reversible process in a sense equivalent to the Carnot engine. It is assumed that the system remains stable for all effects. So, according to the convention proposed by Bennett, every Brownian computer should operate in a thermodynamically reversible way. However, the thermodynamics analysis of the Brownian computer shows that it is equivalent to the process of singlet molecule expansion which represents an irreversible process. So, the Brownian computer executes a thermodynamically irreversible process when it explores the accessible space. During this process, the motion of the system is mechanically reversible, i.e., at any instantaneous moment, we will be unable to distinguish whether the system is moving in the forward or reverse direction. To get an understanding of the evolution of the system (i.e., whether the system is exploring the confined space or reverting back to the initial state), we have to observe the system for a long time.

Bennett misidentifies the Brownian computer as a reversible thermodynamic process due to the tracking of internal energy. Tracking of internal energy to analyze thermodynamic reversibility instead of thermodynamic entropy is not a sufficient criterion. The tracking of the total entropy of the system provides the necessary condition for the verification of whether a process is reversible or not. If the total entropy ($S_{system} + S_{environment}$) remains constant throughout the execution of the process, then it represents a thermodynamically reversible process. Bennett and Landauer didn't track the entropy of the Brownian computer which led to the misidentification of the Brownian computer as a thermodynamically reversible process.

\section{Miscellaneous} \label{sec22}
In this section, some disparate aspects of thermodynamic computation are collated here.

\subsection{Engine as computer}
The study of thermal machines is a key aspect of thermodynamics, both in the classical~\cite{carnot1872reflexions,martini1983stirling,walker1985free,walker1985hybrid,van1985fundamentals,reed1898thermo,barton2019ericsson,scovil1959three,szilard1929entropieverminderung} and quantum regimes~\cite{kim2011quantum,kosloff2013quantum,rossnagel2016single,santos2023pt,martinez2016brownian,chattopadhyay2019relativistic,uzdin2015equivalence,chattopadhyay2021quantum,chattopadhyay2020non,santos2023pt,chattopadhyay2021bound,pandit2021non}.
In a simple sense, one can convey that the Carnot heat engine extracts an amount of heat $Q_{hot}$ from the hot reservoir at temperature $T_{hot}$ and transfers an amount of heat $Q_{cold}$ to the sink at temperature $T_{cold}$. The work done to execute this process is $W_{hot\rightarrow cold} = Q_{hot} - Q_{cold}$. One will observe optimal efficiency if 
\begin{equation}\nonumber
\mathcal{L}_{hot}-\mathcal{L}_{cold} = 0,
\end{equation}
 where $\mathcal{L}_{hot} = - Q_{hot}/T_{hot}$ is the negentropy that is imprinted in the engine. So the loss that occurs in the process is just throwing away the negentropy of amount $Q_{cold}/T_{cold}$. Now we move on to analyze the computer equivalently as a Carnot cycle as shown in the work~\cite{costa1989computer}. We know from the work~\cite{brillouin2013science,prigogine1985self}, that one encounters zero work balance for an ideal computer and the amount of information that is delivered during a process is $ \ln 2 \, dl= \mathcal{L}_{hot} - \mathcal{L}_{cold}$ where $l$ denotes the amount of information. 
 
 So in the case of an ideal computer, we also encounter a loss by throwing out an equivalent amount of negentropy as in the case of the Carnot cycle.

\subsection{Nonergodic systems and its thermodynamics}
A memory device can be designed using a multistable system applied to the condition that, the transition between the states is not allowed. Alternatively, we can say that information can be stored only in a nonergodic system where the `time average' of the system differs from the `phase space average' of the system. In the work~\cite{ishioka2001thermodynamics}, the authors have analyzed the thermodynamics of computation with the notion that information is recorded in a multistable system. These multistable states are considered as the nonergodic state.

\begin{figure}[h]
  \includegraphics[width=1.0\columnwidth]{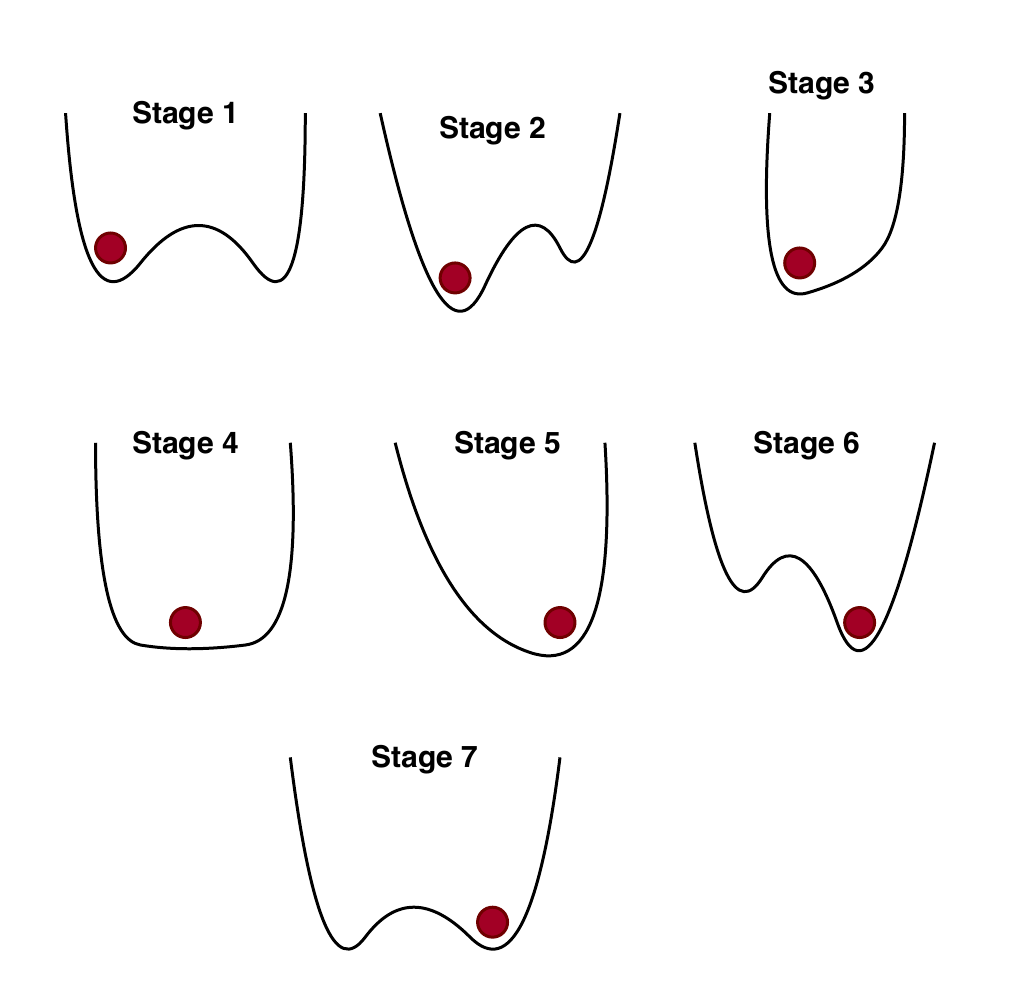}
  \caption{A schematic representation of the thought experiment step by step is depicted. The red solid sphere is the single atom of the system}
  \label{fig15}
  \end{figure}

We will interpret the memory register in terms of thermodynamics. According to the Landauer principle, the system will confront a decrease in entropy due to `restore to one' (RTO) operation. Whereas in the work~\cite{ishioka2001thermodynamics}, the authors have proved that there is no change in the thermodynamic entropy even after the RTO operation. The Clausius definition of thermodynamic entropy for a system in two states $ST_a$ and $ST_b$ is expressed as 
\begin{equation}\nonumber
\Delta S = \int_{ST_a}^{ST_b} dQ/T,
\end{equation} 
where $dQ$ represents the heat change in the system and $T$ is the temperature of the system. So to prove the statement proposed in the work~\cite{ishioka2001thermodynamics} the author considered a thought experiment. The thought experiment is that we will consider a system where the particle is in a bistable-monostable potential well. This system will interact with a heat reservoir.  Goto and his co-workers in their work~\cite{shimizu1989new,goto1996study} coined this model as ``quantum flux parametron". For the analysis, the state of the system will be described as one when the particle is found on the right side of the potential well and zero when one finds the particle to be on the left-hand side of the well. The thought experiment is as follows:
\begin{itemize}
\item The system is considered to be initiated in the state zero. This will be the initial state of the system. The entropy of the system is $S_{zero}$. 

\item The left well of the system is lowered by applying a bias potential. 

\item Now the `double-well potential' is converted to a single well.  This is executed by removing the partition between this well.

\item In the next stage of the process, we will remove the bias potential that was applied to the left well of the system. The system is now in a neutral state and we denote it as $ST_{ne}$. The state of the system of being in one or zero is denoted by $ST_{m}$.  

\item  The bias potential is again applied, but now on the right side of the single potential well. This lowers the right side of the well.

\item In the next stage of the protocol, we then revert back the single potential well into the double potential well just by recreating the partition between the left and right sides of the well. 

\item In the final stage of the process, we remove the bias potential. The system is now in state one, and the entropy of the system is $S_{one}$. 
\end{itemize}
The schematic representation of the thought experiment is shown in Fig.~\ref{fig15}.

The first four steps of the thought experiment are called the erasing process, and the final three steps represent the writing process. If we apply RTO to the system, and the system is in the state zero, we will observe the same configuration even after the RTO operation ($S_{zero}= S_{one}$), subjected to a condition that the state is known before the execution of this experiment.

From an information point of view, entropy is defined as a measure of lack of information, i.e., the lesser the information we gather about the system the higher the entropy of the system. So, the entropy of the unknown state $ST_{ukn}$ should have larger entropy than that of the known state. Based on this notion, Landauer argued that the system will observe a decrease in entropy after RTO operation.

It is a well-known fact that when a computer runs repeated recordings and erasure occurs in a memory device. The physical equivalence of this process can be thought of as a sequence of evolution of the system from an ergodic state to a nonergodic state, and vice versa. Now we will analyze the thermodynamics of this evolution using two different models. The first model that we are going to analyze is the Szilard engine model, and the second one is the bistable-monostable potential well.

The Szilard engine with one atom is the widely studied model for a memory device. This model helps us to understand the physics behind the computation. The Szilard engine model is associated with a heat reservoir at temperature $T$. When the system has the partition wall the state of the system is in nonergodic state $ST_{nerg}$ and it will be in the ergodic state $ST_{erg}$ when one removes the partition. Now we will describe the writing process for this model. The initial state of the system is considered as $ST_{erg}$. Now the partition is inserted in the middle of the chamber, so work is done on the atom in the system. The work done can be expressed as
$ \langle W_{writing} \rangle = k_B T \, \ln 2$ and the heat that is generated to execute the process of writing is

\begin{equation}
\langle Q_{writing} \rangle= T(S_{ST_{erg}} - S_{ST_{nerg}}) = k_B T \, \ln 2.
\end{equation}
\vspace{.02in}

The process is executed a large number of times to evaluate the expectation of the work done and the heat generation during the execution of the process.

Along with the writing process, the system goes through the erasure process. For the writing process, we have two types of erasure processes. One is reversible in nature and the other is irreversible in its form. For the reversible process, the work done and the creation of heat in the surroundings can be expressed as $\langle W_{erasure} \rangle = - \langle W_{writing} \rangle$ and $\langle Q_{erasure} \rangle = - \langle Q_{writing} \rangle$. So for the process, where one executes the writing process as well as the erasure process using the reversible form, one can infer that the total work is done and the heat generation is zero. One will be able to execute the reversible erasure process if the state of the system is known before the process is executed.

The second model of the erasure process is the irreversible process. The initial state of the system is now not known to us. To execute this process one removes the partition from the chamber without any prior information about the state of the memory. The removal of the partition evolves the state of the system from the nonergodic state to the ergodic state. The work done and the heat generation for this erasure process is zero but we encounter an increase in the entropy of the system. The increase of this entropy can be evaluated as $S_{erg} = k_B \ln 2$. So for the total process of writing and as well as the erasure process, the total work done and the heat production can be expressed as 
\begin{equation}\label{ERGD1}
\langle W_{tirr} \rangle = \langle Q_{tirr} \rangle = k_B \, \ln 2,
\end{equation}
where $\langle W_{tirr} \rangle$ and $\langle Q_{tirr} \rangle$ represents the work and heat for the total process.

Now we will explore the second model, i.e., the bistable-monostable potential well model. In this process, the partition of the Szilard engine gets replaced by a potential wall. The writing process and the erasure process for this model are described in the thought experiment. For the writing process, the amount of work done and the heat production can be evaluated as 
\begin{eqnarray}\nonumber
\langle W_{writing} \rangle & = & k_B T \ln \left( \frac{Z_{ST_{ne}}}{Z_{ST_{m}}} \right), \\ \nonumber
\langle Q_{writing} \rangle & = & E_{ST_{ne}} - E_{ST_{m}} + k_B T \ln \left( \frac{Z_{ST_{ne}}}{Z_{ST_{m}}} \right),\\
\end{eqnarray}
where $E$ and $Z$ represent the energy and partition function of the system. The partition  function for the system can be evaluated as 
\begin{eqnarray}\nonumber
Z_{ST_{ne}} & = & \int_{-\infty}^{\infty} e^{-V_{ne}/kT} dx, \\ \nonumber 
Z_{ST_{m}} & = & \int_{0}^{\infty} e^{-V_{m}/kT} dx,
\end{eqnarray}
where $V_{ne}$, $V_{m}$ describes the potential energy of the double and the single potential well. The partition function of the system is defined based on the assumption that we have considered the nonergodicity of the system. The potential wall should be chosen in such a way that it is greater than $k_B T$. The potential wall is chosen in this manner to reduce the error during the operation.

The erasure process considered for this model is the irreversible one. The irreversible erasure process is executed by diminishing the potential wall height (i.e. reducing the height of the hill), and along with that, we will not use the bias potential in this process. Due to this lowering of the height of the wall, we encounter entropy production $S_{ern}$ in the system by an amount of $k_B \ln 2$. The work done and the heat production for the execution of the erasure process can be expressed as 
\begin{eqnarray}\nonumber
\langle W_{erasure} \rangle & = & - \langle W_{writing} \rangle + TS_{ern}, \\ \nonumber
\langle Q_{erasure} \rangle & = & - \langle Q_{writing} \rangle + TS_{ern}.
\end{eqnarray}
 So for the total process of writing and as well as the erasure process, the total work done and the heat production are equivalent to the results (Eq.~\eqref{ERGD1}) obtained in the irreversible process of the Szilard model.

 \subsection{Thermodynamics of algorithm}
It is generally believed that the advent of quantum computers will help to solve some age-old problems in number theory, and physical as well as combinatorial search faster than the existing classical computers. To get a better understanding of quantum speedups, one has to consider a realistic model of computation. The realistic model should consider the time complexity and time-space tradeoffs during analysis. Some works~\cite{banegas2018low,beals2013efficient,bernstein2009cost,fluhrer2017reassessing} have analyzed these types of models. A recent work~\cite{perlner2017thermodynamic} in this direction has analyzed the quantum speedups from a thermodynamic viewpoint. To analyze the cost of the algorithm (in classical and quantum regimes) from the thermodynamic perspective Brownian model of computation is considered.

For the thermodynamic interpretation of algorithms, the authors in their work~\cite{perlner2017thermodynamic} have considered collision finding algorithm and preimage search for their analysis. Before we move on to the thermodynamic analysis of the algorithm, we will describe the algorithms briefly. First, we will learn about the collision finding algorithm and then about the preimage search algorithm.

The parallel collision search algorithm proposed by Van Oorshot and Wiener~\cite{van1999parallel} happens to be the best classical collision finding algorithm. This algorithm can detect a collision in an expected serial depth of $\mathcal{O} (\sqrt{G}/G_{\alpha})$, where $G$ denotes the range of the function and $G_{\alpha}$ depicts the parallel processes with memory $\mathcal{O} (1)$. It is assumed that the communication cost among the thread is negligible than the overall computation costs. Brassard, Hoyer, and Tapp (BHT) extended this algorithm in the quantum realm in their work~\cite{brassard1998quantum}. The operations for this algorithm is $\mathcal{O} (G^{\frac{1}{3}})$ with memory size $G^{\frac{1}{3}}$. One can further generalize this algorithm to a parallel algorithm with $n_\beta$ parallel processors having memory size $G_\beta$. This generalized algorithm satisfies this condition $n_\beta < G_\beta < \mathcal{O} \Big( (Gn_\beta)^{\frac{1}{3}}\Big)$ with a serial complexity $\mathcal{O} \Big(\sqrt{\frac{G}{G_\beta n_\beta}} \Big)$.

Bernstein has shown in their work~\cite{bernstein2009cost} that the BHT algorithm, when measured in terms of memory and serial depth does not encounter any improvement over the classical algorithm, but one encounters an improvement in terms of query complexity. Giovanetti in their work~\cite{giovannetti2008quantum}, counters about the memory cost by proposing a quantum RAM model, where they convey that memory access operation can be executed at logarithmic energy cost despite of large gate complexity. So the question remains whether one can propose a realistic model where one encounters improvement of the complexity of the quantum algorithm over the classical one. Now we will see whether the Brownian model of computation provides a solution to this proposed problem. To do so, we will first analyze the quantum version of the algorithm. For a given time $t$ and memory size $G_\beta$, we are going to calculate the total energy that is required for the collision search. We will consider that in the BHT algorithm, the energy complexity is subjugated by oracle queries over the memory access. So the energy  for per operation is $\mathcal{O} \Big(\frac{\Big(\sqrt{\frac{G}{G_\beta n_\beta}}\Big)}{t} \Big)$. So the total energy for the process can be expressed as
\begin{equation}\nonumber
E_{quantum} = \mathcal{O} \Big(n_\beta \times  \sqrt{\frac{G}{G_\beta n_\beta}} \times \frac{\Big(\sqrt{\frac{G}{G_\beta n_\beta}}\Big)}{t}\Big) = \mathcal{O} \Big(\frac{G G_\beta}{t} \Big).
\end{equation} 
Now  if we consider the classical case the energy per operation is $\mathcal{O} (\frac{\sqrt{G}}{G_\beta t })$, and the total energy for the process is 
\begin{equation}\nonumber
E_{classical} = \mathcal{O} \Big(G_\beta \times \frac{\sqrt{G}}{G_\beta t } \times \frac{\sqrt{G}}{G_\beta } \Big) = \mathcal{O} \Big( \frac{G}{G_\beta t} \Big).
\end{equation}

So we can infer from the results that this realistic model does not provide any advantage of the quantum algorithm over the classical one. So, it requires further investigation to verify whether any realistic model will provide an advantage of the quantum algorithm over the classical algorithm. One can think of applying a similar analysis to the Claw Finding problem. Its objective is to find the collision between two functions with different domain sizes, e.g., $Do_1$ and $Do_2$. The quantum version of this algorithm was explored in the work~\cite{tani2007improved}. The energy cost for finding the collision in the quantum regime is $\mathcal{O} \Big(max \,\Big( \frac{Do_1\, Do_2}{G_beta t}, \frac{Do_1}{t}, \frac{Do_2}{t} \Big)  \Big)$, wherein the case of classical it is $\mathcal{O}\Big(   \frac{(Do_1 + Do_2)^3}{G_\beta^2 t}\Big)$.

Now we will analyze the second algorithm, i.e., the pre-image algorithm.  One can find preimages of a function having a domain size $N_\gamma$ using Grover's algorithm. The serial complexity for this algorithm is $\mathcal{O} (N_\gamma)$. An optimal serial complexity can be obtained if one considers $M_\gamma$ parallel processes with memory $\mathcal{O} (1)$. The serial complexity of this optimal model is $\mathcal{O} \Big(\sqrt{\frac{N_\gamma}{M_\gamma}} \Big)$. This was verified in the work~\cite{zalka1999grover}. If one implements Grover's algorithm with the Brownian computation model the energy per operation is $\mathcal{O} \left( \frac{\sqrt{\frac{N_\gamma}{M_\gamma}}}{t} \right)$, and the total energy of the process for this quantum version is 
\begin{equation}\nonumber
E_{quantum} = \mathcal{O} \left( \frac{\sqrt{\frac{N_\gamma}{M_\gamma}}}{t} \times M_\gamma \times \sqrt{\frac{N_\gamma}{M_\gamma}} \right) = \mathcal{O} \Big(\frac{N_\gamma}{t} \Big).
\end{equation}

If we naively apply the Brownian computation for the classical search the total energy for the process boils down to $E_{classical} = \mathcal{O} \Big(\frac{N_\gamma^2}{M_\gamma t} \Big)$. To improve the complexity of the process, one allows the Brownian motion to steer the system on a random work. If one implements the computation with $M_\gamma$ parallel processes, then the complexity of the algorithm becomes equivalent to that of the quantum version. So we encounter the same asymptotic performance for the algorithm in the classical and quantum regimes.

If one compares Grover's algorithm with the classical search algorithms instead of quantum versus classical collision search, then one will encounter that Grover's algorithm is more efficient than the classical version (where the powered and unpowered Brownian computation model is considered for the analysis). If the assumption of scale invariance for memory size and the energy consumption is removed for unpowered primage search, then one encounters that the preimage algorithm is memory intensive. If one considers Oracle queries then also the unpowered preimage algorithm happens to be less significant than Grover's algorithm.

The results that have been discussed so far suggest that the cost of hardware happens to be the prime factor that determines the advantage of the algorithms. If the cost of the hardware can be reduced, then one can even use the classical search algorithms over Grover's algorithm. One can even encounter that unpowered classical preimage search surpasses Grover's algorithm provided that the memory costs are very close to the fundamental thermodynamic limit. These results also guide us in the field of post-quantum cryptography, especially in choosing the key size of the block ciphers. It can be inferred from this analysis that doubling the key size is unnecessary. Instead of that one can make a small increase from 128 to 192 bits for better protection.

\section{Conclusion and Future prospects}\label{sec24}
The analysis of the computation process from the thermodynamic viewpoint has been one of the central points of attraction for researchers of physics as well as computer science. It was initiated from the physical Church-Turing thesis, where they conveyed that every computational process is physical. Various approaches have been considered for the analysis of the different computational processes. With the advent of modern statistical theory, the research in this area was boosted. Not only in the field of computer science we also encounter its application in different fields from chemical networks~\cite{chen2014deterministic,soloveichik2008computation,murphy2018synthesizing,qian2011scaling}, molecular biology~\cite{prohaska2010innovation,benenson2012biomolecular} and even in neurology~\cite{laughlin2001energy,balasubramanian2001metabolically}.

Further investigation in this direction is required to get a better understanding of the bond between thermodynamics and the computational process. For example, in the case of finite automata, one can investigate the maximum thermodynamic cost that is required to accept a language for an automata.  Also one can calculate the minimal cost for any deterministic automata. One can also work on developing a theory to analyze the non-deterministic finite-state automata in terms of thermodynamics. Models to describe the complex Turing machine, and also network theory from the thermodynamic viewpoint is an open area of research.

Error correction is an important part of any computational process. The analysis of these protocols from a thermodynamic viewpoint is at their baby stage. Modeling of the error correction models by the physical system to explain it thermodynamically needs further investigation. In a recent work~\cite{landi2020thermodynamic},  the authors have shown that there is a similarity between the quantum heat engine and quantum error correction codes. They have strengthened their intuition by making a complete analysis of the thermodynamic properties of the quantum engine-based error correction codes. So the thermodynamic approach to explain the error correction is an open book to read.

Different systems have different heat signatures. One can utilize this property of the system for various purposes such as for security in cryptographic protocols. So one can explore the communications protocols and crypto-systems from a thermodynamic viewpoint. Algorithms in the form of a search algorithm from a thermodynamic viewpoint have already been analyzed. Further exploration in this direction is an open area of research. Thermodynamic analysis of quantum computations needs a rigorous investigation for a better understanding of quantum computers and to develop hardware with lower-cost functions.

%

\end{document}